\documentclass[11pt]{article}
\usepackage{amsfonts,amsmath,amssymb}
\usepackage{enumerate}
\usepackage[hidelinks]{hyperref}
\usepackage{bbm}
\usepackage{nicefrac}
\usepackage[all]{xy}
\usepackage{graphicx}
\usepackage{bm}
\usepackage{makecell}
\usepackage[table]{xcolor}

\usepackage{upgreek}
\usepackage{slashed}

\usepackage{float}			
\usepackage{lscape}                 

\usepackage{booktabs}

\newcommand{\be}{\begin{equation}}
\newcommand{\ee}{\end{equation}}

\newcommand{\bea}{\begin{eqnarray}}
\newcommand{\eea}{\end{eqnarray}}

\newcommand{\bes}{\begin{subequations}}
\newcommand{\ees}{\end{subequations}}

\newcommand{\cN}{{\cal N}}

\def\sst#1{{\scriptscriptstyle #1}}

\def\0{{\sst{(0)}}}
\def\1{{\sst{(1)}}}
\def\2{{\sst{(2)}}}
\def\3{{\sst{(3)}}}
\def\4{{\sst{(4)}}}
\def\5{{\sst{(5)}}}
\def\6{{\sst{(6)}}}
\def\7{{\sst{(7)}}}
\def\8{{\sst{(8)}}}

\def\cV{{{\cal V}}}
\def\cM{{{\cal M}}}

\allowdisplaybreaks  

\usepackage{multirow}
\usepackage{rotating}

\newcommand{\ba}{\begin{align}}
\newcommand{\ea}{\end{align}}

\newcommand{\bse}{\begin{subequations}}
\newcommand{\ese}{\end{subequations}}

\allowdisplaybreaks

\usepackage{lipsum}

\newlength\Colsep
\begin{document}

\makeatletter
\renewcommand{\theequation}{\thesection.\arabic{equation}}
\@addtoreset{equation}{section}
\makeatother

\begin{titlepage}

\begin{flushright}
%
%
\end{flushright}

\vspace{5pt}

   \begin{center}
   \baselineskip=16pt

   \begin{Large}\textbf{
\hspace{-18pt} Consistent subsectors of maximal supergravity \\[8pt]
and wrapped M5-branes
}
   \end{Large}

\vspace{25pt}

{\large Martín Pico$^{1,2}$ and  Oscar Varela$^{3,1}$ }

\vspace{30pt}

	\begin{small}

	{\it $^1$  Instituto de F\'\i sica Te\'orica UAM/CSIC, 28049 Madrid, Spain} 

	\vspace{12pt}

	{\it $^2$  Departamento de F\'\i sica Te\'orica, Universidad Aut\'onoma de Madrid, \\
	Cantoblanco, 28049 Madrid, Spain} 

	\vspace{12pt}
          
   {\it $^3$ Department of Physics, Utah State University, Logan, UT 84322, USA} 
		
	\end{small}

\vskip 70pt

\end{center}

\begin{center}
\textbf{Abstract}
\end{center}

\begin{quote}

A new family of $D=4$ $\mathcal{N}=8$ gauged supergravities is introduced, consisting in a mixture of Scherk-Schwarz and dyonic CSO gaugings that involves the trombone scaling symmetry. A specific theory in this class is shown to admit supersymmetric anti-de Sitter vacua, argued to be related to various wrapped M5-brane configurations. The mass spectrum of these vacua within our maximal theory is computed, extending previous results in smaller gauged supergravity models, and shown to exhibit some exotic features due to the trombone gauging. These vacua are obtained as solutions of certain subsectors of our $\mathcal{N}=8$ supergravity. Such subsectors in turn arise by consistent truncation of the parent maximal supergravity to the invariant fields under specific groups that are not contained in the $\mathcal{N}=8$ gauge group. Prompted by these and other similar subtruncations recently considered, we formalise sufficient  conditions that allow any maximal supergravity to be truncated consistently to invariant subsectors thereof when the invariance group is not necessarily contained in the $\mathcal{N}=8$ gauge group.

\end{quote}

\vfill

\end{titlepage}

\tableofcontents



\section{Introduction}


Lower-dimensional gauged supergravity provides an extremely helpful formalism to model applications of string theory to cosmology, model building or holography in a concrete, hands-on set-up. Using supergravities with maximal supersymmetry for that purpose, as opposed to submaximally supersymmetric models, comes with many benefits and a few difficulties. On the one hand, the advantages of using gauged supergravity in the first place are maximised when computing certain observables. For example, the spectrum of masses about a given vacuum is necessarily complete in gauged supergravity when that vacuum is considered within a maximal model. On the other hand, the maximal supergravity multiplet is very large. For that reason, in many practical applications, as in precisely the search for vacua, one often ends up truncating a maximal theory to more manageable, smaller subsectors, anyway. These considerations apply regardless of whether the gauged supergravity in question, maximal or not, descends itself as a consistent truncation of one of the ten- or eleven-dimensional supergravities. This is particularly so in the context of holography, where multiple top-down or bottom-up gauged supergravity models in various dimensions and with various amounts of supersymmetry have been fruitfully used since the early days of the anti-de Sitter/conformal field theory (AdS/CFT) correspondence \cite{Maldacena:1997re}. 

In the latter context, some gauged supergravities in five and four dimensions, maximal \cite{Gunaydin:1985cu,Pernici:1985ju,deWit:1982ig,deWit:1981eq} or otherwise, have helped develop aspects of the large-$N$ AdS/CFT correspondence for two of the superconformal branes of string/M-theory, the D3- and the M2-brane, whose holography is well established by now \cite{Maldacena:1997re,Aharony:2008ug}. In contrast, the holographic status for the third superconformal brane, the M5-brane, is mixed to date. A complete description of the six-dimensional $(2,0)$ superconformal field theory defined on a stack of $N$ M5-branes in flat space remains an open question. However, placing  the M5-branes on special holonomy manifolds instead, and letting them wrap certain two- or three-dimensional submanifolds therein \cite{Maldacena:2000mw} (see also \cite{Acharya:2000mu,Gauntlett:2006ux}, reference \cite{Gauntlett:2003di} for a review, and \cite{Ferrero:2020laf,Bah:2021mzw,Ferrero:2021wvk} for more recent related developments), is known to lead the worldvolume theory to infrared four- and three-dimensional superconformal phases with various amounts of supersymmetry, some of which are very well understood. In four dimensions, these include $\cN=2$ class ${\cal S}$ \cite{Gaiotto:2009we,Gaiotto:2009gz} and $\cN=1$ \cite{Bah:2011vv,Bah:2012dg} phases and, in three dimensions, $\cN=2$ class ${\cal R}$ \cite{Dimofte:2011ju} phases. 

Various $D=5$ submaximal gauged supergravity models related to the M5-brane AdS$_5$/CFT$_4$ correspondences of \cite{Maldacena:2000mw,Gaiotto:2009we,Gaiotto:2009gz,Bah:2011vv,Bah:2012dg}, have been known for some time \cite{Szepietowski:2012tb,MatthewCheung:2019ehr,Cassani:2019vcl,Faedo:2019cvr,Cassani:2020cod}. More recently, a maximal, $\cN=8$, five-dimensional gauged supergravity has been constructed \cite{Bhattacharya:2024tjw,Varela:2025xeb} that encompasses those submaximal models \cite{Szepietowski:2012tb,MatthewCheung:2019ehr,Cassani:2019vcl,Faedo:2019cvr,Cassani:2020cod} as consistent subsectors. In fact, the $D=5$ $\cN=8$ gauged supergravity of \cite{Bhattacharya:2024tjw,Varela:2025xeb} provides an example of the dichotomy mentioned above that this type of maximal models pose. On the one hand, it proved pivotal to obtain the complete, universal, large-$N$ light operator spectrum of class ${\cal S}$ theories \cite{Bhattacharya:2024tjw}. Well understood as this type of superconformal field theories is, this was a still missing crucial aspect beyond reach of the submaximal gauged supergravity models of \cite{Szepietowski:2012tb,MatthewCheung:2019ehr,Cassani:2019vcl,Faedo:2019cvr,Cassani:2020cod}. On the other hand, the relevant AdS vacua within the larger $\cN=8$ theory were recovered in \cite{Bhattacharya:2024tjw,Varela:2025xeb} by, precisely, truncating it first to the smaller models of \cite{Szepietowski:2012tb,MatthewCheung:2019ehr,Cassani:2019vcl,Faedo:2019cvr,Cassani:2020cod}.

As for the M5-brane-related AdS$_4$/CFT$_3$ examples of \cite{Acharya:2000mu,Gauntlett:2006ux,Dimofte:2011ju}, some associated submaximal gauged supergravity models in $D=4$ dimensions are also known \cite{Gauntlett:2002rv,Donos:2010ax}. In light of the above $D=5$ case, one may wonder whether a similar maximal, $\cN=8$, gauged supergravity exists in $D=4$  that contains those submaximal models \cite{Gauntlett:2002rv,Donos:2010ax} as consistent subsectors. The answer is positive, and we indeed present such $D=4$ $\cN=8$ gauged supergravity in section \ref{sec:New4DSugras}. In fact, this particular theory is just an instance, $\textrm{TCSO} (5,0,0; \textrm{V})$, of a more general family of $D=4$ $\cN=8$ supergravities with gauge group that we dub $\textrm{TCSO} (p,q,r; \mathrm{N})$. Here, $p$, $q$, $r$ are non-negative integers constrained as $p+q+r = 5$, and $\mathrm{N}$ is a Roman numeral, possibly with a subindex. These $\cN=8$ theories can be regarded as mixtures of specific dyonic CSO gaugings \cite{DallAgata:2011aa,Dall'Agata:2014ita}, and a Scherk-Schwarz gauging \cite{Scherk:1979zr} involving a three-dimensional group $G_3$ of Bianchi type $\mathrm{N}$. For unimodular $G_3$, the $\textrm{TCSO} (p,q,r; \mathrm{N})$ theories reduce to concrete dyonic CSO gaugings in the class of \cite{DallAgata:2011aa,Dall'Agata:2014ita}. If $G_3$ is non-unimodular, the $\textrm{TCSO} (p,q,r; \mathrm{N})$ theory is new, and involves a gauging of the trombone scaling symmetry of ungauged $D=4$ $\cN=8$ supergravity.

We have constructed our $D=4$ $\cN=8$ $\textrm{TCSO} (5,0,0; \mathrm{V})$-gauged supergravity by consistent truncation of $D=11$ supergravity on the seven-dimensional internal space of the AdS$_4$ solutions of \cite{Acharya:2000mu,Gauntlett:2006ux}. The latter describe the near-horizon regime of M5-branes wrapped on various negatively curved submanifolds of special holonomy manifolds. We will present the details of this dimensional reduction elsewhere \cite{Pico:2026rji}. Here, we will focus instead on describing the $\textrm{TCSO} (p,q,r; \mathrm{N})$ theories from a strictly four-dimensional perspective, setting them in the general $\cN=8$ gauged supergravity formalism of \cite{deWit:2007mt,LeDiffon:2008sh,LeDiffon:2011wt} (see \cite{Trigiante:2016mnt} for a review). Our $\textrm{TCSO} (p,q,r; \mathrm{N})$ supergravities indeed feature some interesting characteristics that merit this detailed four-dimensional approach. For example, $\textrm{TCSO} (5,0,0; \textrm{V})$ is, to our knowledge, the first trombone-gauged maximal four-dimensional supergravity with an AdS vacuum. As discussed in \cite{LeDiffon:2011wt}, the trombone contribution to the scalar potential is positive, which tends to favour de-Sitter, or Minkowski, vacua in these models.

Short of providing here the full maximally-supersymmetric consistent truncation from $D=11$ to our $\textrm{TCSO} (5,0,0; \textrm{V})$ model \cite{Pico:2026rji}, we will nevertheless offer evidence that the latter is indeed related to the M5-brane configurations of \cite{Acharya:2000mu,Gauntlett:2006ux}. We will do this by showing, in section \ref{eq:AdSvac}, that the submaximal $D=4$ gauged supergravities of \cite{Gauntlett:2002rv,Donos:2010ax}, themselves originally obtained from $D=11$ consistent truncation on the M5-brane solutions of \cite{Acharya:2000mu,Gauntlett:2006ux}, are indeed contained in our $\cN=8$ $\textrm{TCSO} (5,0,0; \textrm{V})$ theory as consistent subsectors. We will retrieve these two smaller models \cite{Gauntlett:2002rv,Donos:2010ax} as SO(3)-invariant sectors of our $\textrm{TCSO} (5,0,0; \textrm{V})$ theory, with SO(3) embedded into the (compact subgroup, SU$(8)$, of the) duality group, $\mathbb{R}^+ \times \mathrm{E}_{7(7)}$, of $D=4$ $\cN=8$ ungauged supergravity \cite{Cremmer:1979up} in two different ways. In our four-dimensional language, the $D=11$ M5-brane solutions of \cite{Acharya:2000mu,Gauntlett:2006ux} manifest themselves as supersymmetric AdS vacua of these SO(3)-invariant subsectors. We will recover those vacua and compute their mass spectrum within the full $\cN=8$ $\textrm{TCSO} (5,0,0; \textrm{V})$ theory, thereby extending previous spectral results \cite{Gauntlett:2002rv,Donos:2010ax}. These spectra exhibit certain exotic features due to the trombone gauging, to be discussed in section \ref{eq:AdSvac}.

Starting with \cite{Warner:1983vz,Warner:1983du}, a time-honoured approach to obtain consistent subsectors of maximal gauged supergravity has been to impose invariance under some group $G_S$. Until recently, all examples of this construction assumed $G_S$ to be contained in the gauge group of the parent $\cN=8$ theory. This has led over time to the folkloric tacit assumption that  $G_S$ should meet that condition for the method to work. However, this is not necessarily the case. As the recent example of \cite{Guarino:2024gke} (see also \cite{Bhattacharya:2024tjw,Varela:2025xeb}) makes manifest, the truncation of a maximal supergravity down to a $G_S \subset \textrm{SU}(8)$ invariant sector may still go through consistently even if $G_S$ is not a subgroup of the $\cN=8$ gauge group. In essence, the constructions of \cite{Guarino:2024gke,Bhattacharya:2024tjw,Varela:2025xeb} replicate within $D=4$ and $D=5$ $\cN=8$ gauged supergravities the supersymmetry-breaking consistent dimensional reductions of \cite{Cassani:2019vcl,Cassani:2020cod}. The latter naturally implement $G_S$ invariance by using the language of generalised $G_S$-structures \cite{Cassani:2019vcl} in exceptional generalised geometry \cite{Coimbra:2011nw,Coimbra:2011ky,Coimbra:2012af}, a duality-covariant rewrite of the higher-dimensional supergravities. 

It is precisely when $G_S$ is not contained in the $\cN=8$ gauge group that the corresponding truncation ``does not seem to arise from symmetry principles'' while, secretly, it does. This was explained in \cite{Guarino:2024gke} using a concrete $D=4$ $\cN=8$ supergravity with higher-dimensional origin, and focusing on truncations thereof to the pure $\cN <8$ supergravity multiplet around a concrete vacuum with the same $\cN <8$ residual supersymmetry. In order to understand our  specific subtruncations of section \ref{eq:AdSvac} in a wider context, we take a fresh look at the $G_S$-invariant subtruncation problem of general maximal supergravities. We take inspiration from \cite{Cassani:2019vcl,Guarino:2024gke} to do this, and borrow tools that we have already employed in \cite{Blair:2024ofc}. Our analysis, however, takes on a strictly four-dimensional perspective, and is completely general for any $D=4$ $\cN=8$ gauged supergravity. It does not depend on whether the maximal theory arises itself as a consistent truncation from higher dimensions. It is also independent of the existence of vacua of the $\cN=8$ theory, let alone on their Minkowski, de Sitter, or anti-de Sitter character or residual supersymmetry. We formulate sufficient kinematic and dynamical conditions for a $G_S$-invariant subtruncation to be consistent at the level of the field equations, regardless of whether $G_S$ is contained in the $\cN=8$ gauge group. These conditions take on the form of linear and quadratic constraints on the $\cN=8$ embedding tensor, similar to those that characterise the full $\cN=8$ theory \cite{deWit:2007mt,LeDiffon:2008sh,LeDiffon:2011wt}, along with suitable $G_S$-invariant tensors. A curious feature of these constraints is that, for the same $\cN=8$ theory, they might be only fulfilled in preferred, not all, duality frames. The details can be found in section \ref{sec:4DSugra}.


\section{$D=4$ $\cN=8$ supergravity and its consistent subsectors} \label{sec:4DSugra}


Let us begin by reviewing general aspects of gauged \cite{deWit:2007mt,LeDiffon:2008sh,LeDiffon:2011wt} maximal four-dimensional supergravity \cite{Cremmer:1979up} before delving into the conditions that allow for consistent truncations to subsectors therein.


\subsection{Field content and embedding tensor} \label{sec:FieldContentET}


Recall that the field content of maximal supergravity in four spacetime dimensions contains bosons and fermions of all spin between 0 and 2, see table \ref{tab:4DSummary} for a summary. The bosonic sector comprises the metric $g_{\mu \nu}$, gauge fields, $A^M$, two-form potentials, $B_\alpha$, and scalars described by a coset representatative $\cV_M{}^{ij}$ of $\mathrm{E}_{7(7)} / \mathrm{SU}(8)$. While the two-forms carry no independent degrees of freedom, they are needed in the embedding tensor formalism \cite{deWit:2007mt,LeDiffon:2008sh,LeDiffon:2011wt}, which we will use. The fermionic sector, in turn, includes the gravitini, $\psi_\mu^i$, and spin-$1/2$ fermions, $\chi^{ijk}$. Bosons and fermions respectively come in the representations of $\mathbb{R}^+ \times \mathrm{E}_{7(7)}$ or $\mathbb{R}^+ \times \mathrm{SU}(8)$ indicated in table \ref{tab:4DSummary}. Here, $\mathbb{R}^+$ is the trombone scaling symmetry, $\mathrm{E}_{7(7)}$ the duality symmetry of the field equations of the ungauged theory, and $\mathrm{SU}(8)$ the maximal compact subgroup of $\mathrm{E}_{7(7)}$. In the table and elsewhere, $\mu=0,1, 2 , 3$ are spacetime indices, whereas $M= 1 , \ldots , 56$, $\alpha = 1, \ldots , 133$ label the fundamental and adjoint representations of $\mathrm{E}_{7(7)}$, and $i=1 , \ldots , 8$ the fundamental of $\mathrm{SU}(8)$. Fundamental and adjoint indices of $\mathrm{E}_{7(7)}$ are respectively raised and lowered with the symplectic form, $\Omega_{MN}$, of Sp$(56)$  and the Killing-Cartan form, $\kappa_{\alpha \beta}$, of $\mathrm{E}_{7(7)}$.

In gauged supergravity, a subgroup $G \subset \mathbb{R}^+ \times \mathrm{E}_{7(7)}$ is promoted to a local symmetry. All the couplings induced by the gauging are codified in the components, $\Theta_M{}^\alpha$ and $\vartheta_M$, of the embedding tensor \cite{deWit:2007mt,LeDiffon:2008sh,LeDiffon:2011wt}
\begin{equation} \label{eq:XSymbolsGen}
X_M = -w \, \vartheta_M \, t_0 + \big( \Theta_{M}{}^\alpha +  8 (t^\alpha)_{M}{}^{Q} \,  \vartheta_{Q} \big)  t_\alpha \; .
\end{equation}
This is a constant object that couples the gauge fields $A^M$ to the $\mathbb{R}^+ \times \mathrm{E}_{7(7)}$ generators $t_0$, $t_\alpha$, in the covariant derivatives $D = d- A^M X_M$. The generators in (\ref{eq:XSymbolsGen}) need to be evaluated in the representations ${\bm{r}_w}$, collected in table \ref{tab:4DSummary}, of $\mathbb{R}^+ \times \mathrm{E}_{7(7)}$ appropriate to the field acted upon by the covariant derivative. For example, acting on gauge fields
\begin{eqnarray} \label{eq:defX}
X_{M N}{}^{P} =
\Theta_{M N}{}^{P}
 +  \Big(  8 \,  \mathbb{P}^Q{}_M{}^P{}_N  - \delta_{M}^{Q}\delta_{N}^{P} \Big)
\vartheta_{Q}  \; , \quad 
\textrm{with \, $\Theta_{M N}{}^{P} \equiv \Theta_M{}^\alpha (t_\alpha)_N{}^P$,}
\end{eqnarray}
so that $X_{M N}{}^{N} = -56 \, \vartheta_{M}$. Here, $\mathrm{E}_{7(7)}$ fundamental representation indices have been introduced, and $\mathbb{P}^K{}_M{}^L{}_N \equiv (t^\alpha)_M{}^{K} (t_\alpha)_N{}^{L}$ is the projector to the adjoint of $\mathrm{E}_{7(7)}$.

\begin{table}[]

\centering


\begin{tabular}{l | cccccccc}
%
%
 	& $g_{\mu\nu} $ &  $\psi_\mu^i $ &  $A_\mu^M$ & $B_{\mu\nu \alpha}$ & $\chi^{ijk}$ & $\cV_M{}^{ij}$ & $\Theta_{M}{}^\alpha$ & $\vartheta_M$ 
\\[2pt]
\hline
\\[-10pt]
$\mathbb{R}^+ \times \mathrm{E}_{7(7)}$ &	$\bm{1}_2$  &  $\bm{1}_{\frac12} $   &	$\bm{56}_1$ & $\bm{133}_2$  & $\bm{1}_{-\frac12}$  & $\bm{56}_0$  & $\bm{912}_{-1}$  & $\bm{56}_{-1}$  
\\[10pt]
$\mathbb{R}^+ \times \mathrm{SU}(8)$ &	$\bm{1}_2$  &  $\overline{\bm{8}}_{\frac12} $   &	$\bm{1}_1$ & $\bm{1}_2$  & $\overline{\bm{56}}_{-\frac12}$  & $\bm{28}_0+\overline{\bm{28}}_0 $ & $\bm{1}_{-1}$ & $\bm{1}_{-1}$ 
\\[10pt]
%
%
%
\end{tabular}


\caption{\footnotesize{Charges ${\bm{r}_w}$ under $\mathbb{R}^+ \times \mathrm{E}_{7(7)}$ and $\mathbb{R}^+ \times \mathrm{SU}(8)$ of the supergravity fields and embedding tensor.
}\normalsize}
\label{tab:4DSummary}
\end{table}

The embedding tensor must obey linear and quadratic constraints. The former restrict it to lie in the $\bm{912}_{-1} + \bm{56}_{-1}$ representation of $\mathbb{R}^+ \times \mathrm{E}_{7(7)}$, and ensures compatibility with maximal supersymmetry. The embedding tensor component $\vartheta_M$ that  couples to the trombone generator $t_0$ in (\ref{eq:XSymbolsGen}) naturally sits in the $\bm{56}_{-1}$, while the remaining components $\Theta_{MN}{}^P$ lie in the $\bm{912}_{-1}$ if
\begin{equation} \label{eq:LinearConst}
\Theta_{(MN}{}^Q \Omega_{P)Q}=0 
\end{equation}
holds. This linear relation, together with the quadratic constraints \cite{deWit:2007mt,LeDiffon:2008sh,LeDiffon:2011wt}
\begin{eqnarray} \label{eq:QCs}
& \bm{133}_{-2} & : \;  \Omega^{MN}\, \Theta_M{}^\alpha\,\vartheta_{N} +16 (t^\alpha)^{MN} \,\vartheta_{M}\, \vartheta_{N} = 0 \; , \nonumber \\[5pt]
& \bm{1539}_{-2} & : \;  (t_{\alpha})_{[M}{}^{P}\, \Theta_{N]}{}^\alpha\,\vartheta_{P} = 0  \; , \\[5pt]
& \bm{133}_{-2} + \bm{8645}_{-2} & : \;  \Theta_M{}^\alpha\,\Theta_N{}^\beta\,\Omega^{MN} -8 \,\vartheta_M\,\Theta_N{}^{[\alpha}\,t^{\beta]}{}^{MN}
+ 4\,c^{\alpha\beta}{}_{\gamma}\,\vartheta_M\,\Theta_N{}^\gamma\,\Omega^{MN} =0 \; , \quad  \nonumber 
\end{eqnarray}
where $c_{\alpha\beta}{}^{\gamma}$ are the E$_{7(7)}$ structure constants, enforce the closure of the embedding tensor (\ref{eq:defX}) into the algebra of the gauge group $G$,
\begin{equation} \label{eq:LieAlg}
[ X_{M} , X_{N}] = - X_{M N}{}^{P} \, X_{P} \; ,
\end{equation}
along with the orthogonality relations 
\begin{equation} \label{eq:QuadOrtho}
\Theta_M{}^\beta  \big(  \Theta^{M\alpha} - 16 \, t^{\alpha MN} \vartheta_N \big) =0  \; , \qquad 
\vartheta_{M}  \big(  \Theta^{M\alpha} - 16 \, t^{\alpha MN} \vartheta_N \big) =0 \; .
\end{equation}
In $\cN=8$ supergravity (but not if $\cN < 8$, see {\it e.g.}~section 3.1.2 of \cite{Trigiante:2016mnt}) the sets of constraints (\ref{eq:LinearConst}), (\ref{eq:QCs}) and (\ref{eq:LinearConst}), (\ref{eq:LieAlg}) are interchangeable, as they carry the same information. 

Maximal supergravities with gaugings strictly contained in the $\bm{912}_{-1}$ are described by a Lagrangian \cite{deWit:2007mt}. In contrast, gauged supergravities whose embedding tensors involve the $\bm{56}_{-1}$ trombone components do not admit Lagrangian descriptions, and must be formulated at the level of the field equations \cite{LeDiffon:2011wt}. For later reference, these have been collected in appendix \ref{sec:4DSugraEoMsMass}. Let us conclude this review by recalling that two $D=4$ $\cN=8$ gauged supergravities whose embedding tensors $X_{MN}{}^P$ and $\tilde{X}_{MN}{}^P$ are tensorially related by an $\mathrm{E}_{7(7)}$ transformation $U_M{}^N$ through
\begin{equation} \label{eq:TransXSymbol}
\tilde{X}_{MN}{}^P = U_M{}^Q \, U_N{}^R \, X_{QR}{}^S  \, (U^{-1})_S{}^P \; , 
\end{equation}
are physically equivalent, though expressed in different duality frames. In sections \ref{sec:New4DSugras} and \ref{eq:AdSvac}, we will find it helpful to describe the same $\cN=8$ supergravity in different duality frames related as in (\ref{eq:TransXSymbol}).


\subsection{Consistent subsectors: kinematics} \label{sec:4DSugraCTKin}


As argued in the introduction, it is often convenient to deal with smaller, more tractable, subsectors obtained by consistent truncation of the parent $\cN=8$ supergravity theory. Namely, it is usually of interest to find a smaller set of fields,
\begin{equation} \label{sec:SubFieldCont}
\textrm{metric:} \;\; g_{\mu\nu} \; , \quad
\textrm{$2n_1$ gauge fields:} \;\; A^I_{\mu} \; , \quad
\textrm{$n_2$ two-forms:} \;\; B_{\mu\nu x} \; , \quad
\textrm{$n_0$ scalars:} \;\; \tilde{\cV}  \; ,
\end{equation}
along with gravitini and spin-$1/2$ fermions, appropriately selected among the $\cN=8$ fields. These do not necessarily need to define a submaximal supergravity, but nevertheless need to have self-consistent dynamics and closed field equations so that, if the field equations are satisfied for the reduced set (\ref{sec:SubFieldCont}), then the $\cN=8$ equations (\ref{eq:EinsteinEOM})--(\ref{eq:ScaalarDuality}) are satisfied as well. Indices $I = 1 , \ldots, 2n_1$ and $x =1 , \ldots, n_2$ in (\ref{sec:SubFieldCont}) label the retained vector and tensors in the truncated subsector. The number of retained vectors must be even, as indicated, so their field strengths can obey a selfduality condition analogue to the first relation in (\ref{eq:SDandBianchi}).

A well known sufficient condition to obtain consistent subsectors of {\it ungauged} supergravity consists in retaining only fields (\ref{sec:SubFieldCont}) that are singlets, and all singlets, under a group $G_S \subset \textrm{SU}(8) \subset \textrm{E}_{7(7)}$. From a four-dimensional perspective, it is natural to require that the invariance group $G_S$ be contained in $\textrm{SU}(8)$ (thus, incidentally, compact) and not merely in $\textrm{E}_{7(7)}$, as fermions typically need to be retained along with bosons to ensure consistency. The prescription $G_S \subset \textrm{SU}(8)$ also guarantees the number of retained vector gauge fields to be even, as required. This is because the $\bm{56}$ of $\textrm{E}_{7(7)}$ splits as two copies of the $\bm{28}$ (or, rather, as the $\bm{28}$ and its transpose) under $\textrm{SU}(8) \subset \textrm{E}_{7(7)}$. From an exceptional geometry perspective \cite{Coimbra:2011nw,Coimbra:2011ky,Coimbra:2012af}, it is also natural to request $G_S \subset \textrm{SU}(8)$, since the generalised metric defines a generalised $\textrm{SU}(8)$-structure, and the retained subsector corresponds to a reduction of the generalised structure group to $G_S$ \cite{Cassani:2019vcl}. While inspired in exceptional generalised geometry, our analysis is purely four-dimensional, though. For simplicity,  we will focus on consistency in the bosonic sector, although our arguments extrapolate straightforwardly to the fermion sector as well. Also for simplicity, we will restrict $G_S$ to be a continuous group, with generators $T_a$, $a = 1 , \ldots , \textrm{dim} \, G_S$.

The $G_S$-invariant prescription entails selecting all vectors and tensors corresponding to singlets in the decomposition of the $\bm{56}$ and $\bm{133}$ of $\textrm{E}_{7(7)}$ under $G_S$, along with the metric, which is already an $\textrm{E}_{7(7)}$ singlet. Explicitly, the retained gauge fields and two-form potentials in (\ref{sec:SubFieldCont}) are selected among the $\cN=8$ ones, $A^M$ and $B_\alpha$, via
\begin{equation} \label{eq:RetainedVecs}
A^M  = A^I \, \hat{K}_I{}^M \; , \qquad 
B_\alpha = B_x \, K^{x}{}_{\alpha} \; .
\end{equation}
Here, $\hat{K}_I{}^M$ and $K^{x}{}_{\alpha}$ are the constant, invariant tensors of $G_S$ that descend from the fundamental and adjoint representations of E$_{7(7)}$, respectively. It is helpful to also introduce their transposed tensors, $K^I{}_M{}$ and $\hat{K}_{x}{}^{\alpha}$, normalised as $ K^I{}_M\hat{K}_J{}^M = \delta^I_J$ and $ K^x{}_\alpha\hat{K}_y{}^\alpha = \delta^x_y$. By definition, all these obey
\begin{equation} \label{eq:InvarFields}
\hat{K}_I{}^M (T_a)_M{}^N = 0 \; , \quad  
(T_a)_M{}^N K^I{}_N{} = 0 \; , \quad  
[ T_a ,  K^{x}{}_{\alpha} \, t^\alpha ] = 0 \; , \quad 
[ T_a , \hat{K}_{x}{}^{\alpha} \,  t_\alpha] = 0 \; ,
\end{equation}
for all $a = 1 , \ldots , \textrm{dim} \, G_S$. Here, $(T_a)_M{}^N$ are the $G_S$ generators in the $\bm{56}$-dimensional, $G_S$-reducible, representation, with the latter indices omitted in the last two relations. The latter signify $t_x \equiv \hat{K}_{x}{}^{\alpha} \,  t_\alpha$ as the generators of the commutant, $\mathrm{C}_{\mathrm{E}_{7(7)}}(G_S)$, of $G_S$ inside E$_{7(7)}$. By duality, this commutant also defines the scalars retained in the subsector. More precisely, $\tilde{\cV}$ in (\ref{sec:SubFieldCont}) is a coset representative of $\mathrm{C}_{\mathrm{E}_{7(7)}}(G_S) / \mathrm{C}_{\mathrm{SU}(8)}(G_S) \,  \subset \, \textrm{E}_{7(7)}/\textrm{SU}(8)$. Below, we will find it convenient to work with the coset $\tilde{\cV}_M{}^N$ in the $\bm{56}$-dimensional, $G_S$-reducible, representation.


\subsection{Consistent subsectors: dynamics} \label{sec:4DSugraCTDyn}


In ungauged supergravity, the $G_S$-invariant truncation specified in section \ref{sec:4DSugraCTKin} is automatically consistent at the level of the field equations. In the gauged case, however, consistency could be obstructed by the embedding tensor. Indeed, we find that further constraints, on top of the $\cN=8$ linear and quadratic constraints, (\ref{eq:LinearConst}), (\ref{eq:LieAlg}), must be imposed on combinations of the embedding tensor and $G_S$-invariant tensors for the truncation to go through consistently in gauged supergravity at the level of the field equations. More concretely, for all $a = 1 , \ldots , \textrm{dim} \, G_S$, the following consistency constraints, linear
\begin{equation} \label{eq:TruncCondXLin}
\textrm{C1}: \; [ T_a , X_I ] = c_{ab}{}^c \,  \Theta_{I}{}^b \, T_c \; , \qquad
\textrm{C2}: \;  (T_a)_P{}^M  \big(  \Theta^{P\alpha} - 16 \, t^{\alpha PN} \vartheta_N \big)  \, K^{x}{}_\alpha = 0 \; , \textrm{ if $\hat{K}_I{}^M \neq 0$,} 
\end{equation}
and quadratic
\begin{equation} \label{eq:TruncCondXQuad}
\textrm{C3}: \; (T_a)_{(M}{}^P \, \tilde{V}_{N) P} = 0 \; , 
\end{equation}
in the embedding tensor, suffice to ensure the consistency of the truncation of the parent $D=4$ $\cN=8$ gauged supergravity to a $G_S$-invariant subsector thereof. Moreover, together with the constraints of the parent $\cN=8$ theory, the relations (\ref{eq:TruncCondXLin}), (\ref{eq:TruncCondXQuad}) guarantee the self-consistency of the subsector as a theory by itself. Please refer to appendix~\ref{sec:ConsProofs}~for a consistency proof. 

The symbol $X_I$ in (\ref{eq:TruncCondXLin}) signifies the generators of the gauge group $\tilde{G}$ of the retained subsector, with representation indices omitted. Restoring representation indices, these generators take on either form
\begin{equation} \label{eq:TruncGen}
X_{IM}{}^N \equiv \hat{K}_I{}^P \, X_{PM}{}^N \; , \qquad
X_{IJ}{}^{K} \equiv \hat{K}_I{}^M \hat{K}_J{}^N X_{MN}{}^{P} K^K{}_P  \;  ,
\end{equation}
in, respectively, the $\bm{56}$-dimensional representation, and the $2n_1$-representation provided by the $G_S$ singlets contained in the $\bm{56}$ of E$_{7(7)}$.  As emphasised in (\ref{eq:TruncCondXLin}), the condition C2 needs to be imposed only in the presence of non-trivial G$_S$-invariant tensors $\hat{K}_I{}^M$, and can be ignored if $\hat{K}_I{}^M=0$, see appendix~\ref{sec:ConsProofs}. In (\ref{eq:TruncCondXQuad}), $\tilde{V}_{MN}$  contains the $\cN=8$ scalar self-interactions defined by (\ref{eq:DerPotTromb}), evaluated on the $G_S$-invariant scalars, and must be satisfied for arbitrary values of those scalars. The constraint C3 enforces these self-interactions to also be $G_S$-invariant. Finally, in (\ref{eq:TruncCondXLin}), $c_{ab}{}^c$ are the $G_S$ structure constants and $\Theta_{I}{}^b$ are constants defined in appendix~\ref{sec:ConsProofs}. Note that the relation C1 in (\ref{eq:TruncCondXLin}) and the first expression in (\ref{eq:TruncGen}) enforce the inclusions
\begin{equation} \label{eq:Gincl}
\tilde{G} \subset \mathrm{C}_{\mathrm{E}_{7(7)}}(G_S) \otimes G_S \; , \qquad
\tilde{G} \subset G \; , \qquad
\end{equation}
where $G$ is the gauge group of the parent $\cN=8$ theory, generated by $X_M$ in (\ref{eq:defX}). In any case, the $G_S$-Lie-algebra-valued piece of the embedding tensor of the retained subsector drops out from all equations in the subsector, and can therefore be effectively disregarded always. In fact, in the examples covered in section~\ref{eq:AdSvac}, we will have a stronger version of C1 given by $[ T_a , X_I ] =0$, which in turn will restrict the gauge group as $\tilde{G} \subset \mathrm{C}_{\mathrm{E}_{7(7)}}(G_S)$.

All consistency conditions, (\ref{eq:TruncCondXLin}) and (\ref{eq:TruncCondXQuad}), are automatically satisfied if the invariance group $G_S$ is a subgroup of the gauge group $G$ of the parent $\cN=8$ supergravity. This situation will be discussed in section~\ref{sec:Subgroup}. The linear conditions (\ref{eq:TruncCondXLin}), but not necessarily the quadratic ones (\ref{eq:TruncCondXQuad}), are satisfied identically for $G_S$-invariant truncations with $\hat{K}_I{}^M=0$, where no vector gauge fields are retained. In fact, as we have just pointed out, C2 does not even need to be imposed in this case, and C1 vanishes at face value by (\ref{eq:TruncGen}). In this case, the quadratic condition C3 in (\ref{eq:TruncCondXQuad}) still needs to be imposed for consistency. In general, the constraints (\ref{eq:TruncCondXLin}), (\ref{eq:TruncCondXQuad}) will not be necessarily satisfied for a given $\cN=8$ embedding tensor $X_M$ and given non-vanishing $G_S$-invariant tensors $\hat{K}_I{}^M$, $K^{x}{}_\alpha$. In this case, the $G_S$-invariant truncation will be inconsistent. We will present worked examples of these situations in section~\ref{eq:SCBianchiV}, which will also illustrate the independence of our three constraints. In any case, we reiterate that, as given in (\ref{eq:TruncCondXLin}), (\ref{eq:TruncCondXQuad}) our constraints are self-consistent and sufficient. Perhaps a simpler set of constraints could be derived for which, in particular, the scalar dependence drops out of C3.

Two slightly unintuitive situations might arise. Firstly, the constraints (\ref{eq:TruncCondXLin}), (\ref{eq:TruncCondXQuad}) might only be satisfied for some, but not all, embedding tensors in the same E$_{7(7)}$-duality orbit (\ref{eq:TransXSymbol}). Namely, it might happen that, given $G_S$-invariants $\hat{K}_I{}^M$, $K^{x}{}_\alpha$, and two $\cN=8$ embedding tensors $X_M$, $\tilde{X}_M$ related by a duality transformation (\ref{eq:TransXSymbol}), the consistency conditions (\ref{eq:TruncCondXLin}), (\ref{eq:TruncCondXQuad}) hold only for one of them, say $\tilde{X}_M$, making the $G_S$-invariant truncation consistent in that frame, but not for the other, $X_M$, rendering the truncation inconsistent in the latter. The $\cN=8$ frame democracy promised by E$_{7(7)}$ duality turns out to have an orwellian character as far as consistent subtruncation is concerned, where ``all frames are equal, but some frames are more equal than others''. Of course, this situation arises because 
(\ref{eq:TransXSymbol}) only affects the embedding tensor, not the $G_S$-invariants, which are fixed. An example of this situation will be discussed in section \ref{eq:AdSvac}. Secondly, it might also happen as a logical possibility that the $G_S$-invariant truncation does go through in two duality-related frames, $X_M$ and $\tilde{X}_M$, but the truncated theories obtained from either one differ.

Finally, we conclude this section with an observation specific to trombone-gauged, $\vartheta_M \neq 0$, maximal supergravities. As reviewed in section \ref{sec:FieldContentET}, the latter must be described at the level of the field equations, as they do not admit a Lagrangian \cite{LeDiffon:2011wt}. Yet, certain consistent subsectors thereof could themselves be described by a perfectly well-defined Lagrangian. This will be the case if the trombone component inherited by the subsector vanishes, 
\begin{equation} \label{eq:TrombVecContr}
\hat{K}_I{}^M \, \vartheta_M \equiv \vartheta_I = 0 \; ,
\end{equation}
 so that the $\mathbb{R}^+$ scaling symmetry is not gauged therein. Some examples will be given in section \ref{eq:AdSvac}. Previously discussed cases in similar contexts include \cite{Bhattacharya:2024tjw,Varela:2025xeb}.


\subsection{Invariance and symmetries under a subgroup of the gauge group} \label{sec:Subgroup}


The conditions (\ref{eq:TruncCondXLin}), (\ref{eq:TruncCondXQuad}) ensure the consistency of a $G_S$-invariant truncation of maximal supergravity regardless of whether $G_S$ is or is not a subgroup of the gauge group $G$ generated by the $\cN=8$ embedding tensor $X_M$. However, as advertised above, the assumption $G_S \subset G$ makes these conditions be identically satisfied. Such truncations are, thus, automatically consistent, a fact that of course has long been known \cite{Warner:1983vz,Warner:1983du}. In order to see this using our language, first note that, if $G_S \subset G$, then the $G_S$ generators $T_a$ can be expressed as linear combinations of generators of $G$. Namely, there must exist constants $\hat{K}_a{}^M$, $a=1 , \ldots , \textrm{dim} \, G_S$, such that
\begin{equation} \label{eq:TasX}
T_a = \hat{K}_a{}^M \, X_M \; .
\end{equation}
Using (\ref{eq:TruncGen}), (\ref{eq:TasX}), the l.h.s.~of the linear constraint C1 in (\ref{eq:TruncCondXLin}) then gives 
$[T_a , X_I ] = \hat{K}_I{}^N \, \hat{K}_a{}^M [X_M , X_N] = -\hat{K}_I{}^N \, \hat{K}_a{}^M \, X_{MN}{}^P \, X_P$, 
where, in the last step, we have used the commutation relations (\ref{eq:LieAlg}) of the $\cN=8$ gauge group $G$. But now, using (\ref{eq:TasX}) in the $\bm{56}$-dimensional representation, we have $\hat{K}_I{}^N \, \hat{K}_a{}^M \, X_{MN}{}^P = \hat{K}_I{}^N (T_a)_{N}{}^P$ which, by the first relation in (\ref{eq:InvarFields}), vanishes due to the $G_S$-invariance of $\hat{K}_I{}^N$. As for C2, this constraint needs to be imposed only if $\hat{K}_I{}^N \neq 0$ but, in any case, it also holds identically. This is because $\Theta^{P\alpha}$ and $\vartheta_N$ are $G$ singlets due to the $\cN=8$ quadratic constraint (\ref{eq:LieAlg}), and $t^{\alpha PN}$ is an E$_{7(7)}$ singlet (and thus also a $G\subset \mathrm{E}_{7(7)}$ singlet). If $G_S \subset G$, then the quantity $ \big(  \Theta^{P\alpha} - 16 \, t^{\alpha PN} \vartheta_N \big)  \, K^{x}{}_\alpha $ is thus also a $G_S$ singlet and is therefore annihilated by the $G_S$ generators $(T_a)_P{}^M$, as required by C2. Finally, since the $G$-singlets $\Theta^{P\alpha}$ and $\vartheta_N$ are also $G_S$ singlets if $G_S \subset G$, then $\tilde{V}_{MN}$ in (\ref{eq:DerPotTromb}) is a $G_S$-singlet, as required by C3, when evaluated on $G_S$-singlet scalars.

In presence of a gauging the (global) symmetry $\mathbb{R}^+ \times \mathrm{E}_{7(7)}$ of maximal four-dimensional ungauged supergravity is reduced to the (local) gauge group $G \subset \mathbb{R}^+ \times \mathrm{E}_{7(7)}$ generated by the embedding tensor $X_M$, along with possible residual global symmetries (see below). Of course, the gauge group $\tilde{G}$ of any $G_S$-invariant subsector is a local symmetry therein. The question of whether $G_S$ is or is not a symmetry of the $G_S$-invariant sector depends on whether $G_S$ is included or not in $G$. If $G_S \not \subset G$, then $G_S$ will only be a (global) symmetry of the $G_S$-invariant subsector if $G_S \subset \mathrm{C}_{\mathrm{E}_{7(7)}}(G) \cap \textrm{SU}(8)$. On the contrary, if $G_S \subset G$, then $G_S$ can never be a (local) symmetry of the $G_S$-invariant subsector by itself: this would require $G_S \subset \tilde{G}$ which, although formally allowed by the first relation in (\ref{eq:Gincl}), the $G_S$ component always drops from the reduced equations of motion, as discussed below that equation. However, if $G_S \subset G$, then $G_S$ will still manifest itself as a (local) symmetry when the subsector is considered inside the full $\cN=8$ theory. In particular, any vacuum that is attained within a $G_S$-invariant subsector will preserve at least $G_S$ local symmetry when regarded as a vacuum of the full $\cN=8$ theory. As a result, $\textrm{dim} \, G_S$ massless gauge fields will always be present in the $\cN=8$ mass spectrum about such vacuum.

The presence of the $G_S$ massless gauge fields can be seen by looking at the $\cN=8$ vector mass matrix, (\ref{eq:VectorMassMat}), as follows. Since the scalar matrix $\tilde{M}_{MN} = (\tilde{\cV} \tilde{\cV}^{\textrm{T}})_{MN}$ in the subsector contains only $G_S$ singlets, it must be annihilated by the $G_S$ generators, $(T_a)_{(M}{}^P \tilde{M}_{N)P} = 0$, analogously to (\ref{eq:InvarFields}) or (\ref{eq:TruncCondXQuad}). If $G_S \subset G$ then, by (\ref{eq:TasX}), $X_{a(M}{}^P \tilde{M}_{N)P}=0$, where we have defined $X_a \equiv \hat{K}_a{}^M X_M \equiv T_a$ to be the projection of the generators $X_M$ of $G$ to the adjoint of $G_S$. Since we are also assuming (see section \ref{sec:4DSugraCTKin}) that $G_S \subset \textrm{SU}(8)$, the trombone components along the $G_S$ directions must vanish as well, $\vartheta_a \equiv - \tfrac{1}{56} \hat{K}_a{}^M X_{MN}{}^N =0$. We thus have that the directions along $G_S$ of the contribution $\left( X_{M(R}{}^T M_{S)T} + \vartheta_M M_{RS} \right) $ to the gauge field mass matrix (\ref{eq:VectorMassMat}) vanish, $\hat{K}_a{}^M \left( X_{M(R}{}^T M_{S)T} + \vartheta_M M_{RS} \right) =0$. Now, the decomposition (\ref{eq:TasX}) implies that the adjoint representation of $G_S$ appears in the branching of the (reducible) $\bm{56}$-dimensional representation of $G$ under $G_S \subset G$. We are indeed using indices $a = 1 , \ldots , \textrm{dim} \, G_S$ to run in the adjoint of $G_S$. Introduce indices $\hat{a}$ to label all other representations different from the adjoint of $G_S$ in the branching of the $\bm{56}$ under $G_S$. The fundamental E$_{7(7)}$ indices thus split as $M=(a , \hat{a})$, and the gauge field mass matrix takes on the block decomposition
\begin{equation}
 (\cM_{\textrm{vector}}^2)_M{}^N = \begin{pmatrix}
 (\cM_{\textrm{vector}}^2)_{a}{}^{b} & (\cM_{\textrm{vector}}^2)_{a}{}^{\hat{b}} \\
   (\cM_{\textrm{vector}}^2)_{\hat{a}}{}^{b} &    (\cM_{\textrm{vector}}^2)_{\hat{a}}{}^{\hat{b}}
 \end{pmatrix} =
 \begin{pmatrix}
0 & 0 \\
   (\cM_{\textrm{vector}}^2)_{\hat{a}}{}^{b} &    (\cM_{\textrm{vector}}^2)_{\hat{a}}{}^{\hat{b}}
 \end{pmatrix}  \; , 
\end{equation}
where the entire row $(\cM_{\textrm{vector}}^2)_{a}{}^M \sim \hat{K}_a{}^P \left( X_{P(R}{}^T M_{S)T} + \vartheta_P M_{RS} \right) $ vanishes as argued above. Such matrix indeed has at least $\textrm{dim} \, G_S$ null eigenvalues.

The above argument fails if $G_S$ is not a subgroup of the gauge group $G$. In this case, a vacuum of the parent $\cN=8$ supergravity that arises within a $G_S$-invariant subsector therein will not preserve a residual local $G_S$ symmetry. Reversing the above argument, no gauge field associated to the $\cN=8$ gauge group $G$ will become massive as a result of spontaneous Higgs symmetry breaking from $G$ to $G_S$. We will see examples of this situation in section \ref{eq:AdSvac}.


\section{A new class of $\cN=8$ trombone-gauged supergravities} \label{sec:New4DSugras}


Let us now change gears and introduce a new family of $D=4$ $\cN=8$ supergravities, with gauge group that we will denote TCSO$(p,q,r;\mathrm{N})$. This is defined as
\begin{equation} \label{eq:ProdTCSO}
\textrm{TCSO} (p,q,r;\mathrm{N}) \equiv \big( G_3 \times \textrm{CSO} (p,q,r) \big)   \ltimes \mathbb{R}^{15}  \,  \subset \, \mathbb{R}^+ \times \mathrm{E}_{7(7)} \; , \qquad p+q+r = 5 \; ,
\end{equation}
and has dimension 
\begin{equation} \label{eq:dimTCSO}
\textrm{dim} \, \textrm{TCSO} (p,q,r;\mathrm{N}) = \tfrac12 \, (p+q)(p+q-1) + (p+q) \, r + 18 \; ,
\end{equation}
less than 28. Here, $\textrm{CSO} (p,q,r) \equiv \textrm{SO} (p,q)  \ltimes \mathbb{R}^{(p+q)r}$, and $G_3$ is a non-unimodular three-dimensional group of Bianchi type $\mathrm{N}$. In the defining duality frame (\ref{eq:ETtrombone}) below, $\textrm{CSO} (p,q,r)$ is gauged electrically, $G_3$ magnetically, and $\mathbb{R}^{15}$ dyonically. The $\cN=8$ theory with gauge group (\ref{eq:ProdTCSO}) can be understood as a generalisation of a specific dyonic CSO gauging in the class of \cite{DallAgata:2011aa,Dall'Agata:2014ita}. Let us start by reviewing the latter.


\subsection{Dyonic CSO gaugings} \label{sec:CSO}


The family of $D=4$ $\cN=8$ dyonic CSO gauged supergravities \cite{DallAgata:2011aa,Dall'Agata:2014ita} is defined by a specific embedding tensor  (\ref{eq:defX}) with no trombone, $\vartheta_M=0$, and only active components in the $\bm{36}+\bm{36}^\prime$ of SL$(8,\mathbb{R})$ descending from the $\bm{912}$ of E$_{7(7)}$, with $\mathbb{R}^+$ scaling charges omitted hereafter. Specifically, splitting the fundamental and adjoint representations of E$_{7(7)}$ under SL$(8,\mathbb{R})$, $\bm{56} \rightarrow \bm{28}+ \bm{28}^\prime$ and $\bm{133} \rightarrow \bm{63}+ \bm{70}$, the embedding tensor and E$_{7(7)}$ generators split, with representation indices omitted, as $X_M = (X_{AB} , X^{AB})$ and $t_\alpha = (t_A{}^B, \,  t_{ABCD}$), with $X_{AB}=X_{[AB]}$, $t_A{}^A=0$, $t_{ABCD}=t_{[ABCD]}$, $A=1, \ldots ,8$, and 
\begin{equation} \label{eq:ETCSO}
X_{AB} = -2 \, \theta_{C[A} \, t_{B]}{}^C \; , \qquad 
X^{AB} = 2 \, \tilde{\theta}^{C[A} \, t_{C}{}^{B]} \; ,
\end{equation}
with signs taken for convenience. We use the conventions of \cite{Guarino:2015qaa} for the E$_{7(7)}$ generators, up to some rescalings. In (\ref{eq:ETCSO}), $\theta_{AB}=\theta_{(AB)}$, $\tilde{\theta}^{AB}=\tilde{\theta}^{(AB)}$ are quadratic forms, in the $\bm{36}$ and $\bm{36}^\prime$ of SL$(8,\mathbb{R})$, subject to either constraint
\begin{equation}
\label{CSOConst}
\theta_{AC} \, \tilde{\theta}^{CB}  = 
\left\{
\begin{array}{ll}
c \, \delta_A^B \; ,  & \textrm{ if both $\theta_{AB}$ and $\tilde{\theta}^{AB}$ are non-degenerate,}  \\[4pt]
0 \; , & \textrm{ if both $\theta_{AB}$ and $\tilde{\theta}^{AB}$ are degenerate, } 
\end{array}
\right.
\end{equation} 
for some non-vanishing constant $c$. Without loss of generality, $\theta_{AB}$ and $\tilde{\theta}^{AB}$ can be taken diagonal, with $p$, $\tilde{p}$ positive, $q$, $\tilde{q}$ negative and $r^\prime$, $\tilde{r}^\prime$ vanishing eigenvalues, with  $p+q +r^\prime= \tilde{p} + \tilde{q} + \tilde{r}^\prime = 8$. Accordingly, the gauge group can be thought of as a(n at most 28-dimensional, dyonic in the defining duality frame (\ref{eq:ETCSO})) combination of $\textrm{CSO} (p,q,r^\prime) \equiv \textrm{SO} (p,q)  \ltimes \mathbb{R}^{(p+q)r^\prime}$ and $\textrm{CSO} (\tilde{p},\tilde{q},\tilde{r}^\prime) \equiv \textrm{SO} (\tilde{p},\tilde{q})  \ltimes \mathbb{R}^{(\tilde{p}+\tilde{q})\tilde{r}^\prime}$. Equation (\ref{eq:ETCSO}) makes apparent that the gauge group is strictly contained in the SL$(8,\mathbb{R})$ maximal subgroup of E$_{7(7)}$. 

The embedding tensor (\ref{eq:ETCSO}) with (\ref{CSOConst}) satisfies the $\cN=8$ linear, (\ref{eq:LinearConst}), and quadratic, (\ref{eq:QCs}), constraints \cite{DallAgata:2011aa,Dall'Agata:2014ita} and thus define viable theories. Some purely electric gaugings, with $\tilde{r}^{\prime}=8$ so that $\tilde{\theta}^{AB}=0$, known prior to \cite{DallAgata:2011aa,Dall'Agata:2014ita} fit in this class. These include the $p=8$ \cite{deWit:1982ig,deWit:1981eq}, $p=7$, $q=1$, \cite{Hull:1984yy}, and the generic $p$, $q$, $r^\prime$ \cite{Hull:1984vg,Hull:1984qz} cases. Among the dyonic theories, with both $\theta_{AB}$ and $\tilde{\theta}^{AB}$ active, some notable instances include the 
$p=\tilde{p}=8$ \cite{Dall'Agata:2012bb}, $p=q= \tilde{p} = \tilde{q} = 4$ \cite{DallAgata:2012plb}, 
$p=\tilde{q}=3$, $q=\tilde{p} =1$ \cite{Catino:2013ppa}, $p=\tilde{r}^\prime = 6$, $r^\prime=2$, $\tilde{p}=\tilde{q}=1$ \cite{Gallerati:2014xra}, or the $p=\tilde{r}^\prime = 7$, $r^\prime=\tilde{p} = 1$ \cite{Guarino:2015qaa} models, to give some examples. Partial though the list \cite{deWit:1982ig,deWit:1981eq}, \cite{Guarino:2015qaa}--\cite{Gallerati:2014xra} of references is, it makes apparent that the dyonic cases with $r^\prime \equiv r+3$, $\tilde{r}^\prime \equiv \tilde{r}+5$ and $p+q+r=5$, $\tilde{p}+\tilde{q}+\tilde{r} = 3$ (or, equivalently after a field redefinition, $r^\prime \equiv r+5$, $\tilde{r}^\prime \equiv \tilde{r}+3$ and $p+q+r = 3$, $\tilde{p}+\tilde{q}+\tilde{r} = 5$) have not attracted much attention. Yet, these cases are somewhat special in that their embedding tensor admits an expression equivalent to, but formally different from, that originally given in \cite{DallAgata:2011aa,Dall'Agata:2014ita} for generic $p , q , r^\prime , \tilde{p} , \tilde{q} , \tilde{r}^\prime$. This alternate expression is amenable to further generalisation, as we will discuss in sec.~\ref{sec:NewGaugingsTromb}.

\begin{table}[]

\centering


\begin{tabular}{lclcl} 
\hline
\hline
Bianchi type && Lie algebra && CSO$(\tilde{p},\tilde{q},\tilde{r})$ \\
\hline \\[-10pt]
I && $\mathbb{R}^3$ (abelian) && \text{CSO}$(0,0,3)$ \\[2pt]
II && Heisenberg && CSO$(1,0,2)$ \\[2pt]
VI$_0$ && $(1+1)$-dimensional Poincar\'e && CSO$(1,1,1)$ \\[2pt]
VII$_0$ && Euclidean, ISO$(2)$ && CSO$(2,0,1)$ \\[2pt]
VIII && $\mathrm{SL}(2,\mathbb{R}) \sim \textrm{SO}(2,1)$ && CSO$(2,1,0)$ \\[2pt]
IX && $\mathrm{SO}(3) \sim \textrm{SU}(2)$ && CSO$(3,0,0)$ \\[2pt]
\hline
\hline
\end{tabular}

%
\caption{\footnotesize{All unimodular three-dimensional Lie algebras in the Bianchi classification and their CSO$(\tilde{p},\tilde{q},\tilde{r})$, $\tilde{p} +\tilde{q} +\tilde{r}=3$, equivalence.}
}\normalsize
\label{tab:UnimodularGroups}
\end{table}

In order to implement this rewrite, first break 
\begin{equation} \label{eq:SL8ToSL3SL5}
\mathrm{SL}(8,\mathbb{R}) \supset \mathrm{SL}(3,\mathbb{R}) \times \mathrm{SL}(5,\mathbb{R}) \times \mathrm{SO}(1,1) \; , 
\end{equation}
so that 
\begin{equation} \label{eq:SL8Split}
{\bf 8} \rightarrow ({\bf 3}, {\bf 1}) + ({\bf 1}, {\bf 5})  \, , \;\;
{\bf 36} \rightarrow ({\bf 6}, {\bf 1}) + ({\bf 3}, {\bf 5})+ \underline{({\bf 1}, {\bf 15})} \, , \;\;
{\bf 36}^\prime \rightarrow \underline{({\bf 6}^\prime, {\bf 1})} + ({\bf 3}^\prime, {\bf 5}^\prime)+ ({\bf 1}, {\bf 15}^\prime) \, ,
\end{equation}
with SO$(1,1)$ charges omitted, and similarly for the ${\bf 8}^\prime$. Thus, fundamental $\mathrm{SL}(8,\mathbb{R})$ indices split as $A=(a,i)$, $a=1,2,3$, $i=4, \ldots , 8$, and $\theta_{AB}$, $\tilde{\theta}^{AB}$ break up as
\begin{equation} \label{eq:Charges1}
\theta_{AB} = \big( \theta_{ab} =0  \, , \, \theta_{ai} = 0  \, , \, \theta_{ij}   \big) \; , \qquad 
\tilde{\theta}^{AB} = \big( \tilde{\theta}^{ab}  \, , \, \tilde{\theta}^{ai} = 0  \, , \, \tilde{\theta}^{ij} =0  \big)   \; ,
\end{equation}
with non vanishing $\theta_{ij} = \theta_{(ij)}$ and $\tilde{\theta}^{ab} = \tilde{\theta}^{(ab)}$ taking values in the symmetric representations $({\bf 1}, {\bf 15})$ and $({\bf 6^\prime}, {\bf 1})$ of $\mathrm{SL}(3,\mathbb{R}) \times \mathrm{SL}(5,\mathbb{R})$, so that the lower row of (\ref{CSOConst}) holds. These representations have been underlined in (\ref{eq:SL8Split}) for easier identification. Both quadratic forms $\theta_{ij}$ and $\tilde{\theta}^{ab}$ are still characterised completely by the sign of their eigenvalues, so one can set \cite{DallAgata:2011aa,Dall'Agata:2014ita}
{\small
\begin{eqnarray} \label{eq:Charges2bisCanonical}
\theta_{ij} = g_1 \eta_{ij} = g_1 \, \textrm{diag} \, \big( 1, \stackrel{p}{\ldots} , 1 , -1 , \stackrel{q}{\ldots} , -1 , 0, \stackrel{r}{\ldots} , 0  \big) ,  \; 
\tilde{\theta}^{ab} = g_2 \, \textrm{diag} \, \big( 1, \stackrel{\tilde{p}}{\ldots} , 1 , -1 , \stackrel{\tilde{q}}{\ldots} , -1 , 0, \stackrel{\tilde{r}}{\ldots} , 0  \big) , \; 
\end{eqnarray}
}\normalsize
with $g_1$, $g_2$ coupling constants and $p+q+r =5$, $\tilde{p}+\tilde{q}+\tilde{r} =3$. Alternatively, the ${\bf 6^\prime}$ of $\mathrm{SL}(3,\mathbb{R})$ can be written in terms of a three-index tensor, $f_{ab}{}^c$,  antisymmetric in its lower indices and traceless, $f_{ab}{}^c = f_{[ab]}{}^c$, $f_{ab}{}^b =0$. That is, the ${\bf 6^\prime}$ is also specified by the structure constants of a unimodular three-dimensional Lie algebra, so we can also take
{\small
\begin{eqnarray} \label{eq:Charges2bis}
\theta_{ij} = g_1 \eta_{ij} = g_1 \, \textrm{diag} \, \big( 1, \stackrel{p}{\ldots} , 1 , -1 , \stackrel{q}{\ldots} , -1 , 0, \stackrel{r}{\ldots} , 0  \big) \; , \; p+q+r =5 \; ,  \quad 
\tilde{\theta}^{ab} = - \tfrac12 \,  g_2 \, f_{cd}{}^{(a} \epsilon^{b)cd} \; , 
\end{eqnarray}
}\normalsize
with $\epsilon^{abc}$ the $\mathrm{SL}(3,\mathbb{R})$ Levi-Civita invariant. The embedding tensor of the $p+q+r =5$, $\tilde{p}+\tilde{q}+\tilde{r} =3$ dyonic CSO gaugings of \cite{DallAgata:2011aa,Dall'Agata:2014ita} is (\ref{eq:ETCSO}) with (\ref{eq:Charges1}), (\ref{eq:Charges2bisCanonical}) or, equivalently, with (\ref{eq:Charges1}), (\ref{eq:Charges2bis}). Of course, either version satisfies the $\cN=8$ linear and quadratic constraints (\ref{eq:LinearConst}), (\ref{eq:QCs}). In the formulation with (\ref{eq:Charges2bis}), the latter are satisfied by virtue of the Jacobi identity
\begin{equation} \label{eq:JIG3}
f_{[ab}{}^d \, f_{c]d}{}^e = 0 \; .
\end{equation}
Recall that three-dimensional Lie algebras, unimodular or otherwise, were classified by Bianchi. The rightmost equality in equation (\ref{eq:Charges2bis}), along with (\ref{eq:JIG3}), is central to derive the Bianchi classification in the unimodular case. That relation also reflects the fact that all unimodular, $f_{ab}{}^b=0$, three dimensional Lie groups $G_3$ in the Bianchi classification are of CSO$(\tilde{p},\tilde{q},\tilde{r})$ type, with $\tilde{p}+\tilde{q}+\tilde{r}=3$. See table \ref{tab:UnimodularGroups} for a review.


\subsection{Tromboneful generalisation} \label{sec:NewGaugingsTromb}


The above construction for $p+q+r =5$, $\tilde{p}+\tilde{q}+\tilde{r} =3$ can be generalised by dropping the unimodularity requirement on $G_3$, namely, by replacing CSO$(\tilde{p},\tilde{q},\tilde{r})$, $\tilde{p}+\tilde{q}+\tilde{r}=3$, with any non-unimodular three-dimensional group $G_3$ of Bianchi type $\mathrm{N}$. The non-unimodular, $f_{ab}{}^b \neq 0$, three-dimensional groups are those of all other Bianchi types, III, IV, V, VI$_h$ with $h\neq 0$, VII$_h$ with $h\neq 0$, not shown in table \ref{tab:UnimodularGroups}. The resulting $D=4$ $\cN=8$ supergravities are new, and we denote the corresponding gauge groups by TCSO$(p,q,r; \mathrm{N}$) as defined in (\ref{eq:ProdTCSO}). These gaugings necessitate of embedding tensor components in the $\bm{912}$ and in the $\bm{56}$ representations of E$_{7(7)}$. For that reason, as reviewed in section \ref{sec:FieldContentET} and indicated in (\ref{eq:ProdTCSO}), the TCSO$(p,q,r; \mathrm{N}$) theories involve a gauging of the $\mathbb{R}^+$ trombone scaling symmetry along with suitable generators of E$_{7(7)}$. The definition of TCSO$(p,q,r; \mathrm{N}$) can be  straightforwardly extended to also include the unimodular Bianchi type $\mathrm{N}$ cases of table \ref{tab:UnimodularGroups}, with the understanding that the latter are tromboneless, dyonic CSO gaugings in the class of \cite{DallAgata:2011aa,Dall'Agata:2014ita}.

Simple as the modularity relaxation prescription for $G_3$ is, it is not entirely obvious from the above construction alone where to turn on the non-vanishing traces $f_{ab}{}^b$ in the $f_{ab}{}^b=0$ embedding tensor (\ref{eq:ETCSO}) with (\ref{eq:Charges1}), (\ref{eq:Charges2bis}), in a respectful way of the $\cN=8$ linear and quadratic constraints (\ref{eq:LinearConst}), (\ref{eq:QCs}). In fact, we obtained the end result, (\ref{eq:ETtrombone}) below, from suitable consistent truncation of $D=11$ supergravity. We will report on that dimensional reduction elsewhere \cite{Pico:2026rji}, and here we will limit ourselves to describing our TCSO$(p,q,r; \mathrm{N})$ theories within the canonical $D=4$ $\cN=8$ gauged supergravity formalism of \cite{deWit:2007mt,LeDiffon:2008sh,LeDiffon:2011wt}.

In order to do this, we still choose to work in the SL$(8, \mathbb{R})$ frame, and consider all possible couplings of the $\bm{912} + \bm{56}$ components of the embedding tensor (\ref{eq:defX}) to the $\bm{133}$ generators of E$_{7(7)}$, all of them branched out in SL$(8, \mathbb{R})$ representations. These couplings are summarised in table \ref{tab:Gaugings}. Out of all these embedding tensor components, we focus on
{\setlength\arraycolsep{0pt}
\begin{eqnarray} \label{eq:ETCompsSL8}
& \bm{36} \, : \; \theta_{AB} = \theta_{(AB)} \; , \qquad
\bm{36}^\prime \, : \; \tilde{\theta}^{AB} = \tilde{\theta}^{(AB)} \; , \qquad
\bm{420}^\prime \, : \; \tilde{\xi}_A{}^{BCD} = \tilde{\xi}_A{}^{[BCD]} \, , \,  \tilde{\xi}_C{}^{ABC} =0 \; , \nonumber \\[4pt]
& \bm{28}^\prime \, : \; \tilde{\vartheta}^{AB} = \tilde{\vartheta}^{[AB]} \; , 
\end{eqnarray}
}and set the $\bm{420}$ and $\bm{28}$ components to zero. The components in the top and bottom rows of (\ref{eq:ETCompsSL8}) respectively branch from the $\bm{912}$, $\Theta_{MN}{}^P$, and the $\bm{56}$, $\vartheta_{M}$, of the embedding tensor. Next, we break SL$(8, \mathbb{R})$ as in (\ref{eq:SL8ToSL3SL5}) and consider the branchings (\ref{eq:SL8Split}) along with
{\setlength\arraycolsep{2pt}
\begin{eqnarray} \label{eq:SL8Split2}
& {\bf 28}^\prime \rightarrow & \underline{({\bf 3}, {\bf 1})} + ({\bf 3}^\prime, {\bf 5}^\prime) + ({\bf 1} , {\bf 10}^\prime)  \, , \nonumber \\[5pt]
& {\bf 420}^\prime \rightarrow & \underline{({\bf 3}, {\bf 1})} + ({\bf 1}, {\bf 5})+ ({\bf 3}^\prime, {\bf 5}^\prime)+ ({\bf 1}, {\bf 10}^\prime)+ ({\bf 6},{\bf 5}^\prime) + ({\bf 3}, {\bf 10}) \nonumber \\
&& + ({\bf 8}^\prime, {\bf 10}^\prime)+ ( {\bf 3}, {\bf 24}^\prime)+ ({\bf 1} , {\bf 40}^\prime)+ ({\bf 3}^\prime, {\bf 45}^\prime) \, ,
\end{eqnarray}
}where SO$(1,1)$ charges have again been omitted. Then, we take $\theta_{AB}$ and $\tilde{\theta}^{AB}$ in (\ref{eq:ETCompsSL8}) as before, namely, as in (\ref{eq:Charges1}), (\ref{eq:Charges2bis}), so that only the underlined $\mathrm{SL}(3,\mathbb{R}) \times \mathrm{SL}(5,\mathbb{R})$ components in (\ref{eq:SL8Split}) are active. Finally, for the remaining components, $\tilde{\xi}_A{}^{BCD}$, $\tilde{\vartheta}^{AB}$ in (\ref{eq:ETCompsSL8}), we only activate the underlined $\mathrm{SL}(3,\mathbb{R}) \times \mathrm{SL}(5,\mathbb{R})$ representations in (\ref{eq:SL8Split2}) and write
\begin{eqnarray} \label{eq:ET28and420}
\tilde{\vartheta}^{ab} = -\tfrac12 \, g_2 \, \epsilon^{abc} f_{cd}{}^{d} \; , \qquad 
\tilde{\xi}_i{}^{jab} = \tfrac23 \, g_2 \, \epsilon^{abc} f_{cd}{}^{d} \, \delta_i^j \; , \qquad 
\tilde{\xi}_a{}^{bcd}  = -\tfrac{10}{3} \, g_2 \, f_{ae}{}^{e} \epsilon^{bcd} , 
\end{eqnarray}
with indices again ranging as above (\ref{eq:Charges1}), and with $g_2$ the same coupling that already appears in (\ref{eq:Charges2bis}). Both components of $\tilde{\xi}_A{}^{BCD}$ that appear in (\ref{eq:ET28and420}) belong to the same underlined representation in (\ref{eq:SL8Split2}), as they are related through $\tilde{\xi}_i{}^{iab} + \tilde{\xi}_c{}^{cab} =0$. It is also helpful to note the identities
\begin{equation} \label{eq:epsSCIds}
\epsilon^{cd[a}f_{cd}{}^{b]} =-\epsilon^{abc}f_{cd}{}^{d} \; , \qquad 
f_{ab}{}^{[c} \epsilon^{de]b} = \tfrac13f_{ab}{}^{b} \epsilon^{cde} \; .
\end{equation}
While the symmetrised combination $\epsilon^{cd(a}f_{cd}{}^{b)}$ goes into $\tilde{\theta}^{AB}$ in (\ref{eq:Charges2bis}) regardless of whether $f_{ab}{}^c$ is traceless or traceful, in the latter case, an antisymmetric contribution $\epsilon^{cd[a}f_{cd}{}^{b]}$ (proportional to the trace by (\ref{eq:epsSCIds})) leaks into $\tilde{\xi}_A{}^{BCD}$ as given in (\ref{eq:ET28and420}). A non-vanishing trace $f_{ab}{}^b$ also turns on the trombone components $\tilde{\vartheta}^{ab}$ given in the latter equation.

\begin{table}[]


\resizebox{\textwidth}{!}{

\scriptsize{

\begin{tabular}{c| c c c c} 
~  &~&~\\[-4mm]
$\Theta_M{}^\alpha$ &${\bf 28}$
&&&${\bf 28}^\prime$
\\ \hline
~&~&~\\[-3.5mm]
${\bf 63}$
&$ {\bf 36}  + {\bf 420} $
&&& $ {\bf 36}^\prime  + {\bf 420}^\prime$
\\
${\bf 70} $
& $ {\bf 420}^\prime$
&&&$ {\bf 420} $
\\ \hline
\end{tabular}

\qquad

\begin{tabular}{c| c c c c} 
~  &~&~\\[-4mm]
$ (t^\alpha)_M{}^N \vartheta_N $ &$ {\bf 28} $
&&&${\bf 28}^\prime  $
\\ \hline
~&~&~\\[-3.5mm]
$ {\bf 63} $
&$ {\bf 28} $
&&& $ {\bf 28}^\prime  $
\\
$ {\bf 70} $
& $ {\bf 28}^\prime  $
&&&${\bf 28} $
\\ \hline
\end{tabular}

}
}

\caption{\footnotesize{ Couplings between gauge fields and $\textrm{E}_{7(7)}$ generators induced by the $\bm{912}$ (left, taken from \cite{deWit:2007mt}) and $\bm{56}$ (right) components of the embedding tensor, branched out into $\textrm{SL}(8,\mathbb{R})$ representations.
}\normalsize}
\label{tab:Gaugings}
\end{table}

The embedding tensor of our TCSO$(p,q,r; \mathrm{N})$ theories is obtained by bringing the above definitions to (\ref{eq:defX}). The result is best expressed by splitting fundamental E$_{7(7)}$ indices under SL$(8,\mathbb{R})$ as above (\ref{eq:ETCSO}), and then further down under (\ref{eq:SL8ToSL3SL5}) using the first branching in (\ref{eq:SL8Split2}) and the analogue for the $\bm{28}$. Omitting again fundamental representation indices, we finally obtain $X_M = (X_{AB} , X^{AB})= (X_{ab} , X_{ai} , X_{ij} , X^{ab} , X^{ai} , X^{ij})$, with 
{\setlength\arraycolsep{0pt}
\begin{eqnarray} \label{eq:ETtrombone}
&& X_{ij} = -2 g_1 \, \eta_{k [i} \, t_{j]}{}^k \; , \qquad 
X_{ai} = g_1 \, t_a{}^j \eta_{ij} -\tfrac{1}{18} g_2 \, f_{ab}{}^{b} \epsilon^{cde} t_{icde}  \; , \qquad 
X_{ab} = X^{ij} = 0  , \quad \\[4pt]
&& 
X^{ai} =  -\frac{1}{2} g_2 \,  \epsilon^{bcd} f_{bc}{}^a \, t_d{}^i    \; , \quad
X^{ab} =  \tfrac12 g_2 \epsilon^{abc} f_{cd}{}^{d} (t_0 - \tfrac43 t_e{}^e)  + g_2 \epsilon^{abc} f_{cd}{}^e ( t_e{}^d - \tfrac13 \delta^d_e t_f{}^f) \; . \nonumber 
\end{eqnarray}
}We have verified that this embedding tensor satisfies all required constraints, linear, (\ref{eq:LinearConst}), and quadratic, (\ref{eq:QCs}), and thus defines viable gaugings of $D=4$ $\cN=8$ supergravity. 
The expression (\ref{eq:ETtrombone}) reduces to (\ref{eq:ETCSO}) with (\ref{eq:Charges1}), (\ref{eq:Charges2bis}) for unimodular $G_3$, and extends the latter embedding tensor to the non-unimodular case.

Let us now argue that (\ref{eq:ETtrombone}) are indeed the generators of the Lie algebra of (\ref{eq:ProdTCSO}). First of all, $X^{ai}$ can be shown to depend linearly on $X_{ij}$, $X_a \equiv \tfrac12 \epsilon_{abc} X^{bc}$ and $X_{ai}$. Thus, only the latter set of generators, the number of which is (\ref{eq:dimTCSO}), should be considered for this purpose. Now, the quadratic constraints (\ref{eq:QCs}) together with the built-in linear constraint guarantee the closure of the general algebra (\ref{eq:LieAlg}). The latter produces the following non-vanishing commutators for the linearly independent generators:
\begin{eqnarray} \label{eq:CommutatorsTCSOpqrN}
& [X_a,X_b] = g_2 \, f_{ab}{}^c X_c \; , \qquad  [ X_{ij} , X_{k\ell} ] = 4 g_1 \, \eta_{[i [k} X_{\ell]  j]}  \; , \nonumber \\[6pt] 
& [X_a,X_{bi}] = g_2 \, f_{ab}{}^c X_{ci} - g_2 f_{ac}{}^{c} X_{bi}  \;  , \qquad
[ X_{ij} , X_{ak} ] = -2 g_1 \, \eta_{k[i|} X_{a|j]} \; . 
\end{eqnarray}
The upper line in (\ref{eq:CommutatorsTCSOpqrN}) makes it apparent that $X_a$ and $X_{ij}$ respectively generate $G_3$ and CSO$(p,q,r)$, $p+q+r =5$. In the duality frame of (\ref{eq:ETtrombone}), the former is gauged magnetically and the latter electrically, as advertised above. The nilpotent generators $X_{ai}$ commute among themselves and thus generate the $\mathbb{R}^{15}$ factor of (\ref{eq:ProdTCSO}). The semidirect action of $G_3 \times \textrm{CSO}(p,q,r)$ on $\mathbb{R}^{15}$ shown in (\ref{eq:ProdTCSO}) is realised by the commutators in the second line of (\ref{eq:CommutatorsTCSOpqrN}). These enforce the $X_{ai}$ generators to transform in the $(\bm{3}, \bm{p}+ \bm{q}) + (\bm{3}, \bm{1}) + \stackrel{r}{\ldots} + (\bm{3}, \bm{1}) $ representation of $G_3 \times \textrm{SO} (p,q) \, \subset \, G_3 \times \textrm{CSO} (p,q,r)$. In the defining frame of (\ref{eq:ETtrombone}), the $\mathbb{R}^{15}$ factor is dyonically generated by $X_{ai}$ and $X^{ai}$.

As anticipated, the $\mathbb{R}^+$ trombone generator $t_0$ is involved in the gauging specified by the embedding tensor (\ref{eq:ETtrombone}). It indeed appears in the generators $X^{ab}$ of $G_3$ affected by the trace $f_{ab}{}^b$, along with the generator $t_a{}^a$ of the $\mathrm{SO}(1,1)$ in (\ref{eq:SL8ToSL3SL5}) that commutes with $\mathrm{SL}(5,\mathbb{R}) \times \mathrm{SL}(3,\mathbb{R})$ inside $\mathrm{SL}(8,\mathbb{R})$. The second contribution to $X^{ab}$ is selected out of the generators, $( t_e{}^d - \tfrac13 \delta^d_e t_f{}^f)$, of $\mathrm{SL}(3,\mathbb{R}) \subset \mathrm{SL}(8,\mathbb{R})$ by using (traceless combinations of) the $G_3 \subset \mathrm{GL}(3,\mathbb{R})$ structure constants $f_{cd}{}^e$ .  Thus, the expression for $X^{ab}$ in (\ref{eq:ETtrombone}) makes it apparent that $G_3$ is realised as a subgroup of $\mathbb{R}^+ \times \mathrm{SO}(1,1) \times \mathrm{SL}(3,\mathbb{R}) \,  \subset \, \mathbb{R}^+ \times \mathrm{SL}(8,\mathbb{R})  \,  \subset \, \mathbb{R}^+ \times \mathrm{E}_{7(7)}$. Similarly, the CSO$(p,q,r)$, $p+q+r=5$, generated by $X_{ij}$ is a subgroup of $\mathrm{SL}(5,\mathbb{R}) \,  \subset \, \mathrm{SL}(8,\mathbb{R})  \,  \subset \, \mathrm{E}_{7(7)}$. However, the nilpotent factor $\mathbb{R}^{15}$ is not contained in the $\mathrm{SL}(8,\mathbb{R})$ maximal subgroup of $\mathrm{E}_{7(7)}$. This is due to the dependence of $X_{ai}$ on the generators 
$t_{ABCD}$, anticipated in table \ref{tab:Gaugings}. Therefore, our $D=4$ $\cN=8$ TCSO$(p,q,r; \mathrm{N})$ gaugings are not contained in the maximal subgroup $\mathbb{R}^+ \times \mathrm{SL}(8,\mathbb{R})$ of $\mathbb{R}^+ \times \mathrm{E}_{7(7)}$. Nor they are contained in other maximal subgroup thereof, like $\mathbb{R}^+ \times \mathrm{SL}(3,\mathbb{R})\times \mathrm{SL}(6,\mathbb{R})$, for that matter. This is no contradiction, as a nilpotent subgroup of E$_{7(7)}$ like $\mathbb{R}^{15}$ does not need to be contained in one of its maximal subgroups. This is in contrast with the standard $D=4$ $\cN=8$ dyonic CSO gaugings \cite{DallAgata:2011aa,Dall'Agata:2014ita}, which are contained in $\mathrm{SL}(8,\mathbb{R})$ as reviewed in section \ref{sec:CSO}. This is also in contrast with the $D=5$ $\cN=8$ TCSO$(p,q,r;\rho)$ counterparts \cite{Varela:2025xeb} of our four-dimensional theories. The latter have trombone components of the embedding tensor active, but are nevertheless contained in the maximal subgroup $\mathbb{R}^+ \times \mathrm{SL}(2,\mathbb{R}) \times \mathrm{SL}(6,\mathbb{R})$ of $\mathbb{R}^+ \times \mathrm{E}_{6(6)}$.

In the reminder of the paper, we will select a specific theory in our TCSO$(p,q,r;\mathrm{N})$ class, and study some of its consistent subsectors containing interesting supersymmetric AdS vacua.


\section{Supersymmetric subsectors and vacua of $\mathrm{TCSO}(5,0,0;\mathrm{V})$} \label{eq:AdSvac}


Let us now focus on the $p=5$, $q=r=0$ theory with three-dimensional group $G_3$ of Bianchi type V. The embedding tensor is (\ref{eq:ETtrombone}), with $\theta_{ij}$ and $f_{ab}{}^c$, respectively taken to be the SO(5)-invariant metric and the Bianchi type V structure constants, namely,
\begin{equation} \label{eq:SCBianchiV}
\eta_{ij} = \delta_{ij} \; , \qquad 
f_{13}{}^1 = -f_{31}{}^1 = f_{23}{}^{2}= -f_{32}{}^{2} = 1  \; ,
\end{equation}
with all other components of $f_{ab}{}^c$ not shown here vanishing.

We will discuss two specific $\cN=2$ and $\cN=1$ subsectors of the $\mathrm{TCSO}(5,0,0;\mathrm{V})$ theory, as an application of the discussion of section \ref{sec:4DSugra}. These subsectors arise by imposing invariance under different $\textrm{SO}(3)_S$ subgroups of SU(8). In either case, $\textrm{SO}(3)_S$ is not contained in the $\cN=8$ gauge group $ G \equiv \mathrm{TCSO}(5,0,0;\mathrm{V}) \equiv \big( \mathrm{SO} (5) \times G_3 \big) \ltimes  \mathbb{R}^{15}$. We will show that these subsectors coincide with, or contain, the $D=4$ models obtained in \cite{Gauntlett:2002rv,Donos:2010ax} by straight reduction from $D=11$ supergravity on the M5-brane geometries of \cite{Acharya:2000mu,Gauntlett:2006ux}, without passing through our  intermediate $D=4$ $\cN=8$ $\mathrm{TCSO}(5,0,0;\mathrm{V})$ theory. We will indeed recover the M5-brane near-horizon AdS solutions of \cite{Acharya:2000mu,Gauntlett:2006ux}, realised here as vacua of our $D=4$ models, and will compute their mass spectra within the full $\cN=8$ $\mathrm{TCSO}(5,0,0;\mathrm{V})$ theory.  We will respectively refer to these subsectors and vacua as special lagrangian (SLAG) and associative three-cycle, borrowing this terminology from their eleven-dimensional interpretation \cite{Acharya:2000mu,Gauntlett:2006ux}.


\subsection{$\cN = 2$ SLAG subsector} \label{sec:SLAGSector}


The first SO$(3)_S$ invariance group that we will consider is embedded in SU(8) via
\begin{equation} \label{eq:SO3SBreaking}
\textrm{SU}(8) \; \supset \; 
\textrm{SO}(8) \; \supset \; 
\textrm{SO}(5) \times \textrm{SO}(3) \; \supset \;  
\textrm{SO}(2) \times \textrm{SO}(3) \times \textrm{SO}(3) \; \supset \; 
\textrm{SO}(3)_S \; ,
\end{equation}
with the diagonal of the two SO(3) factors taken in the last step. Under the intermediate $\textrm{SO}(5) \times \textrm{SO}(3)$ in (\ref{eq:SO3SBreaking}), 
the fundamental representation of $\mathrm{SL}(8 , \mathbb{R})$ breaks up as in the first relation of (\ref{eq:SL8Split}), where now $\bm{5} \rightarrow \bm{3}_0+ \bm{1}_2 + \bm{1}_{-2} $ under
$ \textrm{SO}(5) \supset
\textrm{SO}(2) \times \textrm{SO}(3)
$. It is thus natural to further split the index $i=4, \ldots , 8$ introduced below  (\ref{eq:SL8Split}) as $i = (\alpha, \bar{a})$, with $\alpha = 4,5$ and $\bar{a} = 6,7,8$. Then, the SO$(3)_S$ generators can be taken to be
\begin{equation} \label{eq:SO3SLAGGen}
T_a = \tfrac12 \epsilon_a{}^{bc} R_{bc} +  \tfrac12 \epsilon_a{}^{\overline{b} \overline{c}} R_{\overline{b} \overline{c}}\; , 
\end{equation}
where $R_{AB} = 2 \, t_{[A}{}^C \delta_{B]C}$ are the SO(8) generators, with $A = (a , i) = (a,  \alpha, \bar{a})$. The generators (\ref{eq:SO3SLAGGen}) cannot be written as linear combinations of the embedding tensor (\ref{eq:ETtrombone}) with (\ref{eq:SCBianchiV}), and thus SO$(3)_S$ lies outside the $\mathrm{TCSO}(5,0,0;\mathrm{V})$ gauge group. Interestingly, the SO$(3)_S$ defined by (\ref{eq:SO3SBreaking}), (\ref{eq:SO3SLAGGen}) is the same that preserves the $\cN=1$ AdS solution found in \cite{Comsa:2019rcz} (see also \cite{Bobev:2019dik}) within the purely electric SO(8) gauging \cite{deWit:1982ig}. In the solution of \cite{Comsa:2019rcz,Bobev:2019dik}, unlike in the present case, SO$(3)_S$ does sit inside the SO(8) gauge group.

Let us determine the field content of the subsector invariant under the SO$(3)_S$ defined by (\ref{eq:SO3SBreaking}), (\ref{eq:SO3SLAGGen}), following the process described in section \ref{sec:4DSugraCTKin}. We are instructed to determine the SO$(3)_S$-invariant tensors contained in the E$_{7(7)}$ and SU(8) representations collected in table \ref{tab:4DSummary}. Bringing (\ref{eq:SO3SLAGGen}) to the first equation in (\ref{eq:InvarFields}), we find four such invariant tensors, $\hat{K}_I{}^M= (\hat{K}_I{}^{AB}, \hat{K}_{I\, AB})$, $I=1, \ldots 4$, descending from the $\bm{56}$ of E$_{7(7)}$. Their non-vanishing components are
\begin{equation} \label{eq:SLAGInvVecs}
\hat{K}_1{}^{a \overline{b}} = \tfrac{1}{\sqrt 3} \, \delta^{a \overline{b}} \; , \qquad
\hat{K}_2{}^{\alpha \beta} = \epsilon^{\alpha \beta} \; , \qquad
\hat{K}_{3 \, a \overline{b}} = \tfrac{1}{\sqrt 3} \, \delta_{a \overline{b}} \; , \qquad
\hat{K}_{4  \, \alpha \beta} = \epsilon_{\alpha \beta} \; .
\end{equation}
In turn, SO$(3)_S$ commutes inside E$_{7(7)}$ with $\mathrm{SL}(2,\mathbb{R}) \times \mathrm{G}_{2(2)}$, where the $\mathrm{SL}(2,\mathbb{R})$ factor is generated by
\begin{equation} \label{eq:SL2SLAG}
H_0= -2 (t_4{}^4 + t_5{}^5 )\,, \qquad  E_0= \tfrac12 \delta^{a\overline{b}} \delta^{c\overline{d}}t_{a \overline{b} c \overline{d}} \,, \qquad  F_0=-\delta^{a\overline{b}}t_{a \overline{b} 4 5} \; ,
\end{equation}
and the $\mathrm{G}_{2(2)}$ factor by 
\begin{equation} \label{eq:G2SLAG}
\begin{array}{lclcl}
 H_1= t_5{}^5 - t_4{}^4  \; , && H_2= -   \,  ( 2 t_a{}^a + t_4{}^4 + t_5{}^5) \; , &&  \\[4pt]
E_1= \tfrac{\sqrt{2}}{12} \,  t_5{}^4 \; ,  && E_2=  - \tfrac{1}{\sqrt{6}} \,   \epsilon^{a\overline{b}\overline{c}} t_{a\overline{b}\overline{c}5} \; ,  &&  E_3= \tfrac{1}{\sqrt{6}} \,   \epsilon^{ab\overline{c}} t_{ab\overline{c}5} \; , \\[4pt]
E_4=\tfrac{\sqrt{2}}{12} \,   \delta^{\overline{b}}_a t_{\overline{b}}{}^a  \; ,  && E_5= -  \tfrac{\sqrt{2}}{3!} \,   \epsilon^{\overline{a}\overline{b}\overline{c}} t_{\overline{a}\overline{b}\overline{c}5} \,  \; , &&  E_6= -  \tfrac{\sqrt{2}}{3!} \,   \epsilon^{abc} t_{abc5} \; , \\[4pt]
F_1= \tfrac{\sqrt{2}}{12} \,  t_4{}^5 \; ,  &&  F_2= \tfrac{1}{\sqrt{6}} \,   \epsilon^{ab\overline{c}} t_{ab\overline{c}4}  \; ,  && F_3= \tfrac{1}{\sqrt{6}} \,   \epsilon^{a\overline{b}\overline{c}} t_{a\overline{b}\overline{c}4} \; ,  \\[4pt]
F_4= \tfrac{\sqrt{2}}{12} \,   \delta^a_{\overline{b}} t_a{}^{\overline{b}} \;, && F_5=-  \tfrac{\sqrt{2}}{3!} \,   \epsilon^{abc} t_{abc4} \; ,  && F_6= \tfrac{\sqrt{2}}{3!} \,   \epsilon^{\overline{a}\overline{b}\overline{c}} t_{\overline{a}\overline{b}\overline{c}4} \; .
\end{array}
\end{equation}
Collectively denoting the generators (\ref{eq:SL2SLAG}), (\ref{eq:G2SLAG}) by $t_x$, the SO$(3)_S$ invariant tensors $\hat{K}_{x}{}^{\alpha}$ can be read off from these equations as the coefficients in the r.h.s.~of $t_x=\hat{K}_{x}{}^{\alpha} \,  t_\alpha$. In addition, we find two and six SO$(3)_S$-invariant tensors coming from the $\bm{8}$ and $\bm{56}$ of SU$(8)$. Thus, besides the metric, the SO$(3)_S$-invariant field content comprises four gauge fields (two electric, two magnetic), $A^I$, $I=1, \ldots, 4$, and ten scalars defined on $\mathrm{SL}(2,\mathbb{R})/\mathrm{SO}(2) \times \mathrm{G}_{2(2)}/\mathrm{SO}(4)$, along with two-forms and fermions. This field content corresponds to a $D=4$ $\cN=2$ supergravity coupled to one vector multiplet and two hypermultiples, in agreement with the supergravity model of \cite{Donos:2010ax}. The four gauge fields are selected out of the $\cN=8$ ones by bringing (\ref{eq:SLAGInvVecs}) to (\ref{eq:RetainedVecs}). Denoting the scalars in each factor of the scalar manifold as $(\varphi_0, \chi_0)$ and $(\varphi, \chi, \phi , a, \zeta^0 , \zeta^1, \tilde{\zeta}_0, \tilde{\zeta}_1)$, an SO$(3)_S$-invariant coset representative can be obtained by exponentiating the generators (\ref{eq:SL2SLAG}), (\ref{eq:G2SLAG}) as
\begin{equation} \label{eq:SugraSectorSLAG}
\tilde{\cV} = e^{-\chi_0 E_0}e^{-\tfrac12 \varphi_0 H_0} e^{\tfrac{1}{\sqrt 2} \left( 12 a E_1 + \zeta_0 E_6 - \zeta_1 E_3 + \tilde \zeta_0 E_5 - \tilde \zeta_1 E_2 \right)} e^{-\phi H_1 -\varphi H_2} \; .
\end{equation}

We still need to decide on the consistency of the SO$(3)_S$-invariant truncation at the dynamical level, following section \ref{sec:4DSugraCTDyn}. In order to do this, we choose to duality-rotate the $\cN=8$ embedding tensor (\ref{eq:ETtrombone}), (\ref{eq:SCBianchiV}), through (\ref{eq:TransXSymbol}) with the one-parameter family of E$_{7(7)}$ elements 
\begin{equation} \label{eq:TransMatSLAG}
U_\lambda = e^{ \Upsilon_\lambda}  \; , \quad \textrm{where} \quad \Upsilon_\lambda \equiv - \tfrac12 \lambda \, g_2 \, g_1^{-1} \, \epsilon^{ab\overline{c}} t_{ab\overline{c}8} \; .
\end{equation}
Here, $g_1$ and $g_2$ are the non-vanishing couplings that already appear in (\ref{eq:ETtrombone}), and $\lambda$ is a real parameter such that $\lambda =0$ leaves the embedding tensor in the original duality frame (\ref{eq:ETtrombone}), while $\lambda =1$ rotates it to what will turn out to be the frame of interest. In general, there appears to be no clear aprioristic argument to propose a duality rotation (\ref{eq:TransXSymbol}) by a specific E$_{7(7)}$ element such as (\ref{eq:TransMatSLAG}). In the case at hand, however, this is informed by our $D=11$ construction \cite{Pico:2026rji}. Note that the E$_{7(7)}$ Lie algebra element $\Upsilon_\lambda$ is not SO$(3)_S$-invariant, as it cannot be written as a linear combinations of generators (\ref{eq:SL2SLAG}), (\ref{eq:G2SLAG}). 

Let us leave $\lambda$ in (\ref{eq:TransMatSLAG}) arbitrary for the moment, in order to illustrate the dynamical consistency issues discussed in section \ref{sec:4DSugraCTDyn}. Bringing (\ref{eq:ETtrombone}), (\ref{eq:TransMatSLAG}) to the r.h.s.~of (\ref{eq:TransXSymbol}), and further splitting the indices as explained below (\ref{eq:SO3SBreaking}), the rotated embedding tensor takes on the form 
{\setlength\arraycolsep{2pt}
\begin{eqnarray} \label{eq:ETSLAGChanged}
\tilde{X}_M &=& (\tilde{X}_{AB} , \tilde{X}^{AB}) 
= (\tilde{X}_{ab} , \tilde{X}_{ai} , \tilde{X}_{ij} , \tilde{X}^{ab} , \tilde{X}^{ai} , \tilde{X}^{ij}) \nonumber \\[4pt]
&=& (\tilde{X}_{ab} , \tilde{X}_{a\beta} , \tilde{X}_{a\bar{b}} , \tilde{X}_{\alpha\beta}, \tilde{X}_{\alpha \bar{b}}  , \tilde{X}_{\bar{a} \bar{b}}  , \tilde{X}^{ab} , \tilde{X}^{a\beta} , \tilde{X}^{a\bar{b}} , \tilde{X}^{\alpha\beta}, \tilde{X}^{\alpha \bar{b}}  , \tilde{X}^{\bar{a} \bar{b}} ) \; , 
\end{eqnarray}
}with electric
{\setlength\arraycolsep{2pt}
\begin{eqnarray} \label{eq:ETSLAGChangedElec}
\tilde{X}_{ab} &=& 0 \; , \nonumber \\[4pt]
\tilde{X}_{a \beta} &=& {X}_{a \beta}  \; , \nonumber \\[4pt]
\tilde{X}_{a \overline{b}} &=&  {X}_{a \overline{b}} + \lambda \, g_2 \, \epsilon_{\overline{b}}{}^{c\overline{d}} \, t_{3a c \overline{d}}  \; , \nonumber \\[4pt]
\tilde  X_{\alpha \beta} &=& X_{\alpha \beta} -\lambda ^2  g_2^2 g_1^{-1} \epsilon_{\alpha \beta} \,  t_3{}^8 + \tfrac12 \lambda \, g_2^2 g_1^{-1} \epsilon_{\alpha \beta} \, \epsilon^{abc} f_{bc}{}^{d} \epsilon_{d\overline{e}3} \,  t_a{}^{\overline{e}}  \; , \nonumber \\[4pt]
  \tilde X_{\alpha \overline{a}} &=& X_{\alpha \overline{a}} +  g_2^2 g_1^{-1} \lambda ^2 \epsilon_{\alpha\beta} t_3{}^\beta \delta^8_{\overline{a}} + \tfrac12 g_2 \lambda \epsilon_{\underline{a}}{}^{bc} t_{\alpha bc 8} \nonumber \\[2pt]
  &&  -  \tfrac12 g_2 \lambda \delta^8_{\underline{a}}\epsilon^{bc\overline{d}} t_{\alpha bc \overline{d}} + \tfrac12  g_2^2 g_1^{-1} \lambda \epsilon^{abc} f_{bc}{}^d \epsilon_{\overline{a} d 3} \epsilon_{\alpha \beta} t_{a}^{\beta} \; , \nonumber \\[4pt]
\tilde X_{\overline{a} \overline{b}} &=&  X_{\overline{a} \overline{b}} -  \lambda \, g_2 \, \epsilon_{[\overline{a}}{}^{cd}t_{\overline{b}]bc8} + \lambda \, g_2  \, \delta^8_{[\overline{a}} \epsilon^{c d \overline{e}} t_{\overline{b}]c d \overline{e}} \; ,
\end{eqnarray}
}and magnetic components
{\setlength\arraycolsep{2pt}
\begin{eqnarray} \label{eq:ETSLAGChangedMag}
\tilde{X}^{ab} &=& X^{ab} + \tfrac12 \lambda \, g_2^2 g_1^{-1}  \epsilon^{[a|cd} f_{cd}{}^{\overline{e}} \epsilon^{|b]fg} t_{\overline{e}fg8} +\lambda^2  g_2^2 g_1^{-1} \epsilon^{8 \overline{e} \overline{f} }\epsilon^{3c[a} \delta^{b]d} t_{cd \overline{e} \overline{f}}  \nonumber \\[2pt]
&& + \lambda g_2  \epsilon^{ab \overline{c}} t_{\overline{c}}{}^8 -  \lambda g_2  \epsilon^{ab}{}_{ \overline{c}} t_8{}^{\overline{c}} \; , \nonumber \\[6pt]
\tilde{X}^{a \beta} &=& {X}^{a \beta}  \; , \nonumber \\[4pt]
\tilde{X}^{a \overline{b}} &=&  {X}^{a\overline{b}} - \tfrac14 \lambda \, g_2^2 g_1^{-1}  \epsilon^{acd}f_{cd}{}^{e} \epsilon^{\overline{b}fg} t_{efg8}- \tfrac14 \lambda \, g_2^2 g_1^{-1} \epsilon^{acd}f_{cd}{}^{\overline{e}} \epsilon^{\overline{b}fg} t_{\overline{e}fg3} \nonumber \\[2pt]
&& + \lambda^2 g_2^2 g_1^{-1}   \epsilon^{\overline{b}c3} \epsilon^{d \overline{e} 3}  \delta^{af} t_{fcd \overline{e}} + \lambda \, g_2 \epsilon^{ace} \epsilon^{\overline{b}}{}_{\overline{d}f} \epsilon_{e}{}^{f3} t_c{}^{\overline{d}}  \; , \nonumber \\[6pt]
\tilde X^{\alpha \beta} &=&    \tilde X^{\alpha \overline{b}} = \tilde  X^{\overline{a} \overline{b}} =0   \; .
\end{eqnarray}
}The untilded components, $X_{a \beta}$, etc., that appear on the r.h.s.~of (\ref{eq:ETSLAGChangedElec}), (\ref{eq:ETSLAGChangedMag}) correspond to the original expressions (\ref{eq:ETtrombone}) with branched-out indices as in (\ref{eq:ETSLAGChanged}), and particularised to $\eta_{ij}= \delta_{ij}$. These expressions are general for any Bianchi type $\mathrm{N}$ structure constants, and not merely for the type $\mathrm{V}$ structure constants $f_{ab}{}^c$ in (\ref{eq:SCBianchiV}). We also note, for later reference, that in the rotated frame (\ref{eq:ETSLAGChanged})--(\ref{eq:ETSLAGChangedMag}), the non-vanishing trombone components stay the same as in the original frame (\ref{eq:ETtrombone}), namely, as in the first entry of (\ref{eq:ET28and420}).

The truncation of our $\cN=8$ TCSO$(5,0,0;\mathrm{V})$ supergravity will be consistent to the SO$(3)_S$-invariant sector if the linear, (\ref{eq:TruncCondXLin}), and quadratic, (\ref{eq:TruncCondXQuad}), constraints are satisfied. Combining the invariant tensors (\ref{eq:SLAGInvVecs}) with the embedding tensor (\ref{eq:ETSLAGChanged})--(\ref{eq:ETSLAGChangedMag}) as in (\ref{eq:TruncGen}), we find the following expressions for the $X_{IM}{}^N$, $I=1,  \ldots ,  4$,
\begin{eqnarray} \label{eq:SLAGIntTor}
\tilde{X}_1 = \tfrac{1}{\sqrt 3} \, \delta^{a \overline{b}} \tilde X_{a \overline{b}}  \; , \qquad 
\tilde{X}_2 = \tfrac12 \epsilon^{\alpha \beta} \tilde X_{\alpha \beta}  \; , \qquad 
\tilde{X}_3 =  \tfrac{1}{\sqrt 3} \, \delta_{a \overline{b}} \tilde X^{a \overline{b}}  \; , \qquad 
\tilde{X}_4 = \tfrac12 \epsilon_{\alpha \beta} \tilde X^{\alpha \beta} , \quad 
\end{eqnarray}
with representation indices suppressed. These generators can be checked to commute among themselves for all $\lambda$, thus yielding vanishing $\tilde{X}_{IJ}{}^K $ in (\ref{eq:TruncGen}). This leads to an abelian gauging in the $\cN=2$ vector multiplet sector, and Fayet-Iliopoulos gauging in the hyperscalar sector. Next, we impose the linear constraint C1 in (\ref{eq:TruncCondXLin}) by commuting all three SO$(3)_S$ generators (\ref{eq:SO3SLAGGen}) with all four generators (\ref{eq:SLAGIntTor}) of the truncated gauge group. All such commutators produce linear combinations of E$_{7(7)}$ generators with coefficients proportional to either function $c_1(\lambda)$ or $c_2(\lambda)$ given by
\begin{equation} \label{eq:SLAGLinearConst}
c_1(\lambda) = \lambda -1  \; , \qquad 
c_2(\lambda) = \lambda (\lambda -1)  \; .
\end{equation}
Similarly, briging the SO$(3)_S$ generators (\ref{eq:SO3SLAGGen}), the embedding tensor (\ref{eq:ETSLAGChanged})--(\ref{eq:ETSLAGChangedMag}) and the invariant tensors $\hat{K}_x{}^\alpha$ described below (\ref{eq:G2SLAG}) to the relation C2 in (\ref{eq:TruncCondXLin}), a set of matrix equations is produced where all components are again proportional to the functions of $\lambda$ shown in (\ref{eq:SLAGLinearConst}). Thus, in this example, the constraint C2 adds nothing new with respect to C1, but this is not always the case, see section \ref{sec:A3CSector}. The linear constraints (\ref{eq:TruncCondXLin}) with zero r.h.s.~require (\ref{eq:SLAGLinearConst}) to vanish, which in turn selects the $\lambda=1$ frame for the embedding tensor (\ref{eq:ETSLAGChanged})--(\ref{eq:ETSLAGChangedMag}). The quadratic constraint C3 is also satisfied only if $\lambda=1$. Thus, only in the $\lambda=1$ frame is the SO$(3)_S$-invariant truncation consistent. In particular, it is inconsistent to truncate to SO$(3)_S$-invariant fields in the $\lambda=0$ frame, where the embedding tensor takes on the defining form (\ref{eq:ETtrombone}), (\ref{eq:SCBianchiV}) for our $\cN=8$ TCSO$(5,0,0;\mathrm{V})$ gauged supergravity. 

Of course, there could be other frames, not contained in the class (\ref{eq:ETSLAGChanged}) but related to the original frame (\ref{eq:ETtrombone}) by E$_{7(7)}$ rotations (\ref{eq:TransXSymbol}) outside the family (\ref{eq:TransMatSLAG}), which still render the SO$(3)_S$-invariant truncation consistent. Those truncated SO$(3)_S$-invariant theories may or may not be equivalent to the one just described. We will not explore those possibilities further and, in the next subsection, we will restrict our frame to be (\ref{eq:ETSLAGChanged})--(\ref{eq:ETSLAGChangedMag}) with $\lambda=1$.


\subsection{$\cN = 2$ SLAG AdS vacuum} \label{sec:SLAGVac}


The equations of motion of the SO$(3)_S$-invariant subsector are obtained by 
bringing the retained fields along with the embedding tensor (\ref{eq:ETSLAGChanged})--(\ref{eq:ETSLAGChangedMag}) with $\lambda = 1$ to the $\cN=8$ field equations reviewed in appendix \ref{sec:4DSugraEoMsMass}. This produces the equations of motion (\ref{eq:EinsteinEOMRed})--(\ref{eq:SDandBianchiRed}) for the SO$(3)_S$-invariant sector. Now, the retained generators (\ref{eq:SLAGIntTor}) are traceless, and therefore the trombone components of the embedding tensor vanish, as in (\ref{eq:TrombVecContr}). This can be equivalently seen by contracting the trombone components $\tilde{\vartheta}_M$ of $\tilde{X}_M$, which stay as in (\ref{eq:ET28and420}) as stated below (\ref{eq:ETSLAGChangedElec}), (\ref{eq:ETSLAGChangedMag}), with the invariant tensors (\ref{eq:SLAGInvVecs}). Thus, by the discussion at the end of section \ref{sec:4DSugraCTDyn}, the equations of motion of this subsector integrate into a Lagrangian, in agreement with \cite{Donos:2010ax}. In the parametrisation (\ref{eq:SugraSectorSLAG}), the scalar potential that features in the lagrangian turns out to be rather complicated and not particularly illuminating to be reproduced here. Also, we do not wish to keep track of couplings to gauge fields, as we will be ultimately interested in finding vacuum solutions. 

For these reasons, we choose to present the Lagrangian corresponding to a further consistently truncated subsector. This sector is obtained by imposing $\mathbb{Z}_2$ invariance on top of SO$(3)_S$. While, for simplicity, the discussion about consistent subsectors of section \ref{sec:4DSugra} was limited to continuous invariance groups, we have verified that consistency still holds at the level of the field equations when the $\mathbb{Z}_2$-odd fields are truncated out. The resulting model is thus a bonafide consistent subsector, in the $\lambda=1$ frame of (\ref{eq:ETSLAGChanged})--(\ref{eq:ETSLAGChangedMag}), of our $\cN=8$ TCSO$(5,0,0;\mathrm{V})$ supergravity. Incidentally, the same SO$(3)_S \times \mathbb{Z}_2$--invariant sector has been considered in \cite{Bobev:2019dik} and appendix A of \cite{Guarino:2015qaa} as a subsector of the electric SO(8) and dyonic ISO(7) gaugings, respectively. In those cases, unlike here, SO$(3)_S \times \mathbb{Z}_2$ is a subgroup of the corresponding gauge group.

\begin{table}[]


\resizebox{\textwidth}{!}{

\begin{tabular}{l | l | l } 
\hline
\hline
\textbf{Field} & \textbf{SLAG} & \textbf{Associative 3-cycles} \\[6pt]
\hline
Graviton & $0^{(1)}$ & $0^{(1)}$  \\[6pt]
\hline 
Gravitini & $\pm 2^{(1)}$, $ \pm \sqrt{3}^{(2)}$, $ \pm 1^{(1)}$ &
$ 2^{(1)}$, $ \pm \frac{\sqrt{11}}{2}^{(1)}$, $ \pm \frac{\sqrt{29}}{4}^{(2)}$, $  1^{(1)}$  \\[6pt]
\hline 
\multirow{2}{*}{Vectors} & \multirow{2}{*}{$6^{(4)}$, $(3+\sqrt{3})^{(8)}$, $4^{(1)}$, $2^{(4)}$, $(3-\sqrt{3})^{(8)}$, $0^{(1)}$, $0^{(2)}_*$} &
$6^{(1)}$, $5^{(1)}$, $\tfrac14(11+2\sqrt{11})^{(2)}$, $\frac{15}{4}^{(2)}$, $\tfrac{1}{16}(29+4\sqrt{29})^{(2)}$, $\frac{45}{16}^{(4)}$, \\ [6pt]
&&$2^{(1)}$, $\tfrac14(11-2\sqrt{11})^{(2)}$, $\tfrac{1}{16}(29-4\sqrt{29})^{(2)}$, $\frac{5}{16}^{(4)}$, $0^{(3)}$   \\[6pt]
\hline 
\multirow{2}{*}{Spin-$\tfrac12$} & $\pm 3^{(2)}$, $\pm (1+ \sqrt{3})^{(2)}$, $\pm \left(\frac{1}{2} \left(1+\sqrt{17}\right)\right)^{(1)}$, $\pm 2 ^{(4)}$,  $\pm \sqrt{3}^{(8)}$, &
$\tfrac12(1+\sqrt{21})^{(1)}$, $\pm \tfrac52^{(1)}$, $\pm \sqrt6^{(1)}$, $\pm \tfrac94^{(2)}$, $ 2^{(1)}$, $\pm \tfrac{\sqrt{61}}{4}^{(2)}$, $\tfrac12(-1+\sqrt{21})^{(1)}$,   \\[6pt]  
&$\pm \left(\frac{1}{2} \left(1-\sqrt{17}\right)\right)^{(1)}$, $\pm1^{(3)}$, $\pm (1- \sqrt{3})^{(2)}$, $0^{(4)}$ &
$\pm \tfrac{\sqrt{11}}{2}^{(1)}$, $\pm \tfrac32^{(1)}$, $\pm \tfrac{\sqrt{29}}{4}^{(2)}$, $\pm \tfrac54^{(4)}$, $1^{(5)}$, $\pm \tfrac14^{(2)}$, $0^{(3)}$, $\pm \frac{i}{2}^{(3)}$
\\[6pt]
\hline
\multirow{2}{*}{Scalars} & $10^{(2)}$, $(3+\sqrt{17})^{(1)}$, $4^{(6)}$, $(1+\sqrt{3})^{(8)}$, $-2^{(6)}$ , $-2^{(2)}_*$  ,& 
\multirow{2}{*}{$(4+\sqrt{6})^{(2)}$, $\tfrac{1}{16}(29+4\sqrt{61})^{(4)}$, $3^{(1)}$, $-2^{(5)}$, $\tfrac74^{(2)}$, $-\tfrac{27}{16}^{(4)}$,  $(4-\sqrt{6})^{(2)}$} \\[6pt]
&  $2^{(3)}$, $(3-\sqrt{17})^{(1)}$, $(1-\sqrt{3})^{(8)}$, $0^{(8)}$  &  $\tfrac{13}{16}^{(4)}$, $\tfrac{1}{16}(29-4\sqrt{61})^{(4)}$, $0^{(5)}$, $\left(-\tfrac94 + \frac{i}{2}\right)^{(6)}$, $\left(-\tfrac94 - \frac{i}{2}\right)^{(6)}$ \\[6pt]
\hline
\hline
\end{tabular}

\qquad 
}
\caption{\footnotesize{Squared masses (for bosons) and linear masses (for fermions) for the SLAG (left) and associative three-cycle (right) solutions, in units of their respective AdS radii $L$, within TCSO$(5,0,0;\mathrm{V})$ supergravity. A superscript in parenthesis indicates the multiplicity. A subscript $*$ indicates a generalised eigenvalue.
}\normalsize}
\label{tab:SLAGAndAssCycMass}
\end{table}

The bosonic fields in this further reduced subsector include only the metric and six real scalars, along with the relevant two-forms. The scalars can be taken to be $\varphi_0$, $\chi_0$ in the $\cN=2$ vector multiplet of section \ref{sec:SLAGSector}, along with
\begin{equation} \label{eq:SO3SZ2SLAG}
\varphi_1 \equiv  \phi + \varphi  \; , \qquad 
\chi_1 \equiv - \tilde \zeta_0 \; , \qquad 
\varphi_2 \equiv  \phi - 3 \varphi  \; , \qquad 
\chi_2 \equiv \tfrac{1}{\sqrt{3}} \,  \zeta_1 \; .
\end{equation}
This sector indeed corresponds to an $\cN=1$ supergravity coupled to three chiral multiplets. In this parametrisation, the lagrangian that gives rise to the SO$(3)_S \times \mathbb{Z}_2$-invariant equations of motion (\ref{eq:EinsteinEOMRed})--(\ref{eq:scalarEOMRed}) finally reads
\begin{equation}	\label{eq:LagrangianSLAG}
{\cal L} =   R \, \textrm{vol}_4 + \tfrac32 (d\varphi_0)^2 + \tfrac32 e^{2\varphi_0} (d \chi_0)^2  + \tfrac12 (d\varphi_1)^2 + \tfrac12 e^{2\varphi_1} (d \chi_1)^2 +  \tfrac32 (d\varphi_2)^2 + \tfrac32 e^{2\varphi_2} (d \chi_2)^2  - V \, \textrm{vol}_4   ,
\end{equation}
where $(d\varphi_0)^2 \equiv d\varphi_0^i \wedge * d\varphi_0^j$. The scalar potential is
{\setlength\arraycolsep{2pt}
\begin{eqnarray} \label{eq:PotSLAG}
V &= & \tfrac18 \, g^2 \, e^{-(\varphi_0 + \varphi_1 + 3 \varphi_2)} \Big[
    4 e^{4 \varphi_0}
    - 12 e^{4 \varphi_2}
    - 24 e^{2(\varphi_0 + \varphi_2)}
    - 8 e^{4 \varphi_0 + \varphi_1 + 3 \varphi_2}
    - 24 e^{2 \varphi_0 + \varphi_1 + 5 \varphi_2} \nonumber \\[4pt]
&&\quad
    + 4 e^{4 \varphi_0 + 2 \varphi_1} \chi_1^2
    - 24 e^{2(\varphi_0 + \varphi_1 + \varphi_2)} \chi_1^2
    - 12 e^{2 \varphi_1 + 4 \varphi_2} \chi_1^2
    + 12 e^{6 \varphi_2} \chi_2^2
    + 12 e^{4 \varphi_0 + 2 \varphi_2} \chi_2^2 \nonumber \\[4pt]
&&\quad
    + 48 e^{2 \varphi_0 + 6 \varphi_2} \chi_0^2 \chi_2^2
    + 12 e^{2(2 \varphi_0 + \varphi_1 + \varphi_2)} \chi_1^2 \chi_2^2
    + 12 e^{2 \varphi_1 + 6 \varphi_2} \chi_1^2 \chi_2^2
    \nonumber \\[4pt]
&&\quad
    + 48 e^{2(\varphi_0 + \varphi_1 + 3 \varphi_2)} \chi_0^2 \chi_1^2 \chi_2^2 
    + 24 e^{2 \varphi_0 + 4 \varphi_2} (1 - \chi_2^2)
    + 24 e^{2(\varphi_0 + \varphi_1 + 2 \varphi_2)} \chi_1^2 (1 - \chi_2^2) \nonumber \\[4pt]
&&\quad
    + 3 e^{4(\varphi_0 + \varphi_2)} (1 + 2 \chi_0^2 + 2 \chi_2^2)^2
    + 3 e^{4 \varphi_0 + 2 \varphi_1 + 4 \varphi_2} \chi_1^2 (1 + 2 \chi_0^2 + 2 \chi_2^2)^2 \nonumber \\[4pt]
&&\quad
    + e^{4 \varphi_0 + 6 \varphi_2} \chi_2^2 (3 + 6 \chi_0^2 + 2 \chi_2^2)^2
    + e^{4 \varphi_0 + 2 \varphi_1 + 6 \varphi_2} 
        \big( 2 + \chi_1 \chi_2 (3 + 6 \chi_0^2 + 2 \chi_2^2) \big)^2 
\Big]  , 
\end{eqnarray}
}where we have defined, for simplicity, $g \equiv g_1 = \sqrt{2} \, g_2$. Contact with the dilaton sector of the model of \cite{Donos:2010ax} is obtained through the identifications
\begin{equation} \label{eq:IdentifDil}
\varphi_0 = 4 \lambda^{\text{there}} - 2 \phi^{\text{there}}\, , \quad
 \varphi_1 =  
 3 \lambda^{\text{there}} + \tfrac12 \varphi_2^{\text{there}} + 6 \phi^{\text{there}}\, , 
\quad 
\varphi_2 = 
 - \lambda^{\text{there}} + \tfrac12 \varphi_2^{\text{there}} - 2 \phi^{\text{there}} \; ,
\end{equation}
with $l_{\textrm{there}} = -1$. Further setting $\varphi_2^{\text{there}} =0$, the identifications (\ref{eq:IdentifDil}) also brings our model (\ref{eq:LagrangianSLAG}), (\ref{eq:PotSLAG}) to (2.12), (2.13) of \cite{Gauntlett:2002rv}, with $d_{\textrm{there}} = 3$, $p_{\textrm{there}} = 3$, $q_{\textrm{there}} = 2$, $m_{\textrm{there}} =  \, g$, and also $l_{\textrm{there}} = -1$, as corresponds to the negatively curved SLAG hyperbolic model considered in \cite{Gauntlett:2002rv,Donos:2010ax}.

The scalar potential (\ref{eq:PotSLAG}) attains an AdS extremum at 
{\setlength\arraycolsep{0pt}
\begin{eqnarray} \label{eq:SLAGvac}
& \textrm{SLAG: } \quad e^{2\varphi_0} = 2 \; , \quad 
\varphi_1 = \varphi_2 = \chi_0 = \chi_1 = \chi_2 = 0 \; , \qquad 
L^2=  \sqrt 2 \,  g^{-2}  \,.
\end{eqnarray}
}The AdS radius, $L^2 = -6/V$ with $V$ in (\ref{eq:PotSLAG}) evaluated on the scalar values in  (\ref{eq:SLAGvac}), has also been recorded in the latter equation. This vacuum can be identified with the $\cN=2$  AdS vacuum of \cite{Gauntlett:2002rv,Donos:2010ax}. The latter uplifts \cite{Pico:2026rji} to a $D=11$ solution that describes the near-horizon area of a stack of M5-branes wrapped on a SLAG three-cycle with negative constant curvature inside a Calabi-Yau three-fold \cite{Gauntlett:2006ux}. Regarded as a solution of our $D=4$ $\cN=8$ TCSO$(5,0,0;\mathrm{V})$ supergravity, the vacuum (\ref{eq:SLAGvac}) indeed preserves $\cN=2$ supersymmetry, along a residual $\textrm{U}(1)_R$ bosonic symmetry. This is the $\cN=2$ R-symmetry, and coincides with the intermediate $\textrm{U}(1)_R \sim \textrm{SO}(2)$ shown in the branching (\ref{eq:SO3SBreaking}). There are no other (continuous) bosonic symmetries preserved at the vacuum.

\begin{table}[]

\centering


\small{

\begin{tabular}{l | l  } 
\hline
\hline
\textbf{Multiplet} & \textbf{SLAG}  \\
\hline \\[-10pt]
MGRAV & $\, [2,0]_0$  \\[1pt]

\hline \\[-10pt]
LGINO & $  [\tfrac12 + \sqrt{3}, 1]_{ 2}  
\; \oplus \;
 [\tfrac12 + \sqrt{3}, 1]_{- 2}
\; \oplus \;
[\tfrac12 + \sqrt{3},- 1]_{ 2}   
\; \oplus \;
[\tfrac12 + \sqrt{3},- 1]_{- 2}
$
  \\[1pt]
\hline \\[-10pt] 
SGINO & $  
[\tfrac52, 1]_0
\; \oplus \;
[\tfrac52,- 1]_0  $
  \\[1pt]
\hline \\[-10pt] 
MVEC$_*$ &  $[1,0]_{4}
\; \oplus \;
[1,0]_{-4}
$    \\[1pt]
\hline \\[-10pt] 
LVEC &  $[\tfrac12 (1 + \sqrt{17}),0]_0 $    \\[3pt]
\hline \\[-10pt] 
SVEC &  $[3,2]_0%
\; \oplus \;
[3,-2]_0 $    \\
\hline \\[-10pt] 

HYP & $ [4,4]_0
\; \oplus \;
[4,-4]_0
\; \oplus \;
[2,2]_{4}
\; \oplus \;
[2,2]_{-4}
\; \oplus \;
[2,-2]_{4}
\; \oplus \;
[2,-2]_{-4}  $
\\[1pt]
\hline
\hline
\end{tabular}

\qquad 
}
\caption{\footnotesize{Spectrum of $\textrm{OSp}(4|2) \times \mathrm{U}(1)_S$ supermultiplets for the AdS SLAG solution, within $D=4$ $\cN=8$ TCSO$(5,0,0;\mathrm{V})$ supergravity. The OSp$(4|2)$ supermultiplets are denoted with the acronyms of \cite{Klebanov:2008vq}. The entries collect the superconformal primary dimension $E_0$ and R-charge $y_0$ for each $\textrm{OSp}(4|2)$ supermultiplet, along with the non-conserved charge U$(1)_S$ $q$ (identical for all states in the same supermultiplet) as $[E_0,y_0]_q$. Supermultiplets marked with a $*$ subindex contain generalised mass eigenstates.
}\normalsize}
\label{tab:SLAGMultiplets}
\end{table}

The masses of the individual $D=4$ $\cN=8$ supergravity fields can be computed from the mass matrices (\ref{eq:VectorMassMat}), (\ref{eq:ScalarMassMat}), (\ref{eq:GravitinoMassMat}), (\ref{eq:FermionMassMat}), evaluated at the vacuum (\ref{eq:SLAGvac}) and with the $\lambda=1$ embedding tensor (\ref{eq:ETSLAGChanged})--(\ref{eq:ETSLAGChangedMag}). The resulting masses, obtained after removing spurious and Goldstone states as explained at the end of appendix \ref{sec:4DSugraEoMsMass}, have been brought to table \ref{tab:SLAGAndAssCycMass}. The table also includes, for completeness, the massless graviton state, which we have brought in by hand. Since the mass matrices are no longer symmetric in trombone-gauged supergravity, they are no longer guaranteed to be diagonalisable. Indeed, this turns out to be the case for the gauge field and scalar mass matrices. Both exhibit generalised eigenvalues $0$ and $-2$, respectively, associated to Jordan blocks of size two, and have been marked with an asterisk in table~\ref{tab:SLAGAndAssCycMass}. Recall that generalised eigenvalues in Jordan chains do not have associated eigenvectors in the conventional sense. For this reason, the two $0$ generalised eigenvalues of the gauge field mass matrix do not correspond to symmetries of the vacuum. The associated generalised eigenvectors span an $\mathbb{R}^2$ inside the nilpotent factor $\mathbb{R}^{15}$ of the $\cN=8$ gauge group 
$ \textrm{TCSO} (5,0,0;\mathrm{V}) \equiv \big( G_3 \times \textrm{SO} (5) \big)   \ltimes \mathbb{R}^{15} $.

The mass spectrum accommodates itself, as it must, in representations of OSp$(4|2)$, the $\cN=2$ superisometry group of AdS. In order to see this, one must firstly translate the dimensionless masses $ML$ reported in table \ref{tab:SLAGAndAssCycMass}, referred to the AdS radius $L$ in (\ref{eq:SLAGvac}), to conformal dimensions $\Delta$ via the usual formulae,
\begin{eqnarray} \label{eq:DimMassRel}
& \textrm{gravitons, scalars:} \quad  \Delta (\Delta-3) = M^2L^2 \; , \qquad  
\textrm{vectors:} \quad  (\Delta -1) (\Delta-2) = M^2L^2 \; , \nonumber \\[5pt]
& \textrm{fermions:} \quad  \Delta = \tfrac32 + |ML| \; .
\end{eqnarray}
Secondly, the U$(1)_R$ R-charge of all mass eigenstates must be computed by branching the representations given in table \ref{tab:4DSummary} under the residual $\textrm{U}(1)_R \sim \textrm{SO}(2)$ R-symmetry defined in (\ref{eq:SO3SBreaking}). Finally, supermultiplets must be filled out of states with appropriate patterns of dimensions and R-charges. We follow the notation of \cite{Klebanov:2008vq} for the OSp$(4|2)$ multiplets, appendix A of which comes in  helpful for this exercise. The results are listed in table \ref{tab:SLAGMultiplets}. 

The spectrum not only comes in representations of $\textrm{OSp}(4|2)$ as it must, but, somewhat unexpectedly, it comes in representations of the larger (super)group $\textrm{OSp}(4|2) \times \mathrm{U}(1)_S$, with all states in a given $\textrm{OSp}(4|2)$ multiplet sharing the same $\mathrm{U}(1)_S$ charge. Here, $\mathrm{U}(1)_S$ is the Cartan subgroup of the invariance group SO$(3)_S$ defined in (\ref{eq:SO3SBreaking}), which commutes with $\textrm{U}(1)_R \sim \textrm{SO}(2)$ inside $\textrm{SU}(8)$. This is so even if, as emphasised above, $\mathrm{U}(1)_S$ is not a symmetry, since the only residual compact symmetry is $\textrm{U}(1)_R \subset \textrm{OSp}(4|2)$. While U$(1)_R$ is contained in the $\cN=8$ gauge group 
$ \textrm{TCSO} (5,0,0;\mathrm{V}) \equiv \big( G_3 \times \textrm{SO} (5) \big)   \ltimes \mathbb{R}^{15} $
(through the SO$(5)$ factor), $\mathrm{U}(1)_S$ is not. For that reason, U$(1)_R$ has an associated massless vector state, the proper zero eigenvalue in the gauge field row of table \ref{tab:SLAGAndAssCycMass}. This sits in the MGRAV$[2,0]_0$ mutiplet of table \ref{tab:SLAGMultiplets}, as appropriate to the R-symmetry current. In contrast, no vector in the spectrum becomes massive as a result of the spontanous breaking of the gauge group down to (a residual group containing) $\mathrm{U}(1)_S$, as discussed in section \ref{sec:Subgroup}. Rather puzzlingly,  $\mathrm{U}(1)_S \not\subset \mathrm{C}_{\mathrm{E}_{7(7)}}(G) \cap \mathrm{SU}(8)$, so this symmetry cannot be attributed to the mechanism also discussed in that section.

The two generalised mass eigensates of the gauge field and scalar mass matrices sit in massless vector multiplets $\textrm{MVEC}_* [1,0]_{\pm 4}$. Here and in table \ref{tab:SLAGMultiplets}, an asterisk subindex is used to signify that these multiplets contain these generalised eigenvalues in Jordan chains. Curiously, these states pair up with honest spin-$1/2$ eigenvalues of zero mass to form these multiplets. Note also the presence in table \ref{tab:SLAGAndAssCycMass} of eight massless scalars. These sit in the SGINO$[\tfrac52 ,\pm1]_{0}$ and HYP$[2 ,\pm 2]_{ \pm4}$ multiplets of table \ref{tab:SLAGMultiplets}. Our results for the mass spectrum of the AdS SLAG solution of \cite{Gauntlett:2006ux} are complete within $D=4$ gauged supergravity, in the sense that they have been obtained within the largest possible such theory. Previous partial results on this spectrum were reported in \cite{Donos:2010ax}. The spectrum, computed in that reference within their submaximal theory, was given there as $\textrm{MGRAV} [2,0] \oplus \textrm{LVEC}  [\tfrac12 (1 + \sqrt{17}),0] \oplus \textrm{HYP} [4,\pm4]$. These multiplets are indeed reproduced, further including the $\mathrm{U}(1)_S$ charges, in our table \ref{tab:SLAGMultiplets}.


\subsection{$\cN = 1$ associative three-cycle subsector} \label{sec:A3CSector}


Let us now look at a second SO$(3)_S^\prime$ invariance group, now embedded in SU(8) through
{\setlength\arraycolsep{2pt}
\begin{eqnarray} \label{eq:SO3SA3CBreaking}
\textrm{SU}(8)  & \; \supset \; &
\textrm{SO}(8) \; \supset \; 
\textrm{SO}(5) \times \textrm{SO}(3) \; \supset \;  
\textrm{SO}(4) \times \textrm{SO}(3)  \; \sim \; 
\textrm{SO}(3)_- \times\textrm{SO}(3)_+ \times \textrm{SO}(3) \nonumber \\[4pt]
& \; \supset \; & \textrm{SO}(3)_S^\prime \times \textrm{SO}(3)_+  \; \supset \;  
\textrm{SO}(3)_S^\prime \; , 
\end{eqnarray}
}so that $\textrm{SO}(3)_S^\prime$ is the diagonal subgroup of the $\textrm{SO}(3)_- \times \textrm{SO}(3)$ that appears in the last entry of the first line. We prime it simply to typographically distinguish it from the distinct $\textrm{SO}(3)_S$ of (\ref{eq:SO3SBreaking}). This $\textrm{SO}(3)_S^\prime$ is generated by
\begin{equation} \label{eq:SO3A3CGen}
T_a = - \tfrac12 \epsilon_a{}^{bc} R_{bc} + \tfrac12 (J^-_a){}^{ij} \, R_{ij} \; .
\end{equation}
Here, $R_{ab}$ and $R_{ij}$ are the generators of the $\textrm{SO}(3)$ and $\textrm{SO}(5)$ factors that appear in the second step of (\ref{eq:SO3SA3CBreaking}), suitably selected among the SO(8) generators $R_{AB}$ defined below (\ref{eq:SO3SLAGGen}), and $J^\pm_a$ are the generators of $\textrm{SO}(3)_- \times \textrm{SO}(3)_+$ in the $(\bm{2} , \bm{2} ) +  (\bm{1} , \bm{1} )$ representation, see {\it e.g.} (A.1) of \cite{Guarino:2019jef}. Again, the generators (\ref{eq:SO3A3CGen}) cannot be written as linear combinations of the embedding tensor (\ref{eq:ETtrombone}),  (\ref{eq:SCBianchiV}). For this reason, the SO$(3)_S^\prime$ defined by (\ref{eq:SO3SA3CBreaking}) is not a subgroup of the $\cN=8$ gauge group $\mathrm{TCSO}(5,0,0; \mathrm{V}) \equiv \big( G_3 \times \mathrm{SO} (5) \big) \ltimes  \mathbb{R}^{15}$. To put things in a wider context,
\begin{equation} \label{eq:SO4A3CGen}
\textrm{SO}(4)_S \equiv \textrm{SO}(3)_S^\prime \times \textrm{SO}(3)_+ \; , 
\end{equation}
is also the invariance group of both the subsector of ISO(7)-gauged supergravity addressed in section 5 of \cite{Guarino:2015qaa} and of the $\cN=3$ vacuum \cite{Gallerati:2014xra} contained within it. Unlike here, SO$(3)_S^\prime$ sits inside the ISO(7) gauge group in that case. The $\textrm{SO}(3)_+$ factor does sit inside the $\cN=8$ gauge group, both here (through SO$(5)$) and also in \cite{Gallerati:2014xra,Guarino:2015qaa}.

The characterisation, at both the kinematic and dynamical levels, of the SO$(3)_S^\prime$-invariant subsector of our $\cN=8$ supergravity proceeds as in section \ref{sec:SLAGSector}, so we will now be brief. We find three invariants, in the adjoint of $\mathrm{SO}(3)_+$, descending from the $\bm{56}$ of E$_{7(7)}$. We also find  $\mathrm{C}_{\mathrm{E}_{7(7)}}(\mathrm{SO}(3)_S^\prime) = \mathrm{SL}(2,\mathbb{R}) \times \mathrm{SL}(2,\mathbb{R}) $,  generated by
\begin{equation} \label{eq:CommA3C}
\begin{array}{lclcl}
H_0= t_a{}^a + t_4{}^4  \; , && E_0= \tfrac{1}{6} \epsilon^{abc} t_{a b c 4}  \; , && F_0= \tfrac{1}{3}\tfrac{1}{3!}\tfrac{1}{4!} \epsilon_{abc} \epsilon^{abc4ijkl} t_{ijkl}  \; , \\[5pt]
H_1= -t_a{}^a +3  t_4{}^4  \; ,  && E_1= \tfrac12 L^{aij}  t_{4 a ij} \; , && F_1= \tfrac14 L^{aij} \epsilon_{a}{}^{bc}  \, t_{bcij} \; . 
\end{array}
\end{equation}
Finally, there are one and five singlets in the decompositions of the $\bm{8}$ and $\bm{56}$ of SU$(8)$, respectively. This set of SO$(3)_S^\prime$-invariant tensors leads to a subsector field content compatible with $\cN=1$ supergravity coupled to three vector multiplets and two chiral multiplets. The scalar manifold of the latter is $\mathrm{SL}(2,\mathbb{R}) /\mathrm{SO}(2) \times \mathrm{SL}(2,\mathbb{R}) /\mathrm{SO} (2)$, corresponding to the non-compact generators of (\ref{eq:CommA3C}). We parametrise this space with four real scalars, $\varphi_0, \chi_0, \varphi_1, \chi_1$, sitting in the coset representative
\begin{equation} \label{eq:SugraSectorA3C}
\tilde{\cV} = e^{-\chi_0 E_0}e^{-\tfrac12 \varphi_0 H_0} e^{-\chi_1 E_1}e^{-\tfrac12 \varphi_1 H_1}  \; .
\end{equation}

Dynamically, the SO$(3)_S^\prime$-invariant truncation is obstructed in the original frame $X_M$ of (\ref{eq:ETtrombone}). Instead, the truncation does go through consistently in a frame $\tilde{X}_M$ rotated from the former as in (\ref{eq:TransXSymbol}) with the E$_{7(7)}$ element 
\begin{equation} \label{eq:TransMatA3C}
U_\lambda= e^{ \Upsilon_\lambda}  \; , \quad \textrm{where} \quad \Upsilon_\lambda = - \tfrac12 \,  \lambda  \, g_2 \, g_1^{-1} \, \left( t_{1738} + t_{6238}  + t_{1356} + t_{2357}   \right)\; ,
\end{equation}
with $\lambda = 1$. The constants $g_1$, $g_2$ here are the non-vanishing couplings that already appear in (\ref{eq:ETtrombone}). For generic $\lambda$, the transformation (\ref{eq:TransMatA3C}) defines a one-parameter family of embedding tensors $\tilde{X}_M$. It is interesting to note that the entire family satisfies the linear consistency constraint C1 with vanishing r.h.s.~for all $\lambda$. However, only for $\lambda=1$ are C2 and C3 satisfied in this case. Again, there is no obvious intrinsically four-dimensional argument that selects the $\lambda =1$ frame preferred by the truncation. The  $\lambda =1$ rotation (\ref{eq:TransMatA3C}) was informed by the $D=11$ construction of \cite{Pico:2026rji}. Like in section \ref{sec:SLAGSector}, the element \eqref{eq:TransMatA3C} is not SO$(3)_S^\prime$-invariant.


\subsection{$\cN = 1$ associative three-cycle AdS vacuum} \label{sec:A3CVac}


The SO$(3)_S^\prime$-invariant trombone components of $\tilde{X}_M$ vanish for all $\lambda$ in (\ref{eq:TransMatA3C}), so the truncated $\lambda=1$ SO$(3)_S^\prime$-invariant sector is again described by a Lagrangian. Like in section \ref{sec:SLAGVac}, we are not interested in tracking down vector couplings, so we will get rid of these by imposing invariance under a larger group. In this case, the SO$(4)_S$ defined in (\ref{eq:SO4A3CGen}) does the job of truncating out the gauge fields since these sit in the adjoint of $\mathrm{SO}(3)_+ \subset \mathrm{SO}(4)_S$, as noted in section \ref{sec:A3CSector}. We analysed the SO$(3)_S^\prime$ sector in the previous section both for parallelism with the SO$(3)_S$ analysis of section \ref{sec:SLAGSector}, and to illustrate that our linear consistency constraints C1 and C2 in (\ref{eq:TruncCondXLin}) are independent of each other. 

Directly imposing SO$(4)_S$ invariance is an interesting exercise in its own right, though, as it illustrates that the quadratic constraint C3 is independent of the linear constraints C1 and C2. Repeating the kinematic analysis of section \ref{sec:A3CSector}, we find that $\hat{K}_I{}^M=0$, and that the SO$(4)_S$-invariant field content encompasses only the four real scalars $\varphi_0, \chi_0, \varphi_1, \chi_1$ found above, along with the metric and relevant tensors and fermions. This field content is compatible with $\cN=1$ supergravity coupled to two chiral multiplets with scalar manifold $\mathrm{SL}(2,\mathbb{R}) /\mathrm{SO}(2) \times \mathrm{SL}(2,\mathbb{R}) /\mathrm{SO} (2)$, and coset representative again given by the exponentiation (\ref{eq:SugraSectorA3C}) of the generators (\ref{eq:CommA3C}). As for the dynamical aspects, $\hat{K}_I{}^M=0$ automatically implies the linear constraints (\ref{eq:TruncCondXLin}), as discussed in section \ref{sec:4DSugraCTDyn}, for all $\lambda$. The quadratic constraint C3 in (\ref{eq:TruncCondXQuad}) still needs to be imposed, and some analysis shows that it only holds for $\lambda=1$. This shows the independence of C3 from C1 and C2. Of course, this discussion does not preclude the existence of other preferred frames, still related by duality (\ref{eq:TransXSymbol}) to the original embedding tensor (\ref{eq:ETtrombone}), but outside the class generated by 
(\ref{eq:TransMatA3C}).

\begin{table}[]


\resizebox{\textwidth}{!}{

\begin{tabular}{l | l  } 
\hline
\hline
\textbf{Multiplet} & \textbf{Associative 3-cycle}  \\[6pt]
\hline \\[-10pt]
MGRAV &  $\left[ \tfrac52 \right] \otimes [0]_0$  \\[5pt]
\hline \\[-10pt]
GINO & $  \left[ 1 + \tfrac{\sqrt{29}}{4} \right] \otimes \left[\tfrac12\right]_{\pm\tfrac12} 
\; \oplus \;
\left[  1 + \tfrac{\sqrt{11}}{2} \right] \otimes \left[0\right]_{\pm1}  
\; \oplus \;
[3] \otimes[ 0]_{0}  $
  \\[6pt]
\hline \\[-10pt] 
MVEC &  $\left[ \tfrac32 \right] \otimes [1]_0$ 
    \\[6pt]
\hline \\[-10pt] 
VEC &  $ [3] \otimes[0]_{\pm1}
\; \oplus \;
\left[  1 + \tfrac{\sqrt{21}}{2} \right] \otimes[0]_{0} 
\; \oplus \;
\left[\tfrac{11}4 \right] \otimes \left[\tfrac12 \right]_{\pm \tfrac12}
\; \oplus \;
\left[\tfrac74\right] \otimes \left[\tfrac12\right]_{\pm \tfrac32}
$
    \\[6pt]
\hline \\[-10pt] 
CHIRAL & $ \left( 2 \times  \left[1+\sqrt{6} \right] \otimes [0]_0 \right)
\; \oplus \;
[2] \otimes[0]_{\pm 2}
\; \oplus \;
 [2]\otimes [1]_{0} 
\; \oplus \;
 \left[ 1 + \tfrac{\sqrt{61}}{4} \right] \otimes \left[\tfrac12 \right]_{\pm \tfrac12} 
\; \oplus \;
 \left[ 1 + \tfrac{i}{2} \right] \otimes \left[1 \right]_{\pm 1}   $
\\[6pt]
\hline
\hline
\end{tabular}

\qquad 
}
\caption{\footnotesize{Spectrum of $\textrm{OSp}(4|1) \times \textrm{SO}(3)_+ \times \textrm{U}(1)^\prime_S$ supermultiplets for the associative three-cycle AdS solution, within $D=4$ $\cN=8$ TCSO$(5,0,0;\mathrm{V})$ supergravity. The OSp$(4|1)$ supermultiplets are denoted using the acronyms of table 1 of \cite{Cesaro:2020soq}. The entries collect the superconformal primary dimension $E_0$ for each OSp$(4|1)$ supermultiplet, the spin $s$ of $\textrm{SO}(3)_+$ and the charge $q$ of $\textrm{U}(1)^\prime_S$ in the format $[ E_0 ] \otimes [s]_q$. An entry with $n\times$ in front is $n$ times repeated. 
}\normalsize}
\label{tab:A3CMultiplets}
\end{table}

The SO$(4)_S$-invariant consistent truncation of TCSO$(5,0,0;\mathrm{V})$ supergravity on the frame $\tilde{X}_M$ obtained from the defining $X_M$ in (\ref{eq:ETtrombone}) through rotation with $\lambda=1$ in (\ref{eq:TransXSymbol}), (\ref{eq:TransMatA3C}) is thus described by the Lagrangian
\begin{equation} \label{eq:LagA3C}
{\cal L} =   R \, \textrm{vol}_4 + \tfrac12 (d\varphi_0)^2 + \tfrac12 e^{2\varphi_0} (d \chi_0)^2  + 3 (d\varphi_1)^2 + 3 e^{2\varphi_1} (d \chi_1)^2   - V_\lambda \, \textrm{vol}_4   ,
\end{equation}
where the scalar potential is
{\setlength\arraycolsep{2pt}
\begin{eqnarray} \label{eq:PotA3C}
V & = &
\tfrac{1}{16 }g^2 \, e^{-\varphi_{0}-6\varphi_{1}}
\bigl(1+e^{2\varphi_{1}}\chi_{1}^{2}\bigr)^{2}
\Big[
\; 8
   - 64 e^{\varphi_{0}+3\varphi_{1}}
   + 8 e^{2\varphi_{0}}\chi_{0}^{2}
   + 32 e^{2\varphi_{1}}\chi_{1}^{2}
   - 64 e^{\varphi_{0}+5\varphi_{1}}\chi_{1}^{2}
\nonumber \\[4pt]
&&\quad
   + 48 e^{4\varphi_{1}}\chi_{1}^{4}
   + 32 e^{6\varphi_{1}}\chi_{1}^{6}
   +  8 e^{8\varphi_{1}}\chi_{1}^{8}
   + e^{2(\varphi_{0}+\varphi_{1})}
     \Big(
        3 - 24\chi_{0}\chi_{1}
        + 32\chi_{0}^{2}\chi_{1}^{2}
     \Big)
\nonumber \\[4pt]
&&\quad
   + 2 e^{2\varphi_{0}+8\varphi_{1}} \chi_{1}^{2}
     \Big(
        8 + 3\chi_{1}^{2}
        - 2\chi_{0}\chi_{1}^{3}
     \Big)^{2}
  + 24 e^{2\varphi_{0}+4\varphi_{1}}
     \Big(
        2 + \chi_{1}^{2}
        - 3\chi_{0}\chi_{1}^{3}
        + 2\chi_{0}^{2}\chi_{1}^{4}
     \Big)
\nonumber \\[4pt]
&&\quad
   + e^{2\varphi_{0}+6\varphi_{1}}
     \Big(
        -64
        + 48\chi_{1}^{2}
        - 64\chi_{0}\chi_{1}^{3}
        + 39\chi_{1}^{4}
        - 72\chi_{0}\chi_{1}^{5}
        + 32\chi_{0}^{2}\chi_{1}^{6}
     \Big)
\Big] \; ,
\end{eqnarray}
}with, again, $g \equiv g_1 = \sqrt{2} \, g_2$ for simplicity. This model in fact matches (2.12), (2.13) of \cite{Gauntlett:2002rv} under the scalar identifications
\begin{equation} \label{eq:IdentifScalarA3C}
\varphi_0 = -2 \big( 2 \,  \lambda^{\text{there}} + 3 \, \phi^{\text{there}} \big) \, , \qquad
 \varphi_1 =  -2 \big(   \, \lambda^{\text{there}} -  \, \phi^{\text{there}} \big) \, , 
\qquad 
\chi_0 = \chi_1=0 \; , 
\end{equation}
along with $d_{\textrm{there}} = 3$, $p_{\textrm{there}} = 4$, $q_{\textrm{there}} = 1$, $m_{\textrm{there}} =  \, g$, and $l_{\textrm{there}} = -1$. In our conventions, the supersymmetric AdS vacuum discussed in section 4.2 of \cite{Gauntlett:2002rv} is recovered from the potential (\ref{eq:PotA3C}) at the following scalar values and with AdS radius $L$ given by 
\begin{eqnarray} \label{eq:vacA3C}
& \textrm{Associative 3-cycle: } \quad e^{\varphi_0} =e^{-\varphi_1} =  2 \sqrt{\frac25} \; , \quad
\chi_0 = \chi_1 = 0 \; , \quad 
L^2=\frac{25}{32} \sqrt{\frac{5}{2}}  \, g^{-2}  \, . \quad 
\end{eqnarray}
This vacuum uplifts to a $D=11$ solution that can be identified \cite{Pico:2026rji} as the near-horizon  region of a stack of M5-branes wrapped on an associative three-cycle of constant negative curvature inside a G$_2$-holonomy seven-dimensional manifold \cite{Acharya:2000mu} (see also \cite{Gauntlett:2006ux}). Inside our $D=4$ $\cN=8$ 
supergravity, the vacuum (\ref{eq:vacA3C}) spontaneously breaks the $\cN=8$ supersymmetry down to $\cN=1$, and the gauge group 
$ \textrm{TCSO} (5,0,0;\mathrm{V}) \equiv \big( G_3 \times \textrm{SO} (5) \big)   \ltimes \mathbb{R}^{15} $
down to the $\mathrm{SO}(3)_+ \subset \mathrm{SO}(5)$ subgroup defined in (\ref{eq:SO3SA3CBreaking}). The residual symmetry (super)group of the vacuum (\ref{eq:vacA3C}) within our $D=4$ $\cN=8$ supergravity is therefore $\textrm{OSp}(4|1) \times \textrm{SO}(3)_+$. 

We have obtained the mass spectrum of the $D=4$ $\cN=8$ supergravity fields about the vacuum (\ref{eq:vacA3C}) by diagonalising the mass matrices reviewed in appendix~\ref{sec:4DSugraEoMsMass}. The results are summarised in tables \ref{tab:SLAGAndAssCycMass} and \ref{tab:A3CMultiplets}. Table \ref{tab:SLAGAndAssCycMass} contains the individual mass states, and extends the results of \cite{Acharya:2000mu} to maximal four-dimensional gauged supergravity. Scalar states with masses $M^2L^2= 4 \pm \sqrt{6}$ were identified in their two-scalar model, and we indeed find those values (twice, in fact) in table \ref{tab:SLAGAndAssCycMass}. Table \ref{tab:A3CMultiplets}, in turn, lists the supermultiplet structure. Similarly to the solution of section \ref{sec:SLAGVac}, the spectrum arranges itself in representations of a (super)group, $\textrm{OSp}(4|1) \times \textrm{SO}(3)_+ \times \textrm{U}(1)^\prime_S$, that is larger than the symmetry (super)group of the solution. Here, 
$\textrm{U}(1)^\prime_S$ is the Cartan subgroup of the SO$(3)_S^\prime$ defined in (\ref{eq:SO3SA3CBreaking}). The massless vector multiplet $\textrm{MVEC}\left[ \tfrac32 \right] \otimes [1]_0$, in the adjoint of $\textrm{SO}(3)_+$, includes the only $\cN=8$ gauge fields that remain massless in the model, while all others become massive and go into massive vector, VEC, multiplets. Again, $\textrm{U}(1)^\prime_S$ is not included in the gauge group and, as a result, no gauge field in the spectrum acquires mass by spontaneous symmetry breaking of the gauge group  to $\textrm{U}(1)^\prime_S$. 

Finally, note the presence in table \ref{tab:SLAGAndAssCycMass} of spin-$1/2$ and scalar states of complex mass, despite the supersymmetry of the vacuum. This is again linked with the lack of symmetry of the mass matrices due to the trombone gauging. Intriguingly, the mass of these wayward scalars sits at the Breitenlohner-Freedman bound for stability in four-dimensional AdS, plus or minus an imaginary part. These states go into the CHIRAL multiplets, in the adjoint of $\textrm{SO}(3)_+$, shown with complex dimension in table~\ref{tab:A3CMultiplets}. Further discussion can be found in the next section.


\section{Discussion} \label{sec:Discussion}


We have derived the conditions that render dynamically consistent the truncations of maximal supergravity to $G_S$-invariant subsectors thereof. Here, $G_S$ is a subgroup of $\textrm{SU}(8) \subset \textrm{E}_{7(7)}$ that is not necessarily contained in the gauge group $G$ of the parent $\cN=8$ theory. These conditions take on the form of linear and quadratic constraints involving both the $\cN=8$ embedding tensor and $G_S$-invariant tensors descending from suitable representations of $\textrm{E}_{7(7)}$. These constraints are independent of, and need to be imposed on top of, the usual linear and quadratic constraints of the parent maximal theory. We have shown how these constraints can obstruct or permit a $G_S$-invariant truncation in duality frames that are otherwise equivalent from an  $\cN=8$ perspective. A way to understand this obstruction is to think of the duality transformation from an active point of view (a change in the scalar manifold), rather than from the passive point of view (a change in the duality frame) that we have used in the main text. From an active viewpoint, a subtruncation might only be consistent when performed around preferred points in the $\textrm{E}_{7(7)}/\textrm{SU}(8)$ scalar manifold. This perspective resonates, within a $D=4$ $\cN=8$ context, with the proposal \cite{Gauntlett:2007ma} that a supersymmetric vacuum will always be a preferred point in this sense.

The exceptional generalised geometry proof \cite{Cassani:2019vcl} of the conjecture of \cite{Gauntlett:2007ma} establishes that a $G_S$-invariant truncated dimensional reduction will be consistent if the retained intrinsic torsion is a singlet. 
 Read as an existence proof, the statement of \cite{Cassani:2019vcl} should imply within our strictly four-dimensional context that a set of frames preferred by the truncation should always exist and that, moreover, $G_S$ should determine it. In the subtruncations of the specific $\cN=8$ supergravity that we have discussed in the text, we have indeed been able to find such preferred frames. However, we have used external information, in principle unrelated to $G_S$, to perform the necessary duality transformations, (\ref{eq:SugraSectorSLAG}), (\ref{eq:TransMatA3C}) from the original, defining $\cN=8$ frame. Perhaps these transformations should be looked at more closely to find a relation with $G_S$ more compelling than just the observation that the involved E$_{7(7)}$ elements are not $G_S$-invariant. 

We have illustrated our consistent subsector formalism in the specific context of a new family of maximal four-dimensional supergravities, TCSO$(p,q,r ; \mathrm{N})$, that we have introduced in section \ref{sec:New4DSugras}. This class can be described as a mixture of dyonic CSO gaugings \cite{DallAgata:2011aa,Dall'Agata:2014ita} and the Scherk-Schwarz gauging \cite{Scherk:1979zr} of a three-dimensional group $G_3$ of Bianchi type $\mathrm{N}$. When $G_3$ is non-unimodular, the trombone scaling symmetry $\mathbb{R}^+$ is gauged. We have argued a particular member, TCSO$(5,0,0 ; \mathrm{V})$, of our family to be related to \mbox{$D=11$} configurations of M5-branes wrapped in three-dimensional  submanifolds of negative constant curvature inside special holonomy manifolds. Indeed, we have recovered some AdS near-horizon M5-brane geometries in that class \cite{Acharya:2000mu,Gauntlett:2006ux} as AdS vacua of our $D=4$ $\cN=8$ CSO$(5,0,0 ; \mathrm{V})$ gauged supergravity model. 

Trombone-gauged supergravities must be described at the level of the field equations \cite{LeDiffon:2011wt}, as they do not admit an action. The absence of a lagrangian is prevented, among other things, by mass matrices that are not symmetric. The latter have antisymmetric contributions governed by the trombone components of the embedding tensor. Thus, the well-known theorem that ensures diagonalisability over the reals of a real symmetric matrix does not apply to the mass matrices of trombone-gauged supergravities. Two obstructions might therefore hamper the diagonalisation of these mass matrices over the reals: a defectiveness leading to a Jordan block decomposition, or a diagonalisation over the complex numbers. We have seen both issues appear, in turn, in the mass spectrum of the two AdS vacua studied in section \ref{eq:AdSvac}.

The complex mass modes reported in the spectrum of the associative three-cycle $\cN=1$ AdS vacuum of section \ref{sec:A3CVac} seem at odds with the expected stability of a supersymmetric vacuum. This is a question that will need further analysis beyond the scope of the present paper, which limited itself to present those diagonalisation results. We are inclined to think that, more than a pathology of the supersymmetric vacua of \cite{Acharya:2000mu,Gauntlett:2006ux}, our results invite a more careful analysis of the conditions that lead to the calculation of mass spectra in the context of consistent truncations, in the sense that not all modes seen in a consistently truncated gauged supergravity should uplift to the higher dimension. Whether leading to a subsector of a theory in a fixed dimension, or leading to a dimensional reduction, consistent truncations provide just local staments about the equations of motion of the theories involved. Global aspects and boundary conditions are frequently ignored in this context, but these might play an important  role in the cases discussed in this paper.

An indication that global aspects should be taken into account already lies in our construction. Among other reasons, $D=4$ $\cN=8$ TCSO$(5,0,0 ; \mathrm{V})$ supergravity arises as a consistent truncation \cite{Pico:2026rji} of $D=11$ supergravity on the M5-brane solutions of \cite{Acharya:2000mu,Gauntlett:2006ux} because the wrapped submanifold therein is a hyperboloid, $H^3$, which is locally diffeomorphic to the group manifold $G_3$ of Bianchi type $\mathrm{V}$. Globally, however, $H^3$ and $G_3$ differ. In particular, the former can be compactified by quotienting, $H^3/\Gamma$, by a discrete group of isometries $\Gamma$. This feature enables such compact $H^3/\Gamma$ configuration to provide a bonafide M-theory background. In contrast, a non-unimodular group such as Bianchi type V cannot be similarly compactified. A reasonable expectation is that the delinquent modes of complex mass are in fact unphysical and are projected out of the spectrum by the action of $\Gamma$. We expect to return to these questions in the future.


\section*{Acknowledgements}


We thank Colin Sterckx for suggesting that the consistent subtruncation constraints that we gave in a first draft of this paper should be improved into the final form we now give for these in the main text. We posthumously thank A.~Scriabin for inspiration from his Prelude Op.~11, n$^\textrm{o}$~1, in C major. In spite of being a piece for piano solo, and unfortunately not including a trombone part, this prelude features a 5 (right hand) v.~3 (left hand) polyrhythm that inspired our $p+q+r=5$ (electric) v.~$\tilde{p}+\tilde{q}+\tilde{r}=3$ (magnetic) gaugings. MP is supported by predoctoral award FPU22/02084 from the Spanish Government, and partially by Spanish Government grants CEX2020-001007-S, PID2021-123017NB-I00 and PID2024-156043NB-I00, funded by MCIN/AEI/10.13039/501100011033, and ERDF, EU. OV is supported by NSF grant PHY-2310223.


\appendix

\addtocontents{toc}{\setcounter{tocdepth}{1}}



\section{$D=4$ $\cN=8$ field equations and mass matrices} \label{sec:4DSugraEoMsMass}


The bosonic (Einstein, Maxwell and scalar) equations of motion for general $\bm{912}_{-1}+ \bm{56}_{-1}$ gaugings of $D=4$ $\cN=8$ supergravity read
\cite{LeDiffon:2011wt}, in our conventions, 
{\setlength\arraycolsep{0pt}
\begin{eqnarray}
\label{eq:EinsteinEOM} && R_{\mu \nu} = - \tfrac{1}{48} D_\mu M_{MN} D_\nu M^{MN}  + \tfrac{1}{2} g_{\mu\nu} V + \tfrac{1}{4 }  M_{MN} \big(  H_{\mu \rho}{}^M H_{\nu}{}^\rho{}^N  - \tfrac{1}{4} g_{\mu\nu}  \, H_{\rho \sigma}{}^M \, H^{\rho \sigma N}  \big)  , \quad    \\[8pt]
\label{eq:vectorEOM} && D \big( M_{MN} *H_\2^N \big) - \tfrac{1}{8} \,\big( \Theta_{MN}{}^P - 16 \delta_M^P \vartheta_N  \big)  \, M^{NQ} * D M_{PQ} =0 \,  ,  \\[8pt]
\label{eq:scalarEOM} && (t_\alpha)_{S}{}^{(P} M^{Q)S} \Big( 
D*DM_{PQ}  -M^{RS} DM_{PR} \wedge *DM_{QS}  -12 \, M_{PR} M_{QS} \, H_\2^R \wedge * H_\2^S  \nonumber \\
&& \qquad\qquad\qquad -24 \, \tilde{V}_{PQ} \, \textrm{vol}_4 \Big) =0 \; .
\end{eqnarray}
}These must be supplemented by the following selfduality condition and Bianchi identity,
\begin{equation} \label{eq:SDandBianchi}
H_\2^M = - \Omega^{MN} M_{NP} * H_\2^P \; , \qquad 
D H_\2^{M} =   \big(  \Theta^{M\alpha} - 16 \, t^{\alpha MN} \vartheta_N \big)   H_{\3 \alpha} \; ,
\end{equation}
from which a duality relation between scalars and two-forms follows,
\begin{equation} \label{eq:ScaalarDuality}
 \big(  \Theta^{M\alpha} - 16 \, t^{\alpha MN} \vartheta_N \big)  \big( H_{\3 \alpha} +\tfrac18 (t_{\alpha})_P{}^Q \, M^{PR} * D M_{QR} \big) =0  \, .
\end{equation}
The (self)duality conditions in (\ref{eq:SDandBianchi}), (\ref{eq:ScaalarDuality}) ensure that only the right degrees of freedom propagate: neither the magnetic vectors nor the two-forms carry independent dynamics. 

In equations (\ref{eq:EinsteinEOM})--(\ref{eq:ScaalarDuality}), $M_{MN} = (\cV \cV^{\textrm{T}})_{MN}$ is the usual metric on $\mathrm{E}_{7(7)}/\mathrm{SU}(8)$, with inverse $M^{MN}$. The Hodge dual is taken w.r.t.~the spacetime metric $g_{\mu\nu}$, and all quantities are covariantised. For example, the Ricci tensor $R_{\mu\nu}$ is fully covariant under local rescalings \cite{LeDiffon:2008sh}, and the covariant derivatives are built by putting the embedding tensor (\ref{eq:XSymbolsGen}) in the appropriate representations $\bm{r}_w$ of $\mathbb{R}^+ \times \mathrm{E}_{7(7)}$ recorded in table \ref{tab:4DSummary} of the main text. For example, acting on scalars
\begin{equation} \label{eq:CovDerM}
DM_{MN} \equiv dM_{MN} -2 \,  A^P \,  \big( \Theta_{P}{}^\alpha +  8 (t^\alpha)_{P}{}^{Q} \,  \vartheta_{Q} \big)  (t_\alpha)_{(M}{}^{R} \, M_{N)R} \; ,
\end{equation}
and similarly for the other fields. The gauge field strengths are, in turn,
{\setlength\arraycolsep{0pt}
\begin{eqnarray}
\label{eq:H2Form} && H_\2^M \equiv dA^M + \tfrac12 X_{NP}{}^M A^N \wedge A^P +  \big(  \Theta^{M\alpha} - 16 \, t^{\alpha MN} \vartheta_N \big) B_\alpha \; , \\[5pt]
\label{eq:H3Form} && H_{\3 \alpha}  \equiv  DB_{\alpha} -\tfrac12 t_{\alpha MN} A^M \wedge dA^N - \tfrac{1}{3!} \, t_{\alpha MQ} \,   X_{NP}{}^Q A^M \wedge A^N \wedge A^P \; .
\end{eqnarray}
}The quantities $V$ and $\tilde{V}_{MN}$ that respectively appear in the Einstein, (\ref{eq:EinsteinEOM}), and scalar, (\ref{eq:scalarEOM}), equations, read 
\begin{equation} \label{eq:Cosmo}
V  = \tfrac{8}{3}  \, M^{MN} \vartheta_M \vartheta_N  
+   \tfrac{1}{168} \, M^{MN} \, \big(   M^{PQ} M_{RS} \Theta_{MP}{}^R \Theta_{NQ}{}^S 
 +7 \, \Theta_{MP}{}^Q \Theta_{NQ}{}^P \big) \; ,
\end{equation}
and
{\setlength\arraycolsep{2pt}
\begin{eqnarray} \label{eq:DerPotTromb} 
 \tilde{V}_{MN} & =& 
\tfrac16  \, \vartheta_P M^{PQ} \Theta_{Q(M}{}^R M_{N)R} +\tfrac{1}{168} 
   \Big( M^{PQ} M_{RS} \, \big(  \Theta_{MP}{}^R \Theta_{NQ}{}^S 
  +   \Theta_{PM}{}^R \Theta_{QN}{}^S \big)   \nonumber   \\
 && 
+ 7 \, \Theta_{MP}{}^Q \,  \Theta_{NQ}{}^P 
- M^{PQ} M^{RS} M_{T(M} M_{N)U} \Theta_{PR}{}^{T} \Theta_{QS}{}^{U} 
\Big)\, .
\end{eqnarray}
When $\vartheta_M = 0$, both quantities are related by $\tilde{V}_{MN} =  \partial V / \partial M^{MN}$, and (\ref{eq:Cosmo}) reduces to the scalar potential that features in the Lagrangian of \cite{deWit:2007mt}. Note also that the scalar equation of motion (\ref{eq:scalarEOM}) appears projected throughout with $(t_\alpha)_{S}{}^{(P} M^{Q)S}$, a projector to the coset $\textrm{E}_{7(7)}/\textrm{SU}(8)$. Although the free index $\alpha$ there runs in the adjoint of $\textrm{E}_{7(7)}$, as usual, one may check that $\bm{63}$ components therein, in the adjoint of SU$(8)$, are identically zero. By the $PQ$ symmetrisation, only the coset components, in the $\bm{70}$ of SU$(8)$, survive the projection.

The mass matrices of $D=4$ $\cN=8$ supergravity with generic gauge group $G \subset \mathbb{R}^+ \times \mathbb{E}_{7(7)}$ were obtained in \cite{LeDiffon:2011wt}. We collect them here in our conventions as derived by linearisation of the equations of motion (\ref{eq:vectorEOM}), (\ref{eq:scalarEOM}). Around a given vacuum, attained with maximally symmetric metric, constant scalars subject to
\begin{equation} \label{eq:Vacuum}
(t_\alpha)_{S}{}^{(P} M^{Q)S} \, \tilde{V}_{PQ} = 0 \; ,
\end{equation}
 and all other fields vanishing, the vector and scalar mass matrices are 
{\setlength\arraycolsep{0pt}
\begin{eqnarray}
\label{eq:VectorMassMat} && (\cM_{\textrm{vector}}^2)_M{}^N = 
\tfrac16 M^{NP} \left( X_{PQ}{}^R - 24 \delta_{P}^R \vartheta_Q \right) M^{QS} \left( X_{M(R}{}^T M_{S)T} + \vartheta_M M_{RS} \right)\;, \\[9pt]
\label{eq:ScalarMassMat}  && (\cM_{\textrm{scalar}}^2)_{\alpha}{}^{\beta} = 24 (t_\alpha)_{P}{}^{(M} M^{R)P} \tilde{V}_{M N} \,  M^{N S} \, (t^\beta)_{(R}{}^{Q}M_{S)Q} \\[4pt]
&& \qquad \qquad \qquad  + 24 (t_\alpha)_{P}{}^{(R_1} M^{R_2)P} \big( U_{R_1 R_2 \, S_1 S_2} + W_{R_1 R_2 \, S_1 S_2} + W_{S_1 S_2 \, R_1 R_2 }  \big) \, (t^\beta)_{Q}{}^{(S_1} M^{S_2)Q}  \; .  \nonumber 
\end{eqnarray}
}Here, we have used the shorthands 
{\setlength\arraycolsep{2pt}
\begin{eqnarray}
U_{MN \, PQ } & \equiv & \tfrac{1}{6} \Big( 
M_{MR} \, \vartheta_{P} \,  \Theta_{QN}{}^{R} 
-M_{TP} M_{QM} M^{RS} \vartheta_{R} \Theta_{SN}{}^{T}
 \Big) \; ,  \\[10pt]
W_{MN \, PQ } & \equiv & \tfrac{1}{168} \Big( 
M_{R_1 R_2} \, \Theta_{MP}{}^{R_1} \,  \Theta_{NQ}{}^{R_2} 
 -M^{R_1 R_2} \,M_{P S_1} \,M_{Q S_2} \, \Theta_{MR_1}{}^{S_1} \,  \Theta_{NR_2}{}^{S_2} \nonumber \\[5pt]
&& \quad -M^{R_1 R_2} \,M_{P S_1} \,M_{Q S_2} \, \Theta_{R_1M}{}^{S_1} \,  \Theta_{R_2N}{}^{S_2}  \nonumber \\[5pt]
&& \quad + M^{R_1 R_2} M^{S_1S_2} M_{MP} M_{NT_1} M_{QT_2} \Theta_{R_1 S_1}{}^{T_1} \Theta_{R_2  S_2}{}^{T_2} \Big) \; , 
\end{eqnarray}
}such that $\partial \tilde{V}_{MN}/\partial M^{PQ} =  U_{(MN) (PQ) } + W_{(MN) (PQ)} + W_{(PQ)  (MN) } $. For completeness, let us also quote the fermionic, gravitino and spin-$1/2$, mass matrices 
{\setlength\arraycolsep{-.7pt}
\begin{eqnarray}
\label{eq:GravitinoMassMat} && (\cM_{\textrm{gravitino}})^{ij}   = 2 \big( A_1{}^{ij} - 2B^{ij} \big)  \; ,  \\[5pt]
\label{eq:FermionMassMat} && (\cM_{\textrm{spin-$\frac12$}})_{ijk, \ell mn}   =      \epsilon_{ijkpqr[\ell m|} A_{2\; |n]}{}^{pqr} + 2  \epsilon_{ijk\ell mnpq} B^{pq}   \,,  \qquad 
\end{eqnarray}
}written in terms of the `fermion shifts' $A_{ij}$, $A_{2 i}{}^{jk\ell}$ and $B^{ij}$ \cite{deWit:2007mt,LeDiffon:2011wt}. 

As usual, the eigenvalues of the bosonic and fermionic mass matrices are squared and linear masses, respectively. Not all of these will be physical, though. The vector mass matrix will only contain $ 28$ physical eigenvalues and $28$ spurious zero eigenvalues. For supersymmetry breaking vacua, the fermion mass matrix contains a spurious mode for each Goldstino state eaten by a massive gravitino. At least for the AdS vacua discussed in section \ref{eq:AdSvac} but presumably more generally, a Goldstino mode arises as an eigenvalue $M_{\textrm{Goldstino}}$ of (\ref{eq:FermionMassMat}) related to the corresponding gravitino mass $M_{\textrm{gravitino}}$ as $M_{\textrm{Goldstino}} = 2 \, M_{\textrm{gravitino}}$. This relation does not hold, however,  for `massless' gravitini in a supersymmetric AdS vacuum. Those have $L M_{\textrm{`massless' gravitino}} = \pm 1 $ in units of the AdS radius $L$, and do not undergo superHiggsing. Finally, when trombone gaugings are present, the above mass matrices are not (similar to) symmetric matrices. As a consequence, the usual diagonalisation theorem does not apply, as discussed in the main text.


\section{Consistency proofs} \label{sec:ConsProofs}



In this appendix, we give details of the consistency statements made in section~\ref{sec:4DSugra}. Namely that, if the consistency constraints \eqref{eq:TruncCondXLin}, \eqref{eq:TruncCondXQuad} hold, then 1) a truncation to $G_S$-invariant fields of $D=4$ $\cN=8$ supergravity will be consistent at the level of the field equations, and 2) the retained $G_S$-invariant subsector will correspond to a self-consistent theory in its own right. While that appears to be the conceptually preferred order for our proof, we find it logically more compelling to follow the reverse order instead. The reason is that the subtruncation proof at the level of the field equations flows more easily if the selfconsistency of the retained subsector is shown first. To this aim, we will find it helpful to introduce the following combinations of invariant tensors
\begin{equation} \label{eq:SingletProj}
\mathbb{P}^M{}_N \equiv \hat{K}_I{}^M K^I{}_N \; , \qquad
\mathbb{P}^x{}_y \equiv \hat{K}_y{}^\alpha K^x{}_\alpha \; .
\end{equation}
These respectively behave as projectors of the $\bm{56}$ and $\bm{133}$ representations of E$_{7(7)}$ to the space of singlets in the branchings of those representations under $G_S \subset  \textrm{SU}(8) \subset \textrm{E}_{7(7)}$.

\subsection{Self-consistency of the $G_S$-invariant sector} \label{sec:SelfConsProofs}


A self-consistent subsector must inherit a well-defined embedding tensor from the $\cN=8$ theory. The consistency constraints \eqref{eq:TruncCondXLin} together with the linear, \eqref{eq:LinearConst}, and quadratic, \eqref{eq:LieAlg}, constraints of the parent $\cN=8$ theory guarantee that this is the case. More concretely, \eqref{eq:LinearConst}, \eqref{eq:LieAlg}, \eqref{eq:TruncCondXLin} together imply all the linear and quadratic constraints (equations \eqref{ReducedLinearConstraint}, \eqref{ReducedQuadraticConstraint} and \eqref{CompatibilityReducedTwoForms} below) required for the retained embedding tensor $X_I$ to furnish self-consistent dynamics in the vector/tensor gauge field sector. The reduced subsector may or may not be supersymmetric by itself, as the proof does not rely on supersymmetry. 

We start by noting that equation \eqref{eq:InvarFields} together with C1 in \eqref{eq:TruncCondXLin} allow us to express the truncated embedding tensor $X_I$ as
\begin{equation} \label{ReducedEmbeddingTensor}
X_I = - w \vartheta_I t_0 + \left( \Theta_I{}^x + 8 (t^x)_I{}^J \vartheta_J \right)t_x +\Theta_I{}^a T_a \,.
\end{equation}
Here, we have used that the combination $(t^x)_I{}^M = K^x{}_\alpha \hat{K}_I{}^N (t^\alpha)_N{}^M$ is a $G_S$-singlet, because $(t^\alpha)_N{}^M$  is an E$_{7(7)}$-singlet and $\hat{K}_I{}^M$ and $K^x{}_\alpha$ are both $G_S$-invariant. Thus, the identity $(t^x)_I{}^M=(t^x)_I{}^N \mathbb{P}^M{}_N= (t^x)_I{}^J \hat{K}_J{}^M$ follows. Note that $(t_x)_{I}{}^{J} \equiv \hat{K}_x{}^\alpha \hat{K}_I{}^M K^J{}_N (t_\alpha)_{M}{}^{N}$ are the generators of $\mathrm{C}_{\mathrm{E}_{7(7)}}(G_S)$ in the $2n_1$-dimensional representation. Finally, $\Theta_I{}^a$ are constants that introduce an extra leg for $X_I$ along the $G_S$ generators $T_a$. By virtue of (\ref{eq:InvarFields}), these terms decouple from all expressions (\ref{ReducedLinearConstraint})--(\ref{ZMalphaFactorizedZMx}), and are effectively contained in (\ref{ReducedEmbeddingTensorGeneralCase}) and subsequent expressions through the end of appendix~\ref{sec:SelfConsProofs}. The $\Theta_I{}^a T_a$ terms also decouple and disappear from all field equations of the reduced subsector in appendix \ref{sec:ConsProofsFromN8}. Thus, these terms can be safely disregarded, as we have done in the main text.

The truncated embedding tensor \eqref{ReducedEmbeddingTensor} decomposes into two representations that respectively descend from the $\bm{912}_{-1}$ and $\bm{56}_{-1}$ of the parent $\cN=8$ theory. Compatibility with Sp$(2n_1)$ requires the component $\Theta_I{}^x \equiv \hat{K}_I{}^M\Theta_M{}^\alpha K^x{}_\alpha$ to satisfy the linear constraint (see \cite{Trigiante:2016mnt})
\begin{equation}
\Theta_{(IJ}{}^L \Omega_{K)L}=0\,.
\label{ReducedLinearConstraint}
\end{equation}
Here, $\Omega_{IJ} \equiv \hat{K}_I{}^M \hat{K}_J{}^N \Omega_{MN}$ is the Sp$(2n_1)$ invariant form, which can be used to raise and lower $IJ$ indices. We find it useful to show that (\ref{ReducedLinearConstraint}) holds by using the following alternative form of the $\cN=8$ linear constraint  \eqref{eq:LinearConst}: $\Theta_{(MN)}{}^P = - \tfrac12 \Omega^{PQ} \Theta_Q{}^\alpha (t_\alpha)_{MN}$, see \cite{Trigiante:2016mnt}. Upon contraction with the $G_S$-invariant tensors $\hat{K}_I{}^M$, $\hat{K}_J{}^N$, $K^K{}_P$ the latter reduces to
\begin{equation}
\Theta_{(IJ)}{}^K = - \tfrac12 \Omega^{KL} \Theta_L{}^x (t_x)_{IJ}\,,
\label{LinearConstraintIdentities}
\end{equation}
which, in turn, implies \eqref{ReducedLinearConstraint}. To show that \eqref{LinearConstraintIdentities} holds, note that $K^K{}_P \Omega^{PQ}$ is a $G_S$-singlet because $\Omega^{ MN}$ is an E$_{7(7)}$ singlet and $K^K{}_P$ is $G_S$-invariant. This implies $K^K{}_P \Omega^{PQ}= K^K{}_P \Omega^{PR}\mathbb{P}^Q{}_R$, with $\mathbb{P}^Q{}_R$ defined in (\ref{eq:SingletProj}). Conversely, the combination $t_{\alpha MN} \hat{K}_I{}^M \hat{K}_J{}^N$ is a $G_S$ singlet, because $t_{\alpha MN}$ is an E$_{7(7)}$ singlet and, again, $\hat{K}_I{}^M$ is $G_S$-invariant. Thus, we have the identity:
\begin{equation}
t_{\alpha MN} \hat{K}_I{}^M \hat{K}_J{}^N = t_{\beta MN} \hat{K}_I{}^M \hat{K}_J{}^N \, \mathbb{P}^\beta{}_\alpha= K^x{}_\alpha (t_x)_{IJ}\,.
\label{talphaIJProjectiontxIJ}
\end{equation}
In any case, for further reassurance, the linear constraint \eqref{ReducedLinearConstraint} can be derived directly from \eqref{eq:LinearConst} as
\begin{equation}
\begin{split}
\Theta_{(IJ}{}^L \Omega_{K)L} &= \hat K_{(I}{}^M \hat K_J{}^N \hat K_{K)}{}^P  \Theta_{MN}{}^Q  K^L{}_Q \hat K_L{}^R  \Omega_{PR}\\
&= \hat K_{(I}{}^M \hat K_J{}^N \hat K_{K)}{}^P  \Theta_{MN}{}^Q  \mathbb{P}^R{}_Q  \Omega_{PR}\\
&= \hat K_{(I}{}^M \hat K_J{}^N \hat K_{K)}{}^P  \Theta_{MN}{}^Q  \Omega_{PQ}\\
&= \hat K_{I}{}^{(M} \hat K_J{}^N \hat K_{K}{}^{P)}  \Theta_{MN}{}^Q  \Omega_{PQ}=0\,.
\end{split}
\end{equation}

Moving on to establish the quadratic constraints for the truncated embedding tensor \eqref{ReducedEmbeddingTensor}, the closure of the Lie algebra generators of the retained gauge group $\tilde{G}$ can be proven as follows. The first consistency constraint of \eqref{eq:TruncCondXLin} implies that $X_{IM}{}^N$ takes values on the Lie algebra of the group indicated on the leftmost relation in (\ref{eq:Gincl}). Hence, the quantities $X_{IJ}{}^M= \hat{K}_J{}^P X_{IP}{}^M$ defined in \eqref{eq:TruncGen} are $G_S$-invariant and project as 
\begin{equation}
X_{IJ}{}^M=X_{IJ}{}^N \mathbb{P}^M{}_N=X_{IJ}{}^K \hat{K}_K{}^M\, .
\label{XIJMProjection}
\end{equation}
Projecting the quadratic constraint, \eqref{eq:LieAlg}, of the parent $\cN=8$ theory to the invariant subsector and plugging in \eqref{XIJMProjection}, we obtain
\begin{equation}
[X_I, X_J]=-X_{IJ}{}^K X_K\,,
\label{ReducedQuadraticConstraint}
\end{equation}
so closure of the $\tilde{G}$ algebra indeed holds. 

Consistency of the embedding tensor with the two-forms of the reduced theory requires
\begin{equation} \label{CompatibilityReducedTwoForms}
\Theta_I{}^y Z^{I x}=0=\vartheta_I Z^{I x}\,,
\end{equation}
where
\begin{equation} \label{eq:ZZRed}
Z^{I x} \equiv \big(  \Theta^{Ix} - 16 \, t^{x IJ} \vartheta_J \big)\,.
\end{equation}
This follows form the projection of \eqref{eq:QuadOrtho} to the $G_S$-invariant sector, {\it i.e}, $\Theta_M{}^y Z^{M x}=0=\vartheta_M Z^{M x}$, together with the consistency constraint C2 in \eqref{eq:TruncCondXLin}. To see this, note that (\ref{eq:TruncCondXLin}) says that the combination $\big(  \Theta^{M\alpha} - 16 \, t^{\alpha MN} \vartheta_N \big)  \, \hat{K}^{x}{}_\alpha$ is a $G_S$-singlet, namely, $\big(  \Theta^{M\alpha} - 16 \, t^{\alpha MN} \vartheta_N \big)  \, \hat{K}^{x}{}_\alpha = \mathbb{P}^M{}_P \big(  \Theta^{P\alpha} - 16 \, t^{\alpha PN} \vartheta_N \big)  \, \hat{K}^{x}{}_\alpha$, with the projector $\mathbb{P}^M{}_P$ given by the quadratic combination (\ref{eq:SingletProj}) of invariant tensors. Thus:
\begin{equation} \label{ZMalphaFactorizedZMx}
Z^{M x}= \hat{K}_I{}^M \big(  \Theta^{I\alpha} - 16 \, t^{\alpha IN} \vartheta_N \big)  \, \hat{K}^{x}{}_\alpha = \hat{K}_I{}^M Z^{Ix}\,,
\end{equation}
from which \eqref{CompatibilityReducedTwoForms} follows by plugging it into $\Theta_M{}^y Z^{M x}=0=\vartheta_M Z^{M x}$.

We have therefore proved that the linear, \eqref{ReducedLinearConstraint}, and quadratic, \eqref{ReducedQuadraticConstraint}, \eqref{CompatibilityReducedTwoForms}, constraints that any embedding tensor must satisfy to define a self-consistent theory, follow from our consistency conditions \eqref{eq:TruncCondXLin} together with the linear and quadratic constraints of the parent $\cN=8$ theory. In the remainder of this section, we will derive other useful identities from \eqref{ReducedLinearConstraint}, \eqref{ReducedQuadraticConstraint}, \eqref{CompatibilityReducedTwoForms}, extending some previously known tromboneless identities to also include trombone components of the reduced embedding tensor. Without loss of generality, we will allow for some of the $G_S$-invariant adjoint elements $t_x$ to have a trivial action on vectors, {\it i.e}, $t_x=(t_u, t_l)$ with $(t_l)_I{}^J=0$. This is for instance the case of $\cN=2$ models where the vector kinetic terms are independent of the hyperscalars. The reduced embedding tensor \eqref{ReducedEmbeddingTensor} thus takes the refined form
\begin{equation}
X_I = - w \vartheta_I t_0 + \left( \Theta_I{}^u + 8 (t^u)_I{}^J \vartheta_J \right)t_u + \Theta_I{}^l t_l\,.
\label{ReducedEmbeddingTensorGeneralCase}
\end{equation}

It takes some tedious algebra to show that closure of the algebra \eqref{ReducedQuadraticConstraint} for the embedding tensor \eqref{ReducedEmbeddingTensorGeneralCase} is equivalent to the following three conditions
\begin{align}
\big( \Theta_I{}^u + 8 (t^u)_I{}^K \vartheta_K \big) (t_u)_J{}^L \Theta_L{}^w + \big( \Theta_I{}^u + 8 (t^u)_I{}^K \vartheta_K \big) c_{uv}{}^w \Theta_J{}^v - \vartheta_I \Theta_J{}^w & =0\,,
\label{912Invariance}\\[3pt]
\big( \Theta_I{}^u + 8 (t^u)_I{}^K \vartheta_K \big) (t_u)_J{}^L \vartheta_L -  \vartheta_I \vartheta_J  & =0 \,,
\label{TromboneInvariance}\\[3pt]
\big( \Theta_I{}^u + 8 (t^u)_I{}^K \vartheta_K \big) (t_u)_J{}^L \Theta_L{}^n +\Theta_I{}^l  c_{lm}{}^n \Theta_J{}^m - \vartheta_I \Theta_J{}^n & =0\, .
\label{912ExtraInvariance}
\end{align}
These are reminiscent of those for tromboneful maximal supergravity \cite{LeDiffon:2008sh,LeDiffon:2011wt}. To further simplify these equations, we need the identity
\begin{equation}
(t_\alpha)_I{}^K (t^\alpha)_J{}^L = (t_x)_I{}^K (t^x)_J{}^L = \frac{1}{24} \delta_I^K \delta_J^L + \frac{1}{12} \delta_I^L \delta_J^K + (t_x)_{IJ} (t^x)^{KL} -\frac{1}{24} \Omega_{IJ} \Omega^{KL}\,,
\label{ProjectorDecomposition}
\end{equation}
which stems from the reduction of the formally identical E$_{7(7)}$ identity. Upon use of \eqref{LinearConstraintIdentities}, \eqref{ProjectorDecomposition}, equation \eqref{TromboneInvariance} can be shown to split as
\begin{align}
(t_u)_{[I}{}^K \Theta_{J]}{}^u \vartheta_K & =0\,, \label{eq:QCs1} \\
\Omega^{IJ}\vartheta_J \Theta_I{}^u + 16( t^u)^{IJ}\vartheta_I\vartheta_J & =0\,. \label{eq:QCs2}
\end{align}
Also, note that the quadratic constraint \eqref{ReducedQuadraticConstraint} implies $X_{(IJ)}{}^K X_K=0$. From here, using \eqref{LinearConstraintIdentities} and again \eqref{ProjectorDecomposition}, the following relations arise:
\begin{align}
\Omega^{IJ}\Theta_I{}^u \Theta_J{}^v  - 8 \vartheta_I \Theta_J{}^{[u} (t^{v]})^{IJ}+ 4 c^{uv}{}_w \vartheta_I \Theta_J{}^w \Omega^{IJ} & =0\,, \label{eq:QCs3}
\\
\Omega^{IJ}\Theta_I{}^l \Theta_J{}^u  - 4 \vartheta_I \Theta_J{}^{l} (t^{u})^{IJ} & =0\,. \label{eq:QCs4}
\end{align}
Since the reduced linear and quadratic constraints, \eqref{ReducedLinearConstraint}, \eqref{ReducedQuadraticConstraint} and \eqref{CompatibilityReducedTwoForms}, follow from \eqref{eq:TruncCondXLin} together with the corresponding linear and quadratic constraints of the $\cN=8$ theory, the constraints \eqref{eq:QCs1}--\eqref{eq:QCs4} are satisfied by the embedding tensor \eqref{ReducedEmbeddingTensor}, \eqref{ReducedEmbeddingTensorGeneralCase} of any consistent subsector. When all $G_S$-invariant generators have no trivial action on vectors, equation \eqref{eq:QCs4} drops and equations \eqref{eq:QCs1}-\eqref{eq:QCs3} are formally identical to \eqref{eq:QCs}. Thus, we recover for the subsector the same set of quadratic constraints as the one exhibited by the $\cN=8$ theory. 

To conclude this section, let us note that when there are generators with trivial action on vectors, $(t_l)_I{}^J=0$, equation \eqref{CompatibilityReducedTwoForms} implies:
\begin{equation}
\Omega^{IJ}\vartheta_I \Theta_J{}^l=0\,, \quad\Omega^{IJ}\Theta_I{}^l \Theta_J{}^m=0\,.
\end{equation}
These two equations recover, and extend to tromboneful gaugings, condition (3.66) of \cite{Trigiante:2016mnt}, which must be imposed as an extra quadratic constraint on the embedding tensor when such generators with trivial action on vectors are present in the theory.


\subsection{Consistency of the $G_S$-invariant truncation of $\cN=8$ supergravity} \label{sec:ConsProofsFromN8}


Having established the self-consistency as a theory in its own right of the $G_S$-invariant subsector of a maximal supergravity, we will now move on to show that the consistency conditions (\ref{eq:TruncCondXLin}), (\ref{eq:TruncCondXQuad}) also guarantee the consistency of the truncation from the parent $\cN=8$ supergravity at the level of the equations of motion. Again, the proof does not rely on supersymmetry, or on the retained subsector being itself an $\cN <8$ supergravity. More concretely, we will show that, under the conditions (\ref{eq:TruncCondXLin}), (\ref{eq:TruncCondXQuad}), the parent $\cN=8$ field equations (\ref{eq:EinsteinEOM})--(\ref{eq:ScaalarDuality}) reduce consistently to the following field equations for the $G_S$-invariant fields:
{\setlength\arraycolsep{0pt}
\begin{eqnarray}
\label{eq:EinsteinEOMRed} && R_{\mu \nu} = - \tfrac{1}{48} D_\mu \tilde{M}_{MN} D_\nu \tilde{M}^{MN}  + \tfrac{1}{2} g_{\mu\nu} V + \tfrac{1}{4 }  \tilde{M}_{IJ} \big(  H_{\mu \rho}{}^I H_{\nu}{}^\rho{}^J  - \tfrac{1}{4} g_{\mu\nu}  \, H_{\rho \sigma}{}^I \, H^{\rho \sigma J}  \big)  , \quad    \\[8pt]
\label{eq:vectorEOMRed} &&  D \big( \tilde{M}_{IJ} *H_\2^J \big) - \tfrac{1}{8}   \,\big( \Theta_{IN}{}^P - 16 \hat{K}_I{}^P \vartheta_N  \big)  \, \tilde{M}^{NQ} * D \tilde{M}_{PQ}=0\, ,
  \\[8pt]
\label{eq:scalarEOMRed} && \hat{K}_{x}{}^{\alpha} (t_\alpha)_{S}{}^{(P} \tilde{M}^{Q)S} \Big( 
D*D\tilde{M}_{PQ}  -\tilde{M}^{RS} D\tilde{M}_{PR} \wedge *D\tilde{M}_{QS}  -12 \, \tilde{M}_{PR} \tilde{M}_{QS} \, H_\2^R \wedge * H_\2^S  \nonumber \\
&& \qquad\qquad\qquad \qquad\quad -24 \, \tilde{V}_{PQ}  \, \textrm{vol}_4 \Big) =0  \; .
\end{eqnarray}
}As for the $\cN=8$ theory, these must be supplemented by the following selfduality condition and Bianchi identity for the retained gauge fields
\begin{equation} \label{eq:SDandBianchiRed}
H_\2^I = - \Omega^{IJ} \tilde{M}_{JK} * H_\2^K \; , \qquad 
D H_\2^{I} =    \big(  \Theta^{I x} - 16 \, t^{x IJ} \vartheta_J \big)   H_{\3 x}  \; .
\end{equation}
In~(\ref{eq:EinsteinEOMRed})--(\ref{eq:SDandBianchiRed}), $\tilde{M}_{MN} = (\tilde{\cV} \tilde{\cV}^{\textrm{T}})_{MN}$ is the usual metric on $\mathrm{C}_{\mathrm{E}_{7(7)}}(G_S) / \mathrm{C}_{\mathrm{SU}(8)}(G_S)$, built out of the $G_S$-invariant coset $\tilde{\cV}$ in (\ref{sec:SubFieldCont}) in the ($G_S$-reducible) $\bm{56}$-dimensional representation, and $\tilde{M}_{IJ} \equiv \hat{K}_I{}^M \hat{K}_J{}^N \tilde{M}_{MN}$ the same metric in the $2n_1$-dimensional representation provided by the space of $G_S$-singlets that descend from the $\bm{56}$. These matrices have respective inverses $\tilde{M}^{MN}$ and $\tilde{M}^{IJ}$. The covariant derivative of $\tilde{M}^{MN}$ is
\begin{equation} \label{eq:CovDerMRed}
D \tilde{M}_{MN} = d \tilde{M}_{MN} - 2 A^I X_{I(M}{}^{P} \tilde{M}_{N)P} -2  A^I \vartheta_I \tilde{M}_{MN} \; , 
\end{equation}
and the invariant gauge field strengths of the retained vectors and two-forms in (\ref{sec:SubFieldCont}) are
{\setlength\arraycolsep{0pt}
\begin{eqnarray}
\label{eq:H2FormRed} && H_\2^I \equiv dA^I + \tfrac12 X_{JK}{}^I A^J \wedge A^K +  \big(  \Theta^{I x} - 16 \, t^{x IJ} \vartheta_J \big)\, B_x  \; , \\[5pt]
\label{eq:H3FormRed} && H_{\3 x}  \equiv  DB_x -\tfrac12 t_{x IJ} A^I \wedge dA^J - \tfrac{1}{3!} \, t_{x IL} \,   X_{JK}{}^L A^I \wedge A^J \wedge A^K \; .
\end{eqnarray}
}Finally, the quantities $V$ and $\tilde{V}_{MN}$ that respectively appear in the Einstein, (\ref{eq:EinsteinEOMRed}), and scalar, (\ref{eq:scalarEOMRed}), equations of motion, formally remain as in (\ref{eq:Cosmo}), (\ref{eq:DerPotTromb}), now written in terms of the $G_S$-invariant $\tilde{M}_{MN}$. The reduced field equations (\ref{eq:EinsteinEOMRed})--(\ref{eq:SDandBianchiRed}) naturally feature quantities in either, or both, the $\bm{56}$ and $2n_1$-dimensional representations. The former is inherited from the $\cN=8$ theory, while the latter is intrinsic to the subsector.

Let us first see how the field strengths (\ref{eq:H2FormRed}), (\ref{eq:H3FormRed}) of the reduced vector and tensors follow from their $\cN=8$ counterparts (\ref{eq:H2Form}), (\ref{eq:H3Form}). In order to do this, we will build on the results derived in appendix~\ref{sec:SelfConsProofs} on self-consistency of the retained subsector. Under (\ref{eq:RetainedVecs}), the gauge field strength (\ref{eq:H2Form}) factorises as 
\begin{equation} \label{eq:H2MFactorizationH2i}
H_\2^M = H_\2^I \hat{K}_I{}^M \; .
\end{equation}
The factorisation is straightforward for the term linear in $dA^M$. For the $A^M \wedge A^N$ term, it follows from \eqref{XIJMProjection}, and for the term linear in $B_\alpha$, it follows from \eqref{ZMalphaFactorizedZMx}. Thus, $\hat{K}_I{}^M$ factorises across all three terms that make up $H_\2^M$ in (\ref{eq:H2Form}). Similarly, under (\ref{eq:RetainedVecs}), the three-form field strength (\ref{eq:H3Form}) factorises in terms of (\ref{eq:H3FormRed}) as 
\begin{equation} \label{eq:H3alphaFactorizationH3x}
H_{\3 \alpha}  = H_{\3 x} \, K^x{}_\alpha \; .
\end{equation}
This is again straightforward for the $dB_\alpha$ term. For the $A^M \wedge dA^N$ contribution, \eqref{talphaIJProjectiontxIJ} guarantees that $K^x{}_\alpha$ factorises from the second term leaving behind the combination $t_{xIJ}$ defined below \eqref{ReducedEmbeddingTensor}. Finally, in the $A^M \wedge A^N \wedge A^P$ term, $K^x{}_\alpha$ factorises by virtue of \eqref{talphaIJProjectiontxIJ} and \eqref{XIJMProjection}. Direct substitution of \eqref{eq:H2MFactorizationH2i}, \eqref{eq:H3alphaFactorizationH3x} into \eqref{eq:SDandBianchi} immediately leads to the self-duality conditions and Bianchi identities \eqref{eq:SDandBianchiRed} for the retained gauge fields. The result follows from the fact that $\hat{K}_I{}^M$ and $\hat{K}_x{}^\alpha$ respectively form a basis of the vector spaces spanned by the $G_S$ singlets coming from the $\bm{56}$ and $\bm{133}$ representations. 

The consistency of the reduced equations of motion proper, \eqref{eq:EinsteinEOMRed}--\eqref{eq:scalarEOMRed}, can be similarly argued as follows. The Einstein equation \eqref{eq:EinsteinEOM} is already an E$_{7(7)}$-singlet, so its consistency is guaranteed if expressed in terms solely of $G_S$-invariant fields. Simply substituting in the expression \eqref{eq:H2MFactorizationH2i} for the retained gauge field strengths, the Einstein equation can be cast as in \eqref{eq:EinsteinEOMRed}. The Maxwell equation \eqref{eq:vectorEOM} reduces, in turn, to \eqref{eq:vectorEOMRed} by again using the factorisation \eqref{eq:H2MFactorizationH2i} of the gauge field strengths. To see this, first note that the kinetic term in \eqref{eq:EinsteinEOM} factorises as $ K^I{}_MD \big( \tilde{M}_{IJ} *H_\2^J \big)$ upon substitution of \eqref{eq:H2MFactorizationH2i}. This happens, firstly, because $\tilde{M}_{MN} \hat{K}_J{}^N H_\2^J$ is $G_S$-invariant, so $\mathbb{P}^M{}_N$ in \eqref{eq:SingletProj} accordingly acts on it as $\delta^M_N$; and, secondly, because $ D \big( \tilde{M}_{MJ} *H_\2^J \big)$ contains only the $G_S$-invariant components of the embedding tensor $X_I$ in \eqref{ReducedEmbeddingTensor}, \eqref{ReducedEmbeddingTensorGeneralCase}, along with $G_S$-invariant gauge fields. The scalar current term in \eqref{eq:EinsteinEOM} factorises in turn as $ \tfrac{1}{8} K^I{}_M \hat{K}_I{}^R \,\big( \Theta_{RN}{}^P - 16 \delta_R^P \vartheta_N  \big)  \, \tilde{M}^{NQ} * D \tilde{M}_{PQ}$ by virtue of C2 in \eqref{eq:TruncCondXLin} and the observation that $\tilde{M}^{NQ} * D \tilde{M}_{PQ}$ is adjoint projected to $\mathrm{C}_{\mathrm{E}_{7(7)}}(G_S)$, namely, $\tilde{M}^{NQ} * D \tilde{M}_{PQ} = (t_x)_P{}^N (t^x)_R{}^S \tilde{M}^{RQ} * D \tilde{M}_{SQ}$. 

Finally, let us argue that the $\cN=8$ scalar equation of motion \eqref{eq:scalarEOM} reduces to its counterpart \eqref{eq:scalarEOMRed} in the reduced subsector. To see this, note that from C1 in \eqref{eq:TruncCondXLin} it follows that the scalar covariant derivative \eqref{eq:CovDerMRed} is a $G_S$ singlet. We have already seen that the factorised $H_\2^M$ in (\ref{eq:H2MFactorizationH2i}) is $G_S$-invariant. The term $\tilde{V}_{MN}$ containing scalar self-interactions only contains $G_S$-singlets by the condition C3 in \eqref{eq:TruncCondXQuad}. Under these circumstances, \eqref{eq:scalarEOM} contains only $G_S$ singlets and can be explicitly projected with $\hat{K}_{x}{}^{\alpha}$ as in \eqref{eq:scalarEOMRed}. Some, but not all, terms in the reduced scalar equation \eqref{eq:scalarEOMRed} admit a rewrite in terms of quantities in the $2n_1$-dimensional representation inherent to the subsector. This is the case of the term quadratic in gauge field strengths, which can be written in terms of $H_2^I$ and $\tilde{M}_{IJ}$. However, the term quadratic in $D\tilde{M}_{MN}$ does not admit such rewrite and must stay in the $\bm{56}$ inherited from $\cN=8$. In order to avoid a hybrid notation, all quantities in \eqref{eq:scalarEOMRed} have been placed in the $\bm{56}$.

We conclude this section by commenting on the consistency of the truncation when there are no singlets contained in the branching of the $\bm{56}$ under $G_S \subset \textrm{E}_{7(7)}$, so that $\hat{K}_I{}^M=0$. In this case, the field strengths $H_\2^I$ in (\ref{eq:H2FormRed}) are not well defined and the need to impose C2 in (\ref{eq:TruncCondXLin}) does not even arise. One must instead work with the full $\cN=8$ field strength $H_\2^M$, now defined as
\begin{equation} \label{eq:HNoVecRed}
H_\2^M \equiv   \big(  \Theta^{M\alpha} - 16 \, t^{\alpha MN} \vartheta_N \big)\, K^x{}_\alpha \,  B_x \; ,
\end{equation}
and still subject to the selfduality relation (\ref{eq:SDandBianchi}), now written in terms the singlet $\tilde{M}_{NP}$,
\begin{equation}
H_\2^M = - \Omega^{MN} \tilde{M}_{NP} * H_\2^P \; .
\end{equation}
The Bianchi identity in (\ref{eq:SDandBianchi}) enforces the duality of two-forms and scalars,
\begin{equation}
\label{eq:vectorEOMRedNoVectors}    \big(  \Theta^{Mx} - 16 \, t^{x MN} \vartheta_N \big)  \big( d B_{ x} +\tfrac18 (t_x)_P{}^Q \, \tilde{M}^{PR} * d \tilde{M}_{QR} \big) =0  \, , 
\end{equation}
while the $\cN=8$ Maxwell equation (\ref{eq:vectorEOM}) becomes redundant, as it simply becomes the Hodge dual version of (\ref{eq:vectorEOMRedNoVectors}). 
The reduced scalar equation formally stays the same, \eqref{eq:scalarEOM}, as in the $\hat{K}_I{}^M \neq 0$ case, only with covariant derivatives replaced with ordinary derivatives, gauge field strenghts given by (\ref{eq:HNoVecRed}), and still with $G_S$-singlet scalar self-interactions by virtue of C3. Finally, the following Einstein equation replaces (\ref{eq:EinsteinEOMRed}):
\begin{equation}
\label{eq:EinsteinEOMRedNoVectors} R_{\mu \nu} = - \tfrac{1}{48} \partial_\mu \tilde{M}_{MN} \partial_\nu \tilde{M}^{MN}  + \tfrac{1}{2} g_{\mu\nu} V  +  \tfrac{1}{4 }  \tilde{M}_{MN} \big(  H_{\mu \rho}{}^M H_{\nu}{}^\rho{}^N  - \tfrac{1}{4} g_{\mu\nu}  \, H_{\rho \sigma}{}^M \, H^{\rho \sigma N}  \big) , \quad   
\end{equation}
with gauge field strenghts again as in (\ref{eq:HNoVecRed}).


\bibliography{references}

\providecommand{\href}[2]{#2}\begingroup\raggedright\begin{thebibliography}{10}

\bibitem{Maldacena:1997re}
J.~M. Maldacena, {\it {The Large N limit of superconformal field theories and
  supergravity}},  {\em Int. J. Theor. Phys.} {\bf 38} (1999) 1113--1133,
  [\href{http://arxiv.org/abs/hep-th/9711200}{{\tt hep-th/9711200}}]. [Adv.
  Theor. Math. Phys.2,231(1998)].

\bibitem{Gunaydin:1985cu}
M.~Gunaydin, L.~J. Romans, and N.~P. Warner, {\it {Compact and Noncompact
  Gauged Supergravity Theories in Five-Dimensions}},  {\em Nucl. Phys. B} {\bf
  272} (1986) 598--646.

\bibitem{Pernici:1985ju}
M.~Pernici, K.~Pilch, and P.~van Nieuwenhuizen, {\it {Gauged N=8 D=5
  Supergravity}},  {\em Nucl. Phys. B} {\bf 259} (1985) 460.

\bibitem{deWit:1982ig}
B.~de~Wit and H.~Nicolai, {\it {N=8 Supergravity}},  {\em Nucl.Phys.} {\bf
  B208} (1982) 323.

\bibitem{deWit:1981eq}
B.~de~Wit and H.~Nicolai, {\it {N=8 Supergravity with Local SO(8) x SU(8)
  Invariance}},  {\em Phys. Lett.} {\bf B108} (1982) 285.

\bibitem{Aharony:2008ug}
O.~Aharony, O.~Bergman, D.~L. Jafferis, and J.~Maldacena, {\it {N=6
  superconformal Chern-Simons-matter theories, M2-branes and their gravity
  duals}},  {\em JHEP} {\bf 10} (2008) 091,
  [\href{http://arxiv.org/abs/0806.1218}{{\tt arXiv:0806.1218}}].

\bibitem{Maldacena:2000mw}
J.~M. Maldacena and C.~Nunez, {\it {Supergravity description of field theories
  on curved manifolds and a no go theorem}},  {\em Int. J. Mod. Phys. A} {\bf
  16} (2001) 822--855, [\href{http://arxiv.org/abs/hep-th/0007018}{{\tt
  hep-th/0007018}}].

\bibitem{Acharya:2000mu}
B.~S. Acharya, J.~P. Gauntlett, and N.~Kim, {\it {Five-branes wrapped on
  associative three cycles}},  {\em Phys. Rev. D} {\bf 63} (2001) 106003,
  [\href{http://arxiv.org/abs/hep-th/0011190}{{\tt hep-th/0011190}}].

\bibitem{Gauntlett:2006ux}
J.~P. Gauntlett, O.~A.~P. Mac~Conamhna, T.~Mateos, and D.~Waldram, {\it {AdS
  spacetimes from wrapped M5 branes}},  {\em JHEP} {\bf 11} (2006) 053,
  [\href{http://arxiv.org/abs/hep-th/0605146}{{\tt hep-th/0605146}}].

\bibitem{Gauntlett:2003di}
J.~P. Gauntlett, {\it {Branes, calibrations and supergravity}},  {\em Clay
  Math. Proc.} {\bf 3} (2004) 79--126,
  [\href{http://arxiv.org/abs/hep-th/0305074}{{\tt hep-th/0305074}}].

\bibitem{Ferrero:2020laf}
P.~Ferrero, J.~P. Gauntlett, J.~M. P{\'e}rez~Ipi{\~n}a, D.~Martelli, and
  J.~Sparks, {\it {D3-Branes Wrapped on a Spindle}},  {\em Phys. Rev. Lett.}
  {\bf 126} (2021), no.~11 111601, [\href{http://arxiv.org/abs/2011.10579}{{\tt
  arXiv:2011.10579}}].

\bibitem{Bah:2021mzw}
I.~Bah, F.~Bonetti, R.~Minasian, and E.~Nardoni, {\it {Holographic Duals of
  Argyres-Douglas Theories}},  {\em Phys. Rev. Lett.} {\bf 127} (2021), no.~21
  211601, [\href{http://arxiv.org/abs/2105.11567}{{\tt arXiv:2105.11567}}].

\bibitem{Ferrero:2021wvk}
P.~Ferrero, J.~P. Gauntlett, D.~Martelli, and J.~Sparks, {\it {M5-branes
  wrapped on a spindle}},  {\em JHEP} {\bf 11} (2021) 002,
  [\href{http://arxiv.org/abs/2105.13344}{{\tt arXiv:2105.13344}}].

\bibitem{Gaiotto:2009we}
D.~Gaiotto, {\it {N=2 dualities}},  {\em JHEP} {\bf 08} (2012) 034,
  [\href{http://arxiv.org/abs/0904.2715}{{\tt arXiv:0904.2715}}].

\bibitem{Gaiotto:2009gz}
D.~Gaiotto and J.~Maldacena, {\it {The Gravity duals of N=2 superconformal
  field theories}},  {\em JHEP} {\bf 10} (2012) 189,
  [\href{http://arxiv.org/abs/0904.4466}{{\tt arXiv:0904.4466}}].

\bibitem{Bah:2011vv}
I.~Bah, C.~Beem, N.~Bobev, and B.~Wecht, {\it {AdS/CFT Dual Pairs from
  M5-Branes on Riemann Surfaces}},  {\em Phys. Rev. D} {\bf 85} (2012) 121901,
  [\href{http://arxiv.org/abs/1112.5487}{{\tt arXiv:1112.5487}}].

\bibitem{Bah:2012dg}
I.~Bah, C.~Beem, N.~Bobev, and B.~Wecht, {\it {Four-Dimensional SCFTs from
  M5-Branes}},  {\em JHEP} {\bf 06} (2012) 005,
  [\href{http://arxiv.org/abs/1203.0303}{{\tt arXiv:1203.0303}}].

\bibitem{Dimofte:2011ju}
T.~Dimofte, D.~Gaiotto, and S.~Gukov, {\it {Gauge Theories Labelled by
  Three-Manifolds}},  {\em Commun. Math. Phys.} {\bf 325} (2014) 367--419,
  [\href{http://arxiv.org/abs/1108.4389}{{\tt arXiv:1108.4389}}].

\bibitem{Szepietowski:2012tb}
P.~Szepietowski, {\it {Comments on a-maximization from gauged supergravity}},
  {\em JHEP} {\bf 12} (2012) 018, [\href{http://arxiv.org/abs/1209.3025}{{\tt
  arXiv:1209.3025}}].

\bibitem{MatthewCheung:2019ehr}
K.~C. Matthew~Cheung, J.~P. Gauntlett, and C.~Rosen, {\it {Consistent KK
  truncations for M5-branes wrapped on Riemann surfaces}},  {\em Class. Quant.
  Grav.} {\bf 36} (2019), no.~22 225003,
  [\href{http://arxiv.org/abs/1906.08900}{{\tt arXiv:1906.08900}}].

\bibitem{Cassani:2019vcl}
D.~Cassani, G.~Josse, M.~Petrini, and D.~Waldram, {\it {Systematics of
  consistent truncations from generalised geometry}},  {\em JHEP} {\bf 11}
  (2019) 017, [\href{http://arxiv.org/abs/1907.06730}{{\tt arXiv:1907.06730}}].

\bibitem{Faedo:2019cvr}
A.~F. Faedo, C.~Nunez, and C.~Rosen, {\it {Consistent truncations of
  supergravity and $\frac{1}{2}$-BPS RG flows in $4d$ SCFTs}},  {\em JHEP} {\bf
  03} (2020) 080, [\href{http://arxiv.org/abs/1912.13516}{{\tt
  arXiv:1912.13516}}].

\bibitem{Cassani:2020cod}
D.~Cassani, G.~Josse, M.~Petrini, and D.~Waldram, {\it {$\mathcal{N} $ = 2
  consistent truncations from wrapped M5-branes}},  {\em JHEP} {\bf 02} (2021)
  232, [\href{http://arxiv.org/abs/2011.04775}{{\tt arXiv:2011.04775}}].

\bibitem{Bhattacharya:2024tjw}
R.~Bhattacharya, A.~Katyal, and O.~Varela, {\it {Class S Superconformal Indices
  from Maximal Supergravity}},  {\em Phys. Rev. Lett.} {\bf 134} (2025), no.~18
  181601, [\href{http://arxiv.org/abs/2411.16837}{{\tt arXiv:2411.16837}}].

\bibitem{Varela:2025xeb}
O.~Varela, {\it {Trombone gaugings of five-dimensional maximal supergravity}},
  \href{http://arxiv.org/abs/2509.12391}{{\tt arXiv:2509.12391}}.

\bibitem{Gauntlett:2002rv}
J.~P. Gauntlett, N.~Kim, S.~Pakis, and D.~Waldram, {\it {M theory solutions
  with AdS factors}},  {\em Class. Quant. Grav.} {\bf 19} (2002) 3927--3946,
  [\href{http://arxiv.org/abs/hep-th/0202184}{{\tt hep-th/0202184}}].

\bibitem{Donos:2010ax}
A.~Donos, J.~P. Gauntlett, N.~Kim, and O.~Varela, {\it {Wrapped M5-branes,
  consistent truncations and AdS/CMT}},  {\em JHEP} {\bf 12} (2010) 003,
  [\href{http://arxiv.org/abs/1009.3805}{{\tt arXiv:1009.3805}}].

\bibitem{DallAgata:2011aa}
G.~Dall'Agata and G.~Inverso, {\it {On the Vacua of N = 8 Gauged Supergravity
  in 4 Dimensions}},  {\em Nucl.Phys.} {\bf B859} (2012) 70--95,
  [\href{http://arxiv.org/abs/1112.3345}{{\tt arXiv:1112.3345}}].

\bibitem{Dall'Agata:2014ita}
G.~Dall'Agata, G.~Inverso, and A.~Marrani, {\it {Symplectic Deformations of
  Gauged Maximal Supergravity}},  {\em JHEP} {\bf 1407} (2014) 133,
  [\href{http://arxiv.org/abs/1405.2437}{{\tt arXiv:1405.2437}}].

\bibitem{Scherk:1979zr}
J.~Scherk and J.~H. Schwarz, {\it {How to Get Masses from Extra Dimensions}},
  {\em Nucl. Phys. B} {\bf 153} (1979) 61--88.

\bibitem{Pico:2026rji}
M.~Pico and O.~Varela, {\it {Maximal trombone supergravity from wrapped
  M5-branes}},  \href{http://arxiv.org/abs/2601.07960}{{\tt arXiv:2601.07960}}.

\bibitem{deWit:2007mt}
B.~de~Wit, H.~Samtleben, and M.~Trigiante, {\it {The Maximal D=4
  supergravities}},  {\em JHEP} {\bf 0706} (2007) 049,
  [\href{http://arxiv.org/abs/0705.2101}{{\tt arXiv:0705.2101}}].

\bibitem{LeDiffon:2008sh}
A.~Le~Diffon and H.~Samtleben, {\it {Supergravities without an Action: Gauging
  the Trombone}},  {\em Nucl. Phys. B} {\bf 811} (2009) 1--35,
  [\href{http://arxiv.org/abs/0809.5180}{{\tt arXiv:0809.5180}}].

\bibitem{LeDiffon:2011wt}
A.~Le~Diffon, H.~Samtleben, and M.~Trigiante, {\it {N=8 Supergravity with Local
  Scaling Symmetry}},  {\em JHEP} {\bf 04} (2011) 079,
  [\href{http://arxiv.org/abs/1103.2785}{{\tt arXiv:1103.2785}}].

\bibitem{Trigiante:2016mnt}
M.~Trigiante, {\it {Gauged Supergravities}},  {\em Phys. Rept.} {\bf 680}
  (2017) 1--175, [\href{http://arxiv.org/abs/1609.09745}{{\tt
  arXiv:1609.09745}}].

\bibitem{Cremmer:1979up}
E.~Cremmer and B.~Julia, {\it {The SO(8) Supergravity}},  {\em Nucl. Phys. B}
  {\bf 159} (1979) 141--212.

\bibitem{Warner:1983vz}
N.~Warner, {\it {Some New Extrema of the Scalar Potential of Gauged $N=8$
  Supergravity}},  {\em Phys.Lett.} {\bf B128} (1983) 169.

\bibitem{Warner:1983du}
N.~P. Warner, {\it {Some Properties of the Scalar Potential in Gauged
  Supergravity Theories}},  {\em Nucl. Phys. B} {\bf 231} (1984) 250--268.

\bibitem{Guarino:2024gke}
A.~Guarino, C.~Sterckx, and M.~Trigiante, {\it {Consistent N=4, D=4 truncation
  of type IIB supergravity on $S^1 \times S^5$}},  {\em Phys. Rev. D} {\bf 111}
  (2025), no.~4 046019, [\href{http://arxiv.org/abs/2410.23149}{{\tt
  arXiv:2410.23149}}].

\bibitem{Coimbra:2011nw}
A.~Coimbra, C.~Strickland-Constable, and D.~Waldram, {\it {Supergravity as
  Generalised Geometry I: Type II Theories}},  {\em JHEP} {\bf 11} (2011) 091,
  [\href{http://arxiv.org/abs/1107.1733}{{\tt arXiv:1107.1733}}].

\bibitem{Coimbra:2011ky}
A.~Coimbra, C.~Strickland-Constable, and D.~Waldram, {\it {$E_{d(d)} \times
  \mathbb{R}^+$ generalised geometry, connections and M theory}},  {\em JHEP}
  {\bf 02} (2014) 054, [\href{http://arxiv.org/abs/1112.3989}{{\tt
  arXiv:1112.3989}}].

\bibitem{Coimbra:2012af}
A.~Coimbra, C.~Strickland-Constable, and D.~Waldram, {\it {Supergravity as
  Generalised Geometry II: $E_{d(d)} \times \mathbb{R}^+$ and M theory}},  {\em
  JHEP} {\bf 03} (2014) 019, [\href{http://arxiv.org/abs/1212.1586}{{\tt
  arXiv:1212.1586}}].

\bibitem{Blair:2024ofc}
C.~D.~A. Blair, M.~Pico, and O.~Varela, {\it {Infinite and finite consistent
  truncations on deformed generalised parallelisations}},  {\em JHEP} {\bf 09}
  (2024) 065, [\href{http://arxiv.org/abs/2407.01298}{{\tt arXiv:2407.01298}}].

\bibitem{Guarino:2015qaa}
A.~Guarino and O.~Varela, {\it {Dyonic ISO(7) supergravity and the duality
  hierarchy}},  {\em JHEP} {\bf 02} (2016) 079,
  [\href{http://arxiv.org/abs/1508.04432}{{\tt arXiv:1508.04432}}].

\bibitem{Hull:1984yy}
C.~Hull, {\it {New Gauging of $N=8$ Supergravity}},  {\em Phys.Rev.} {\bf D30}
  (1984) 760.

\bibitem{Hull:1984vg}
C.~M. Hull, {\it {Noncompact Gaugings of $N=8$ Supergravity}},  {\em Phys.
  Lett. B} {\bf 142} (1984) 39.

\bibitem{Hull:1984qz}
C.~M. Hull, {\it {More Gaugings of $N=8$ Supergravity}},  {\em Phys. Lett. B}
  {\bf 148} (1984) 297--300.

\bibitem{Dall'Agata:2012bb}
G.~Dall'Agata, G.~Inverso, and M.~Trigiante, {\it {Evidence for a family of
  SO(8) gauged supergravity theories}},  {\em Phys.Rev.Lett.} {\bf 109} (2012)
  201301, [\href{http://arxiv.org/abs/1209.0760}{{\tt arXiv:1209.0760}}].

\bibitem{DallAgata:2012plb}
G.~Dall'Agata and G.~Inverso, {\it {de Sitter vacua in N = 8 supergravity and
  slow-roll conditions}},  {\em Phys. Lett. B} {\bf 718} (2013) 1132--1136,
  [\href{http://arxiv.org/abs/1211.3414}{{\tt arXiv:1211.3414}}].

\bibitem{Catino:2013ppa}
F.~Catino, G.~Dall'Agata, G.~Inverso, and F.~Zwirner, {\it {On the moduli space
  of spontaneously broken $N = 8$ supergravity}},  {\em JHEP} {\bf 09} (2013)
  040, [\href{http://arxiv.org/abs/1307.4389}{{\tt arXiv:1307.4389}}].

\bibitem{Gallerati:2014xra}
A.~Gallerati, H.~Samtleben, and M.~Trigiante, {\it {The $ \mathcal{N}>2 $
  supersymmetric AdS vacua in maximal supergravity}},  {\em JHEP} {\bf 12}
  (2014) 174, [\href{http://arxiv.org/abs/1410.0711}{{\tt arXiv:1410.0711}}].

\bibitem{Comsa:2019rcz}
I.~M. Comsa, M.~Firsching, and T.~Fischbacher, {\it {SO(8) Supergravity and the
  Magic of Machine Learning}},  {\em JHEP} {\bf 08} (2019) 057,
  [\href{http://arxiv.org/abs/1906.00207}{{\tt arXiv:1906.00207}}].

\bibitem{Bobev:2019dik}
N.~Bobev, T.~Fischbacher, and K.~Pilch, {\it {Properties of the new $
  \mathcal{N} $ = 1 AdS$_{4}$ vacuum of maximal supergravity}},  {\em JHEP}
  {\bf 01} (2020) 099, [\href{http://arxiv.org/abs/1909.10969}{{\tt
  arXiv:1909.10969}}].

\bibitem{Klebanov:2008vq}
I.~Klebanov, T.~Klose, and A.~Murugan, {\it {AdS(4)/CFT(3) Squashed, Stretched
  and Warped}},  {\em JHEP} {\bf 03} (2009) 140,
  [\href{http://arxiv.org/abs/0809.3773}{{\tt arXiv:0809.3773}}].

\bibitem{Guarino:2019jef}
A.~Guarino, J.~Tarr\'\i{}o, and O.~Varela, {\it {Halving ISO(7) supergravity}},
   {\em JHEP} {\bf 11} (2019) 143, [\href{http://arxiv.org/abs/1907.11681}{{\tt
  arXiv:1907.11681}}].

\bibitem{Cesaro:2020soq}
M.~Ces{\`a}ro and O.~Varela, {\it {Kaluza-Klein fermion mass matrices from
  exceptional field theory and $ \mathcal{N} $ = 1 spectra}},  {\em JHEP} {\bf
  03} (2021) 138, [\href{http://arxiv.org/abs/2012.05249}{{\tt
  arXiv:2012.05249}}].

\bibitem{Gauntlett:2007ma}
J.~P. Gauntlett and O.~Varela, {\it {Consistent Kaluza-Klein reductions for
  general supersymmetric AdS solutions}},  {\em Phys.Rev.} {\bf D76} (2007)
  126007, [\href{http://arxiv.org/abs/0707.2315}{{\tt arXiv:0707.2315}}].

\end{thebibliography}\endgroup


\end{document}